\documentclass[aps,prd,reprint,groupedaddress,nofootinbib]{revtex4-1}
\usepackage{amsmath}
\usepackage{graphicx}
\usepackage{color}
\usepackage{bm}
\usepackage{comment}
\usepackage{hyperref}

\newcommand{\Mf}{M_{\rm f}}
\newcommand{\Qf}{Q_{\rm f}}

\newcommand{\Jf}{J_{\rm f}}
\newcommand{\eps}{\epsilon}
\newcommand{\calE}{{\cal E}}
\newcommand{\calL}{{\cal L}}
\newcommand{\calW}{{\cal W}}
\newcommand{\calR}{{\cal R}}
\newcommand{\Einf}{E_{\infty}}
\newcommand{\Linf}{L_{\infty}}

\newcommand{\beq}{\begin{equation}}
\newcommand{\eeq}{\end{equation}}

\definecolor{red  }{rgb}{1,0,0}
\definecolor{blue }{rgb}{0,0,1}
\definecolor{green}{rgb}{0,1,0}

\begin{document}


\title{Overspinning a Kerr black hole: the effect of self-force}


\author{Marta Colleoni and Leor Barack}
\affiliation{Mathematical Sciences, University of Southampton,
Southampton, SO17 1BJ, UK}


\date{\today}

\begin{abstract}
We study the scenario in which a massive particle is thrown into a rapidly
rotating Kerr black hole in an attempt to spin it up beyond its extremal
limit, challenging weak cosmic censorship. We work in black-hole
perturbation theory, and focus on non-spinning, uncharged particles sent
in on equatorial orbits. We first identify the complete parameter-space
region in which overspinning occurs when back-reaction effects from the
particle's self-gravity are ignored. We find, in particular, that
overspinning can be achieved only with particles sent in from infinity.
Gravitational self-force effects may prevent overspinning by radiating
away a sufficient amount of the particle's angular momentum
(``dissipative effect''), and/or by increasing the effective centrifugal
repulsion, so that particles with suitable parameters never get captured
(``conservative effect''). We analyze the full effect of the self-force,
thereby completing previous studies by Jacobson and Sotiriou (who
neglected the self-force) and by Barausse, Cardoso and Khanna (who
considered the dissipative effect on a subset of orbits). Our main
result is an inequality, involving certain self-force quantities, which
describes a necessary and sufficient condition for the overspinning
scenario to be overruled. This ``censorship'' condition is formulated on a
certain one-parameter family of geodesics in an extremal Kerr
geometry. We find that the censorship condition is insensitive to the
dissipative effect (within the first-order self-force approximation used
here), except for a subset of perfectly fine-tuned orbits, for which a
separate censorship condition is derived. We do not obtain here the
self-force input needed to evaluate either of our two conditions, but
discuss the prospects for producing the necessary data using
state-of-the-art numerical codes.
\end{abstract}


\maketitle

\section{Introduction}


The cosmic censorship conjecture \cite{wcc} has over the years become a cornerstone of classical general relativity. In its weak version it states, in essence, that curvature singularities arising in solutions to the Einstein's field equations must be cloaked behind event horizons, so that they are prevented from being in causal contact with distant observers. Despite being strongly motivated on physical grounds, the conjecture's precise extent of validity remains unclear. A notable counterexample involves finely tuned initial conditions \cite{chop}.
The formulation of the conjecture may be refined to exclude such examples \cite{waldrev}.


In a 1974 paper \cite{wald} Wald proposed a simple but powerful framework for testing weak cosmic censorship, using the gedanken experiment of a particle thrown into a Kerr--Newman black hole. If parameters can be chosen such that the post-capture mass $\Mf$, charge $\Qf$ and spin $\Jf$ satisfy $\Mf^2<(\Jf/\Mf)^2+\Qf^2$, then a naked singularity would presumably form, in direct violation of weak censorship. Whether the equations of classical general relativity permit such a process has since been subject of much investigation. It is usually assumed that the particle's energy and electric charge are much smaller than those of the black hole, which then places the problem within the realm of black-hole perturbation theory.


In \cite{wald} Wald showed that the over-extremality scenario is ruled out when the configuration is that of a pointlike test particle captured by an extremal Kerr--Newman black hole. Electrostatic and centrifugal repulsion, he showed, would prevent a particle carrying sufficient charge and/or angular momentum from entering the black hole. The same conclusion was shown to hold true also for a spinning test particle dropped from rest at infinity along the symmetry axis of an extremal Kerr black hole, with its spin aligned along the axis. In this case, it is the repulsion force from spin-spin coupling that prevents suitable particles from ever entering the black hole.

However, later work has demonstrated that over-extremality is achievable when the initial black hole is taken to be \textit{nearly} extremal---if back-reaction effects on the particle's trajectory are ignored. This was first shown by Hubeny \cite{hub} for a nearly extremal Reissner-Nordstr\"om black hole, and more recently by Jacobson and Sotiriou \cite{js} for a nearly extremal Kerr black hole (``overcharging'' and ``overspinning'' scenarios, respectively). The nearly extremal Kerr-Newman case was subsequently studied in Ref.\ \cite{Saa}. In all cases, all orbits identified as capable of driving the black hole beyond the extremal limit lie very close, in the relevant parameter space, to the separatrix between orbits that are captured by the black hole and ones that are scattered off it. In Hubeny's analysis of a radially falling electric charge, electrostatic repulsion only marginally fails to prevent the particle from falling into the hole: The particle's radial velocity upon crossing (what would have been) the event horizon is proportional to the ratio $\tilde\eta\ll 1$ between the particle's energy and the black hole's mass. The amount of post-capture excess charge, $\Qf-\Mf$, is found to be {\it quadratic} in $\tilde\eta$. Similarly, in Ref.\ \cite{js}'s analysis of equatorial-plane captures, overspinning particles clear the peak of the effective potential barrier with radial velocities $\propto\tilde\eta$, and the post-capture excess spin, $\Jf-\Mf^2$, is quadratic in $\tilde\eta$.

This suggests strongly that back-reaction effects cannot be ignored and may well change the outcome of the gedanken experiment. Heuristically, effects of the (electromagnetic and/or gravitational) self-force enter the analysis in two ways. First, the {\it dissipative} piece of the self-force continually removes some of the particle's energy and angular momentum, sending them to infinity and down the event horizon in gravitational waves. In the Kerr case, dissipative effects may accumulate as the particle ``hovers'' above the peak of the effective potential on a nearly circular orbit. Second, the {\it conservative} piece of the self-force might supply just the right amount of additional repulsive force to prevent would-be overcharging/overspinning particles from ever entering the black hole. 
For particles sent in from infinity in the Kerr case, this second effect may be formulated in terms of a shift in the critical impact parameter for capture: If the gravitational self-force (GSF) shifts the critical impact parameter inward by a sufficient amount (for a given energy-at-infinity), then would-be overspinning particles may end up being scattered away rather than captured. 

There have been several recent attempts to quantify the effect of back-reaction in the problem. Focusing on the Reissner-Nordstr\"om case, Isoyama, Sago and Tanaka \cite{soich} argued that the full effect can be properly taken into account by considering the quasi-equilibrium configuration of a charged particle placed precisely on the capture--scatter separatrix. An exact solution is known for this configuration---the static double Reissner-Nordstr\"{o}m spacetime---and the authors calculated that its total energy is always greater than its total charge. They have also established that radiative losses during the final plunge are negligible, hence concluding that (under the assumption that the true capture system does indeed go through a quasi-equilibrium state) the final configuration cannot be a naked singularity. 

In a later work, Zimmerman, Vega and Poisson \cite{zimm} took up the challenge of directly calculating the charged particle's trajectory including the full effect of the electromagnetic self-force. Analyzing numerically a large sample of orbits within the domain identified by Hubeny, the authors found no example of successful overcharging: All particles with a combination of charge and energy suitable for overcharging the black hole were found to be repelled  before reaching the horizon. This analysis, however, neglected the potentially important effect of back-reaction from the gravitational perturbation sourced by the particle's electromagnetic energy-momentum. A complete analysis would require calculation of the corresponding GSF, but techniques for calculating self-forces in the coupled problem are only now starting to be developed \cite{zimmerman,tlinz}.

In that respect, the Kerr setup provides a cleaner environment, in which the perturbative problem is purely gravitational (at the obvious cost of abandoning spherical symmetry). Barausse, Cardoso and Khanna \cite{bck1,bck2} studied the dissipative GSF effect in the Kerr overspinning problem, focussing on ultra-relativistic particles on equatorial orbits.
Using analytic arguments backed by a numerical calculation of the energy and angular-momentum carried away in gravitational waves, they showed that dissipation averts the overspinning for some but not all of Jacobson--Sotiriou's orbits. For sufficiently small $\tilde\eta$, the dissipative effect is always negligible and cannot prevent overspinning. 
This result highlights the importance of accounting for the {\em full} effect of GSF. To reach a definitive conclusion necessitates an actual calculation of the full local GSF acting on the captured particles.  

In the past few years, rigorous methods for GSF calculations in Kerr spacetime have advanced enough to allow a more systematic and complete treatment of the overspinning problem. The program initiated with this paper revisits the problem from this new vantage point. It seeks to obtain a more conclusive answer to the question of whether it is indeed the self-force that provides the mechanism by which black holes protect themselves from being overspun. 

Our current paper lays the necessary groundwork. Concentrating on equatorial orbits, we first identify the complete ``window'' in the parameter space in which overspinning occurs if the GSF is ignored. We then formulate a condition for this window to be eliminated by the effect of the full GSF. The condition takes the form of an inequality that is required to hold for each member of a certain 1-parameter family of geodesics, and it involves the GSF calculated along such orbits. Here we do not obtain the necessary GSF data, but we discuss methods for computing it (numerically) using existing codes. With collaborators we have began work to obtain the GSF data, and we intend to present the results in a follow-up paper. 

The rest of this introduction summarizes our analysis (also in relation to previous work) and describes its main results.

\subsection{This work: overview and results}


Jacobson and Sotiriou \cite{js} assumed that overspinning occurs if two conditions are met: (i) the geodesic trajectory of the test particle is timelike at the horizon, and (ii) $\Jf>\Mf^2$. The first
condition is very lax. It allows for low-energy orbits that are deeply bound to the black hole and confined to the immediate neighborhood of the horizon. The physics of such orbits becomes very subtle, especially when self-gravity and finite-size effects are included [consider that deeply-bound orbits below the innermost stable circular orbit (ISCO) plunge into the black hole within an amount of proper time that shrinks to zero in the extremal limit \cite{Jacobson}], so one would preferably avoid such orbits as candidates (see, however, Hod \cite{hod} for a heuristic treatment). 
Jacobson and Sotiriou acknowledge this issue, and to address it they supplement their analysis with two specific numerical examples of overspinning orbits that are sent in from afar. They stop short of determining the full range of overspinning orbits when deeply bound orbits are disallowed.

Our capture condition will be more stringent, and more in the spirit of Ref.\ \cite{zimm}: We will send our particle in from ``sufficiently far'' (this condition will be made precise in the next section), and deem it ``captured'' if it has no inner radial turning point outside the black hole. Thus, for a legitimate capture we demand that the particle ``clears'' the peak of the effective potential on its inward journey. 



In Sec.\ \ref{Sec:geodesics} we revisit the overspinning problem in the geodesic approximation. We identify the precise region in the parameter space of equatorial orbits around a nearly-extremal Kerr black hole (excluding deeply bound orbits) where overspinning occurs, and give analytic expressions for the boundaries of that region. The overspinning window is illustrated in Figs.\ \ref{bubble} and \ref{elos}. Perhaps unexpectedly, we find that only particles sent in from infinity are capable of overspinning the black hole. This fact has somehow gone unnoticed in previous work, to the best of our knowledge. We find that for any given value of the particle's energy at infinity, there exists an open range of orbital angular momenta and particle's rest masses for which overspinning occurs. That only orbits coming from infinity are potential overspinniners is somewhat fortuitous, because for such orbits it is straightforward to identify the system's total [Arnowitt--Deser--Misner (ADM)] energy and angular momentum even when GSF effects are included.



We then turn to analyze the GSF effect. In Sec.\ \ref{Sec:GSF} we first review essential results from GSF theory, and then discuss the determination of the ``critical orbit'' that separates (in the relevant parameter space) between plunging and scattered orbits. We do this in two steps. First, we ignore the dissipative piece of the GSF, and calculate the correction due to the conservative GSF to the critical value of the angular momentum for a fixed value of the energy-at-infinity. Then we restore dissipation and consider its consequences. Under the full GSF, all critical orbits merge into a ``global attractor'' that takes the system adiabatically along a sequence of quasi-circular unstable orbits ending at the ISCO, before plunging into the black hole. By fine-tuning the initial value of the angular momentum (for a given initial energy), an orbit can be made to evolve arbitrary far along the global attractor. We make a formal distinction between ``generic'' and ``fine-tuned'' captured orbits, based on how the difference between the initial and final values of the particle's specific energy scales with the particle's mass $\mu$ (in a procedure whereby $\mu$ is taken to zero while holding fixed the initial specific energy and angular momentum). {\it Generic} orbits are ones for which that difference vanishes for $\mu\to 0$ (this includes, for example, all of the orbits considered in Refs.\ \cite{bck1,bck2}); {\it fine-tuned} orbits are ones for which the difference does not vanish even for $\mu\to 0$. 


With this preparatory work in place, we move on, in Sec.\ \ref{Sec:OSwSF}, to obtain the overspinning condition as modified by the full GSF. The end result are two inequalities, one for generic orbits [Eq.\ (\ref{OSfinal})] and another for fine-tuned ones [Eq.\ (\ref{OSfinalFT})], which describe conditions for overspinning to be averted under the effect of the full GSF. In the generic case, the condition involves only the conservative piece of the GSF, evaluated along critical geodesics in the extremal Kerr limit. Overspinning can be ruled out if and only if the condition is met for each member of this one-parameter family. The condition for fine-tuned orbits requires, in addition to the conservative GSF, also knowledge of the fluxes of energy and angular momentum radiated to infinity by particles on unstable circular orbits, in the extremal Kerr limit.  Overspinning can be ruled out if and only if the condition is met for any values of the initial and final energies. 



In Sec.\ \ref{Sec:redshift} we propose an alternative form of the overspinning conditions, based on the framework of the ``first law of binary black-hole mechanics'', as recently applied to orbits in Kerr \cite{isoyama14}. The alternative form, given (for the generic case) in Eq.\ (\ref{OSfinalZ}), involves only perturbative quantities calculated along (unstable) circular orbits, which should be more easily computable with existing GSF codes. This simplified formulation relies on explicit expressions, given in \cite{isoyama14}, for the ADM-like energy and angular momentum of circular-orbit configurations in Kerr, including leading-order self-interaction terms. The underlying theoretical framework is yet to be firmly established and tested, however, so we regard the simplified condition (\ref{OSfinalZ}) as somewhat less rigorous than our direct condition (\ref{OSfinal}). Our stance is that it would be desirable to evaluate both forms of the condition, for the sake of establishing confidence in the result. 

In Sec.\ \ref{Sec:num} we discuss the numerical input required for evaluating our overspinning conditions, and the prospects for obtaining it through adaptation of existing codes. Evaluation of the direct conditions (\ref{OSfinal}) and (\ref{OSfinalFT}) involves GSF calculations along unbound orbits on a nearly-extremal Kerr background, which has not been attempted so far. However, we think the basic computational infrastructure for such calculations is well in place. 

Section \ref{Sec:conclusions} summarizes our results and speculates on what a numerical evaluation of our censorship conditions might yield. The Appendices contain some of the details of calculations done in Secs.\ \ref{Sec:GSF} and \ref{Sec:OSwSF}: a derivation of the ADM energy and angular momentum for the system under consideration; a calculation of the GSF-induced shift in the critical value of the angular momentum; and an evaluation of radiative loses during the final plunge into the black hole. 


Throughout this paper we set $G=c=1$ and use the metric signature $(-,+,+,+)$.

\section{Overspinning orbits in the geodesic approximation}\label{Sec:geodesics}

Our initial configuration features a Kerr black hole of mass $M$ and angular momentum $J=a\,M<M^2$. A pointlike test particle of rest mass $\mu\ll M$ is sent in on a geodesic of the background Kerr geometry. As in \cite{js}, we restrict attention to prograde orbits in the equatorial plane, so that the orbital angular momentum is aligned with the spin of the black hole. (Intuitively, this configuration seems most favourable for a successful overspinning.) 
We denote the particle's {\em specific} energy and angular momentum by $E$ and $L$, respectively; these are constants of the geodesic motion. For the geodesic approximation to make sense, we must assume $\mu E\ll M$ and $\mu L\ll J$. Then, clearly, overspinning could only be possible, in principle, if the black hole is nearly extremal. We write
\begin{equation}\label{eps}
a/M=1-\eps^2,
\end{equation}
where $\eps\ll 1$.\footnote{Note, to avoid confusion, that \cite{js} has instead $a/M=1-2\epsilon^2$.}


Below we study the overspinning scenario in the above setup, but we begin with a survey of some essential properties of timelike equatorial geodesics of the Kerr metric.

\subsection{Relevant results for Kerr geodesics}

Let $u^{\alpha}$ denote the particle's four-velocity. In Boyer-Lindquist coordinates $\{t,r,\theta,\phi\}$ we have $u^\theta\equiv 0$, and 
\begin{equation}
\dot{u}_{t}=0,\ \ \ \ \dot{u}_{\phi}=0,
\end{equation}
where an overdot denotes differentiation with respect to proper time. The two equalities express the conservation of energy $E=-\xi_{(t)}^{\alpha}u_{\alpha}=-u_{t}$ and angular momentum $L=\xi_{(\phi)}^{\alpha}u_{\alpha}=u_{\phi}$, where $\xi_{(t)}^{\alpha}:=\partial_{t}^{\alpha}$ and $\xi_{(\phi)}^{\alpha}:=\partial_{\phi}^{\alpha}$ are Killing vectors associated with the time-translation and rotational symmetries of the Kerr background. The pair $\{E,L\}$ parametrizes the family of equatorial geodesics (up to initial conditions).

The normalization $u_{\alpha}u^{\alpha}=-1$ now gives the radial equation of motion, which we write in the form
\begin{eqnarray}
\label{rdot}
\dot{r}^{2}&=&B(r)\left[{E}-V_{-}({L},{r})\right] \left[{E}-V_{+}({L},{r})\right].
\end{eqnarray}  
Here $r$ is the Boyer-Lindquist radius of the orbit,
$B(r):=1+ a^2(r+2M)/r^3$, and (for $M a L\ne 0$)
\begin{equation}
\label{vplus}
V_{\pm}({L},{r}):=\frac{2 M {a} {L}}{B {r}^3}\left( 1\pm\sqrt{1+\frac{ B {r}^3 [L^2(r-2M)+{r}\Delta]}{4M^2 a^2 L^2}}\right)\,,
 \end{equation} 	
with 
$\Delta:=r^2-2Mr+{a}^2$.
For prograde orbits, the potential $V_{-}$ is manifestly negative definite, so the factor $B(r)(E-V_{-})$ in Eq.\ (\ref{rdot}) is manifestly {\it positive} definite.
Thus, $V_{+}$ plays the role of an effective potential for the radial motion, which is allowed for $E\geq V_{+}(L,{r})$, with an equality identifying radial turning points.

Stationary points of $V_{+}(L,r)$ outside the black hole, when they exist, correspond to circular orbits. 
These satisfy the simultaneous conditions  
\begin{equation} \label{circ}
E=V_+, \quad\quad \partial_r V_+=0 \quad\text{(circular orbits)}.
\end{equation}
Substituting from Eq.\ (\ref{vplus}) and solving for $E$ and $L$ in terms of the circular-orbit radius, $r=R$, gives $E=E_c(R)$ and $L=L_c(R)$, with 
\begin{align}
\label{Ec}
E_{c}(R)&=\frac{1-2 \tilde{R}^{-1}+\tilde a\tilde{R}^{-3/2}}{\sqrt{1-3\tilde{R}^{-1}+2\tilde{a}\tilde{R}^{-3/2}}}\,,\\ 
\label{Lc}
\tilde L_{c}({R})&=\frac{\tilde{R}^{1/2}(1-2 \tilde{a} \tilde{R}^{-3/2}+\tilde{a}^2\tilde{R}^{-2})}{\sqrt{1-3\tilde{R}^{-1}+2\tilde{a}\tilde{R}^{-3/2}}}.
\end{align}
Here an overtilde denotes a-dimensionalization using $M$, i.e., $\tilde{R}:=R/M$, $\tilde{a}:=a/M$ and $\tilde{L}:=L/M$; we shall adopt this notation throughout the rest of the paper.
Timelike circular orbits exist only for $R>R_{\rm ph}(a)$, the radius of a photon's unstable circular orbit (``light ring''). $R_{\rm ph}(a)$ is the (unique) root of ${1-3\tilde{R}^{-1}+2\tilde{a}\tilde{R}^{-3/2}}$ greater than the event horizon's radius, $\tilde{R}_{\rm eh}(a)=1+(1-\tilde a^2)^{1/2}$. 
The angular velocity $\Omega:=u^\phi/u^t$ of any circular geodesic orbit reads
\begin{equation}\label{Omega}
\tilde\Omega(R)=(\tilde a+\tilde R^{3/2})^{-1}.
\end{equation}

The number of stationary points of $V_+$ and their location depend on $L$. There are none outside the black hole when $L$ is below a certain critical value $L_{\rm isco}(a)$, and there are two for $L>L_{\rm isco}(a)$: a maximum representing an unstable circular orbit, and, further out, a minimum representing a stable one. The critical value $L_{\rm isco}(a)$ marks the innermost stable circular orbit (ISCO). It is given by $L_{\rm isco}=L_c(R_{\rm isco})$, where the ISCO radius $R_{\rm isco}$ is found
by solving Eqs.~(\ref{circ}) simultaneously with $\partial^{2}_{{r}}V_+(L,r)=0$. 
The ISCO may also be said to represent the outer boundary of the region of unstable circular orbits. 

The radii of unstable circular geodesic orbits span the interval $R_{\rm ph}(a)<R<R_{\rm isco}(a)$. This 1-parameter family of orbits will feature dominantly in our analysis, because it defines the capture--scatter threshold where much of the relevant physics occurs. Members of the family may be parametrized by either $E$ or $L$, both being monotonically decreasing functions of $R$ between $R_{\rm ph}$ (where $E,L\to\infty$) and $R_{\rm isco}$ for any $\tilde{a}<1$. This monotonicity can be readily established from Eqs.\ (\ref{Ec}) and (\ref{Lc}). Hence, the radius $R$ itself is also a valid parameter.

To each unstable circular orbit there correspond non-circular homoclinic-like geodesic orbits \cite{Levin} that join the circular orbit asymptotically in either their infinite past or their infinite future, or both. Nearly-homoclinic orbits exhibit a ``zoom-whirl'' behavior \cite{glamp}: an episode of prolonged rotation (``whirl'') about the location of the associated unstable circular orbit. We will see that all orbits relevant to the overspinning problem fall in that category. 
Based on the correspondence with homoclinic orbits, unstable circular orbits may be divided into ``bound'' ($E<1$) and ``unbound'' ($E\geq 1$). The radius of the innermost bound circular orbit (IBCO) is obtained by solving $E_c(R)=1$, giving $\tilde R_{\rm ibco}=[1+(1-\tilde{a})^{1/2}]^2$. 

\begin{figure}
  \begin{center}
    \includegraphics[scale=0.46]{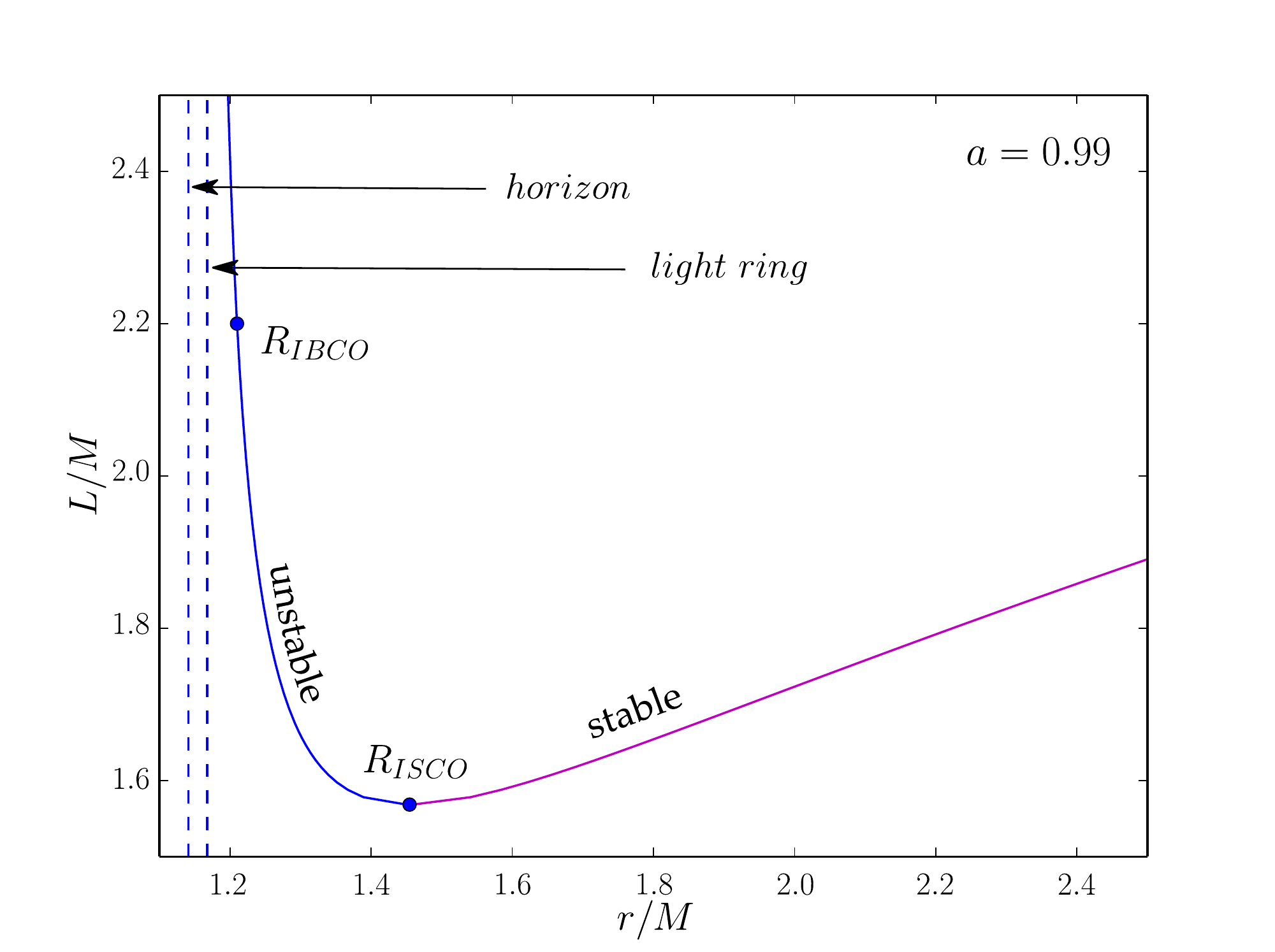}
    \caption{\label{radii} 
    Timelike circular equatorial geodesics around a nearly extremal Kerr black hole, shown here for $a=0.99M$. The plot shows specific angular momentum versus Boyer-Lindquist radius. Orbits with $r>R_{\rm isco}$ (magenta) are stable, while these with $r<R_{\rm isco}$ (blue) are unstable. Also indicated are the innermost bound circular orbit (IBCO, $E=1$) and the photon orbit (``light ring'', $E,L\to\infty$). In the extremal limit, $a\to M$, the radii $R_{\rm isco}$, $R_{\rm ibco}$ and $R_{\rm ph}$ all coincide with the horizon radius $R_{\rm eh}$. } 
  \end{center}
\end{figure}

Figure \ref{radii} illustrates the range of stable and unstable circular orbits, and the location of the various special orbits mentioned, in a particular example ($\tilde a=0.99$). We note the ordering 
\begin{equation}\label{ordering}
R_{\rm eh}<R_{\rm ph}<R_{\rm ibco}<R_{\rm isco},
\end{equation}
which applies for any $\tilde a<1$.

Let us now specialize to a near-extremal Kerr background with spin as in Eq.\ (\ref{eps}). One finds 
\begin{align}
\label{reh}
\tilde{R}_{\rm eh}&=1+\sqrt{2}\, \epsilon+O(\epsilon^2)\,,\\
\label{rph}
\tilde{R}_{\rm ph}&=1+\sqrt{8/3}\, \epsilon+O(\epsilon^2)\,,\\
\label{ribco}
\tilde{R}_{\rm ibco}&=(1+\eps)^2 \quad \text{(exact)}\,,\\
\label{risco}
\tilde{R}_{\rm isco}&=1+\left(2\epsilon\right)^{2/3}+O(\epsilon^{4/3})\,.
\end{align}
The function $E_c(R)$ in Eq.\ (\ref{Ec}) can be inverted perturbatively in $\eps$ to obtain the radius of an arbitrary unstable circular orbit in terms of its energy $E$. We find
\begin{equation}
\label{rhopar}
\tilde{R}= 1+\eps \rho_1(E) +\eps^2 \rho_2(E) +O(\eps^3) ,  
\end{equation}
where the first two coefficients, needed below, read
\begin{equation}
\label{rho12}
\rho_1= \frac{2\sqrt{2}E}{\sqrt{3E^2-1}}, 
\quad\quad
\rho_2= \frac{2(2E^4-E^2+1)}{(3E^2-1)^2}.
\end{equation}
Equation (\ref{ribco}) is the special case of (\ref{rhopar}) with $E=1$, giving $\rho_1=2$ and $\rho_2=1$.

It follows that, in the extremal limit $\eps\to 0$, the Boyer-Lindquist radii of the light-ring and the ISCO both coincide with the horizon radius, and so do the radii of all unstable circular orbits enclosed between them. 
Also peculiar is the fact that the ratio of coordinate differences $(\tilde{R}_{\rm isco}- \tilde{R}_{\rm eh})/(\tilde{R}_{\rm ibco}- \tilde{R}_{\rm eh})$ diverges as $\eps\to 0$.
A closer look reveals \cite{bard} that the light ring, IBCO and ISCO remain separated from the horizon, and from each other, when examined on a Boyer-Lindquist $t$=const slice: On that slice, the proper radial distance between the light-ring and the horizon is finite, and so is the distance between any fixed-$E$ unstable circular orbit and the light ring. The proper radial distance between the ISCO and any fixed-$E$ unstable circular orbit {\em diverges} on the $t$=const slice; 
the geometry of the $t$=const hypersurface appears to ``stretch'' infinitely around the ISCO location \cite{bard}. The situation, however, is rather different when examined on a horizon-crossing time slice. As emphasized recently by Jacobson \cite{Jacobson}, on any such slice, the light ring, IBCO and ISCO all actually coincide with horizon generators. From that perspective, they---and all unstable orbits in between them---are ``at the same place'' in the extremal limit.

These subtleties will not affect our analysis directly: $\eps$ will be kept small but nonzero, and the strict ordering (\ref{ordering}) will therefore apply on any time slice. However, we must take note of the degeneracy of $R$ as a parameter for unstable circular orbits when $\eps\to 0$. The energy $E$, on the other hand, remains a good parameter even in this limit, spanning the entire range $\infty>E>\frac{1}{\sqrt{3}}$. We will thus generally adopt $E$ for labelling unstable circular orbits. 
Given $E$, the angular momentum $L_c(R(E))$, which we henceforth write as $L_c(E)$, is obtained by substituting Eqs.\ (\ref{eps}) and (\ref{rhopar}) in Eq.\ (\ref{Lc}) and then expanding in $\eps$. The result is 
\begin{equation}\label{LofE}
\tilde L_c(E)=2E+(6E^2-2)^{1/2} \eps +O(\eps^2).
\end{equation}
We note that to determine the $O(\eps)$ term here required the explicit values of both $\rho_1$ and $\rho_2$ of Eq.\ (\ref{rho12}).

\subsection{Exclusion of deeply bound orbits}

Heuristically, if we assume our point particle represents a compact object---say, a Schwarzschild black hole---then its effective proper ``diameter'' is $\sim\mu$. Below it will become clear that a successful overspinning requires $\mu\sim\eps$, and so relevant objects have proper diameters  $\sim\eps$. Now consider placing such an object in a deeply bound orbit with an {\em outer} turning point at $r<R_{\rm isco}$ [and with $L>L_c(E)$]. Such an object (it can be checked) will plunge through the horizon within a proper time of $O(\eps)$ (at most), comparable to its own ``light-crossing time''. It is not clear whether the object can be made to initially ``fit'' in its entirety outside the hole. At the very least, it is not clear if the simple model of a point particle and a stationary horizon provides a faithful description of the physics in this case.

To avoid such subtleties, we wish to exclude deeply bound orbits from our analysis. 
We achieve this by requiring that, if the orbit possesses an outer radial turning point at some $r=r_{\rm out}$, then 
\begin{equation}\label{nodeep}
r_{\rm out}> R_{\rm isco}(\eps).
\end{equation}
It can be checked that, under this condition, the proper-time interval along any timelike  equatorial geodesic connecting $r=r_{\rm out}$ to $r=R_{\rm eh}$ is finite (nonzero) even in the limit $\eps\to 0$ (taken with fixed $E,L$). 
The condition (\ref{nodeep}) demands that eligible 
particles must clear the peak of the effective potential (when such a peak exists) as they plunge into the black hole.

\subsection{Overspinning domain}

Given the restriction (\ref{nodeep}), a necessary and sufficient condition for a falling particle of specific energy $E$ to be captured by the black hole is 
\begin{equation}\label{capture}
L<L_c(E).
\end{equation}
A captured particle would overspin the black hole if and only if
\begin{equation}\label{OScondition}
(M+\mu E)^2<a M+\mu L.
\end{equation}
Using $\tilde a=1-\eps^2$ and introducing the small mass ratio 
$\eta:=\mu/M$, this condition becomes 
\begin{equation}
\label{OScond}
\epsilon^2+\eta W+\eta^2 E^2<0\,,
\end{equation}
where we have introduced\footnote{Heuristically, $W/2$ may be interpreted as the specific energy in a co-rotating frame with $\tilde\Omega=1/2$, i.e., the common angular velocity of all unstable circular geodesics in the extremal limit.}
\begin{equation}\label{calE}
W:=2E-\tilde{L}.
\end{equation}
Note that Eq.\ (\ref{OScond}) sets a lower bound on $L$ (for given $E,\eta,\eps$), while Eq.\ (\ref{capture}) sets an upper bound. Also note that Eq.\ (\ref{OScond}) implies the necessary condition $W<0$ for overspinning to occur.

Our goal now is to identify the complete domain in the space of $\{\eta,E,L\}$ for which the conditions (\ref{capture}) and (\ref{OScond}) are simultaneously satisfied, assuming $\eps\ll 1$ and the condition (\ref{nodeep}). For easy reference, let us call this domain ``OS'', for ``overspinning''. 

We first show that orbits with $L\leq L_{\rm isco}$ all fall {\it outside} OS. To this end, consider first the ISCO itself, where $W=2E_{\rm isco}-\tilde L_{\rm isco}=:W_{\rm isco}$. Using Eqs.\ (\ref{Ec}), (\ref{Lc}) and (\ref{risco}) we obtain $W_{\rm isco}= -\hat{c}\eps^{4/3}+O(\eps^2)$, where $\hat{c}=2^{1/3}\sqrt{3}>0$. Thus, $W_{\rm isco}$ is negative as required, but it can be easily checked that (\ref{OScond}) is always violated for sufficiently small $\eps$: Replacing $W\to -\hat{c}\eps^{4/3}$ in Eq.\ (\ref{OScond}) and considering the left-hand side as a quadratic function of $\eta$, we find this function is positive definite for any $\eps<(2E_{\rm isco}/\hat c)^3$. [Since $E_{\rm isco}$ is bounded from below by $E_{\rm isco}(\eps=0)=\frac{1}{\sqrt{3}}$, we find that (\ref{OScond}) is always violated for $\eps<\frac{4}{27}$.]
This rules out the ISCO itself, and it immediately rules out also all orbits with $\{E>E_{\rm isco},L=L_{\rm isco}\}$, for which $W>W_{\rm isco}$. Orbits with $\{E<E_{\rm isco},L_{\rm isco}\}$ {\it can} potentially satisfy Eq.\ (\ref{OScond}), but they are always deeply bound in the sense of failing to satisfy Eq.\ (\ref{nodeep}): For any $E<E_{\rm isco}$, the orbit has an outer radial turning point at a radius $r_{\rm out}<R_{\rm isco}$.

The upshot is that orbits with $L=L_{\rm isco}$ are all outside OS. For orbits with $L<L_{\rm isco}$ we would need to require $E<E_{\rm isco}$ in order for $W$ to be sufficiently negative. But, once again, such orbits are excluded on account of their being deeply bound. We conclude that orbits with $L\leq L_{\rm isco}$ are all outside OS.




Let us focus therefore on orbits with $L>L_{\rm isco}$. For such an orbit to be in OS, we require that (given $E,\eta,\eps$) $L$ is bounded from above by $L_c(E)$ and simultaneously from below via Eq.\ (\ref{OScond}):
\begin{equation}
\label{Lrange}
\epsilon^2+2\eta E+\eta^2 E^2< \eta\tilde{L} < \eta\tilde L_c(E;\eps). 
\end{equation}
We have made here the $\eps$ dependence of $L_c$ explicit, for clarity.
The span of the permissible range is $\eta\Delta_L:=-\epsilon^2-\eta[2E-\tilde L_c(E;\eps)]-\eta^2 E^2$, or, using Eq.\ (\ref{LofE}),
\begin{equation}
\label{Delta_L}
\eta\Delta_L=-\epsilon^2+\eta\eps \sqrt{6E^2-2}-\eta^2 E^2,
\end{equation}
where we have omitted terms of $O(\eta\eps^2)$.
OS is populated if and only if we can find $E,\eta,\eps$ for which
$\Delta_L>0$.


A few conclusions can be drawn immediately. First, considering $\eta\Delta_L$ in Eq.\ (\ref{Delta_L}) as a quadratic function of $\eta$, we find it has a maximum value 
\begin{equation}\label{maxDeltaL}
\max_\eta \eta\Delta_L = \frac{\eps^2(E^2-1)}{2 E^2}.
\end{equation}
This is positive only for $E>1$. Therefore, all orbits with $E\leq 1$ fall outside OS. {\it Bound orbits cannot overspin.}

Second, for any $E>1$, we can obtain $\Delta_L>0$ by choosing the mass ratio $\eta$ from within the interval 
\begin{equation}\label{mrange}
\eps \eta_-(E) <\eta <  \eps \eta_+(E),
\end{equation}
where
\begin{equation}\label{etaplusminus}
\eta_{\pm}=\frac{1}{\sqrt{2}\, E^2}\left[\sqrt{3E^2-1}\pm \sqrt{E^2-1}\right].
\end{equation}
In other words, {\it overspinning can be achieved for any $E>1$, as long as $\eta$ satisfies (\ref{mrange})}. Since the condition $\Delta_L>0$ is both necessary and sufficient, the converse also holds: All orbits in OS satisfy $E>1$ with Eq.\ (\ref{mrange}). 

\begin{figure}
\includegraphics[scale=0.45]{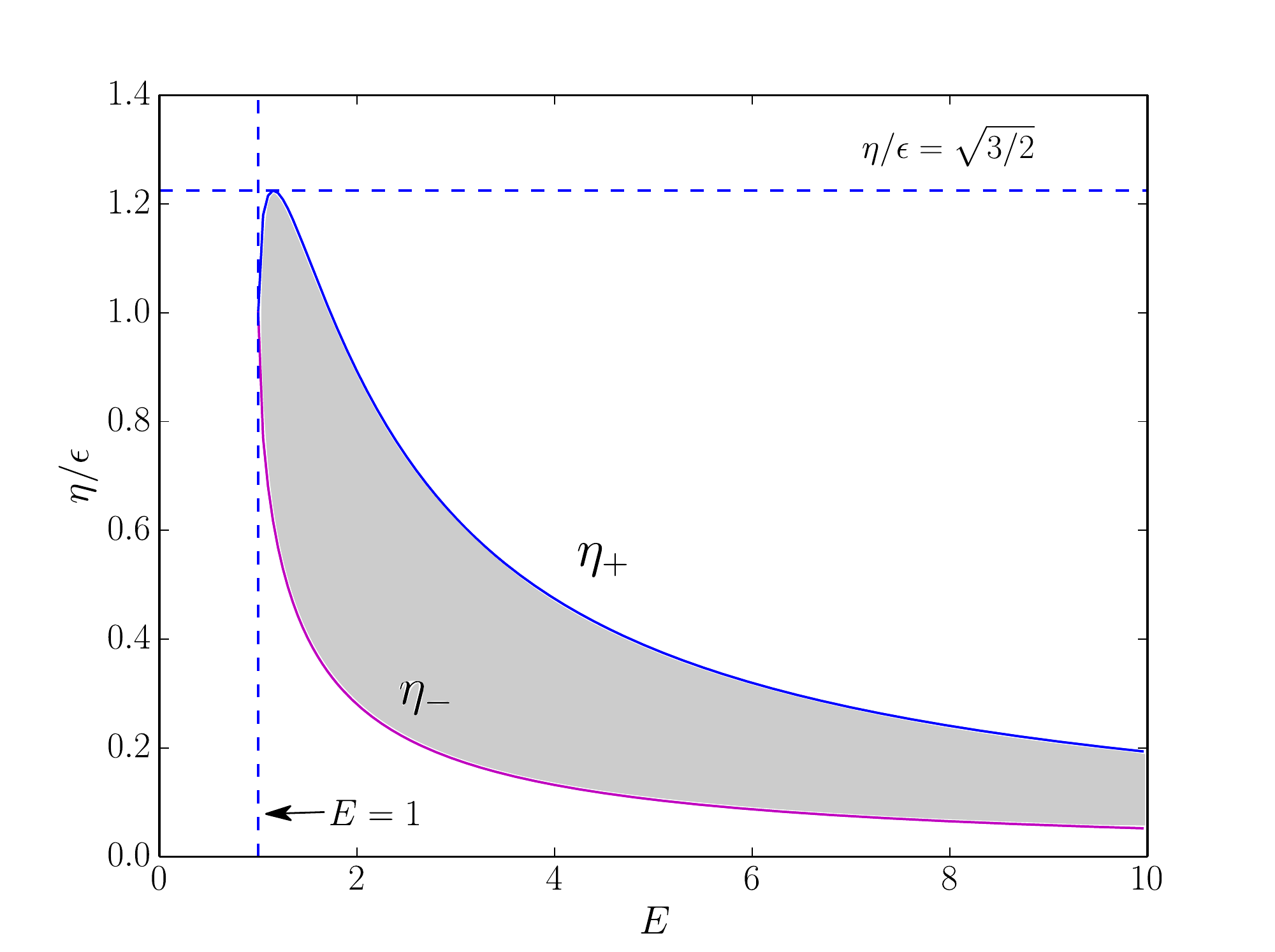}
\caption{\label{bubble} 
Domain of mass ratios $\eta$ for which overspinning is possible in the geodesic approximation. $\eta$ is shown divided by the near-extremality parameter $\eps=(1-a/M)^{1/2}$, and $E$ is the particle's specific energy. The boundaries $\eta_{\pm}(E)$ are given in Eq.\ (\ref{etaplusminus}). Overspinning is not possible for $E<1$ or $\eta>\sqrt{3/2}\,\eps$. However, for any value $E>1$ there is a range of $\eta$ for which the black hole may be overspun. This happens if the particle's angular momentum is chosen from within the range indicated in Eq.\ (\ref{calErange}).
}
\end{figure}

Third, from Eq.\ (\ref{mrange}) it follows that $\eta$ must be chosen to be of $O(\eps)$ (assuming $E\ll 1/\eps$). One can check that $\eta_+$ has a maximal value of
\begin{equation}
\max_E \eta_+ = \sqrt{3/2},
\end{equation}
obtained for $E=2/\sqrt{3}$. Therefore, {\it the range $\eta\geq \sqrt{3/2}\, \eps$ lies outside OS}.
The bandwidth of admissible mass ratios, for given $E$ and $\eps$, is 
\begin{equation}\label{Deltaeta}
\Delta_\eta:=\eps\eta_{+}-\eps\eta_{-}=\eps\sqrt{2(E^2-1)}/E^2,
\end{equation} 
which is maximal for $E=\sqrt{2}$.
Figure \ref{bubble} depicts the permissible range of $\eta/\eps$ as a function of $E$. 

Fourth, from Eqs.\ (\ref{maxDeltaL}) and (\ref{Deltaeta}) we learn that an $E=\text{const}(>1)$ slice of OS has maximal dimensions $\Delta_L\sim \eps^2/\eta\sim\eps$ and $\Delta_\eta\sim\eps$. OS is thus a narrow ``tube'' in the $\{E,L,\eta\}$ parameter space, of a cross section $\sim\eps\times\eps$, whose boundary is tangent to the surface of unstable circular orbits, $L=L_c(E)$. 

\begin{figure}
\includegraphics[scale=0.45]{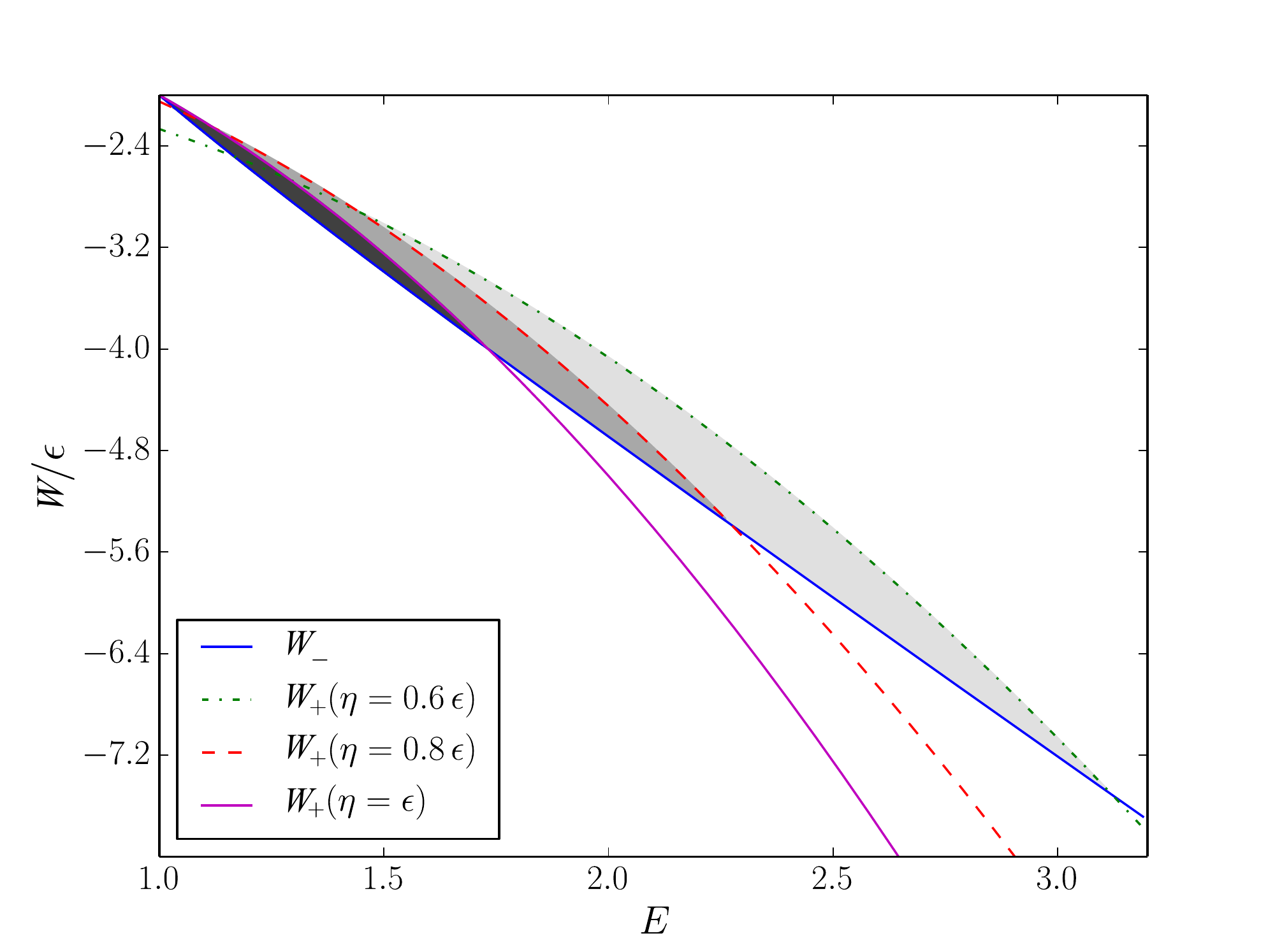}
\caption{\label{elos} 
The overspinning window, shown in the plane of $E,W$ (where  $W = 2E-L/M$) for several values of $\eta/\epsilon$. Note $W$ is shown divided by $\eps$. The boundaries $W_{\pm}$ are given in Eq.\ (\ref{calEmp}). The lower boundary $W_{-}(E)$ (which does not depend on $\eta$) arises from the requirement that the particle is captured by the black hole. The upper boundary $W_{+}(E,\eta/\eps)$ comes from the requirement that the final object is an over-extremal black hole. Overspinning is possible with any $E>1$, provided $\eta$ is chosen from within the range shown in Eq.\ (\ref{mrange}).
}
 \end{figure}

To summarize, we have found that OS is a narrow tube-like region of the $\{E,L,\eta\}$ space, described by $E>1$, $L_c(E;\eps)-\Delta_L(E,\eta;\eps)<L<L_c(E;\eps)$ and $\eps\eta_-(E)<\eta<\eps\eta_+(E)$, where $\Delta_L$ and $\eta_{\pm}$ are given in Eqs.\ (\ref{Delta_L}) and (\ref{etaplusminus}), respectively. A neater description of the OS window is obtained in terms of the quantity $W$ defined in Eq.\ (\ref{calE}): Rearranging Eq.\ (\ref{Lrange}) and using (\ref{LofE}), we find
\begin{equation}\label{calErange}
\eps W_-(E) < W <\eps W_+(E,\eta/\eps), 
\end{equation}
where
\begin{equation}\label{calEmp}
W_- = -\sqrt{6E^2-2}, \quad\ \ 
W_+ = -\left(\frac{\eps}{\eta} +\frac{\eta}{\eps}E^2\right).
\end{equation}
This domain is illustrated in Figure \ref{elos} for a sample of $\eta/\eps$ values.
To overspin a black hole of given $M$ and $\eps\ll 1$, one should pick an $E$ greater than $1$, choose any $\eta$ from within the interval (\ref{mrange}), and then choose $W$ (hence $L$) from within the interval (\ref{calErange}).


\section{Self-force preliminaries}\label{Sec:GSF}

Because the width of the overspinning window is of $O(\eta)$, self-gravity effects may potentially close this window, and they must therefore be included in the analysis.  Specifically, the GSF modifies the capture condition (\ref{capture}) by changing the functional relation $L_c(E)$ at $O(\eta)$. It also modifies the overspinning condition (\ref{OScondition}) by dissipating away some of the system's initial energy and angular momentum. In this section we introduce relevant results from the theory of self-forced motion. In Sec.\ \ref{Sec:OSwSF} we will then use these results to derive conditions for capture and overspinning under the full GSF effect.

\subsection{Equation of motion with self-force}

There now exists a rigorous formulation of the equations of motion for compact objects in curved spacetime, valid through first post-geodesic order in perturbation theory---see \cite{Gralla,Pound:2009sm,Harte, Gralla} and references therein, and \cite{Poisson,Harte:2014wya} for recent reviews. The formulation applies in situations where all lengthscales associated with the compact object are much smaller than the typical curvature radius of the background geometry. The motion of the compact object is then determined via a systematic procedure of matched asymptotic expansions, and interpreted as an accelerated motion in the background spacetime, subject to an effective GSF ($\propto\eta^2$).  One of the results is that the internal structure of the object does not affect the self-acceleration at $O(\eta)$ (except, if the particle is spinning, through the familiar Mathisson--Papapetrou spin term).

The GSF formalism should be applicable in our setup, since we work under the assumption $\mu E\ll M$. The introduction of the small background-related parameter $\eps$ should not pose a problem, because the background's curvature radius remains much larger than $\mu E$ even in the limit $\eps\to 0$, and even as the particle approaches the horizon. We will indeed proceed under the assumption that the standard first-order GSF formalism is applicable anywhere along the particle's trajectory until it crosses the horizon. 

The equation of motion, including the leading-order GSF, may be written in the form
\begin{equation}\label{EOM}
\mu \hat u^{\beta}\nabla_{\beta}\hat u^{\alpha}=F^{\alpha}.
\end{equation}
Here $\hat u^{\alpha}$ is the particle's four-velocity, tangent to the (accelerated) trajectory in the background spacetime (Kerr, in our case) and normalized using 
\begin{equation}\label{normalization}
g_{\alpha\beta}\hat u^{\alpha}\hat u^{\beta}=-1,
\end{equation}
where $g_{\alpha\beta}$ is the background (Kerr) metric. The covariant derivative in (\ref{EOM}) is taken with respect to $g_{\alpha\beta}$, and $F^{\alpha}$ is the first-order GSF, proportional to $\mu^2$. The GSF is normal to the four-velocity, $g_{\alpha\beta}\hat u^{\alpha}F^{\beta}=0$, so that the rest mass $\mu$ remains constant.
Methods to compute $F^{\alpha}$ in Kerr spacetime are reviewed in \cite{barack}. In should be noted that $F^{\alpha}$ itself is a gauge-dependent notion: A full, gauge-invariant information about the motion is contained only in the combination of the GSF and the metric perturbation with which it is associated  \cite{gsfandgt}.



Now consider a particle sent in along the equator of the Kerr black hole, i.e.~with $\theta=\pi/2$ and $\hat {u}^{\theta}=0$ at the initial moment, where hereafter $\tau$ is proper time along the self-accelerated orbit in $g_{\alpha\beta}$.  In any reasonable gauge, the component $F^{\theta}$ would vanish from symmetry and the motion will remain equatorial. 
Let us then define 
\begin{equation}
\label{E&L}
\hat E:=-\hat u_{t}, \quad\quad \hat L:=\hat u_{\phi},
\end{equation}
in analogy with $E$ and $L$ of the geodesic case. Here, however, $\hat E$ and $\hat L$ are not constants of the motion. Rather, Eq.\ (\ref{EOM}) tells us they evolve (slowly) according to
\begin{equation}
\label{E&Ldot}
\mu\frac{d\hat{E}}{d\tau}=-{F}_{t}, \quad\quad \mu\frac{d\hat L}{d\tau}={F}_{\phi},
\end{equation}
where $F_{\alpha}=g_{\alpha\beta}F^{\beta}$.
With these definitions, Eq.\ (\ref{normalization}) produces the radial equation of motion 
\begin{equation}
\label{rdotGSF}
\dot{r}^{2}=B(r)\left({\hat E}-V_{-}(\hat{L},{r})\right) \left(\hat{E}-V_{+}(\hat{L},{r})\right),
\end{equation}  
whose form is identical to that of Eq.\ (\ref{rdot})---except that here $\hat E$ and $\hat L$ are slow functions of $\tau$ along the orbit. 


The results of the previous section lead us to focus attention on particles sent in from infinity, i.e., ones with $r(\tau\to-\infty)\to\infty$. For such particles, we define 
\begin{equation}
\Einf:=\hat E(\tau\to-\infty),\quad\quad \Linf:=\hat L(\tau\to-\infty).
\end{equation}
From Eq.\ (\ref{E&Ldot}) we have
\begin{equation}
\label{hatEL}
\hat{E}(\tau) = E_{\infty}+\Delta E(\tau), \quad \quad
\hat{L}(\tau) = L_{\infty}+\Delta L(\tau),
\end{equation}
where
\begin{equation} 
\label{DeltaEL}
\mu\Delta E(\tau)=-\int_{-\infty}^{\tau}\!\!\!\! F_t\, d\tau,
\quad\quad
\mu\Delta L(\tau)=\int_{-\infty}^{\tau}\!\!\!\!  F_\phi\, d\tau.
\end{equation}
In principle, the coupled set (\ref{rdotGSF}) with (\ref{hatEL}) determines the self-accelerated orbit, given the initial values $\Einf,\Linf$ and a method for calculating the GSF along the orbit.

\subsection{Dissipative and Conservative pieces of the self-force}

The quantities $\Delta E(\tau)$ and $\Delta L(\tau)$ encapsulate both conservative and dissipative effects of the GSF. This terminology refers to a splitting of the GSF in the form
\begin{equation}
F^{\alpha}=F^{\alpha}_{\rm cons}+F^{\alpha}_{\rm diss},
\end{equation}
where the first and second terms are the self-forces exerted, respectively, by the ``time-symmetric'' and ``time-antisymmetric'' pieces of the (regularized) metric perturbation (cf.\ \cite{barack} for a more precise definition). For  geodesic motion in the equatorial plane of a Kerr black hole, $F^{\alpha}$ can be thought of as a function of only $r$ and $\dot{r}$ along the orbit. The particular time symmetry of such geodesics then implies \cite{barack}
 \begin{eqnarray}
 \label{Fcons}
 F^{\alpha}_{\rm cons}(r,\dot{r})&=&\frac{1}{2}\left[F^{\alpha}(r,\dot{r})+s_{(\alpha)}F^{\alpha}(r,-\dot{r})\right],\\
 \label{Fdiss}
 F^{\alpha}_{\rm diss}(r,\dot{r})&=&\frac{1}{2}\left[F^{\alpha}(r,\dot{r})-s_{(\alpha)}F^{\alpha}(r,-\dot{r})\right]
 \end{eqnarray}
(no summation over $\alpha$), where $s_{(t)}=-1=s_{(\phi)}$ and $s_{(r)}=+1$.
This gives a simple prescription for constructing $F^{\alpha}_{\rm cons}$ and $F^{\alpha}_{\rm diss}$ along geodesics, given the full GSF.

For circular orbits we have $F^{\alpha}(r,\dot{r})=F^{\alpha}(r,-\dot{r})$, meaning $F^t,F^{\phi}$ are purely dissipative while $F^r$ is purely conservative. In general, however, each component has both dissipative and conservative pieces. Of particular interest to us will be nearly-circular orbits with $|\dot{r}|\ll 1$. Along such orbits we may write, to leading order in $|\dot{r}|$,
\beq \label{Ftnearcirc}
F^{\alpha}_{\rm cons} \simeq \dot{r} F^{\alpha}_{1}(r),\quad\quad
F^{\alpha}_{\rm diss} \simeq F^{\alpha}_{0}(r)
\eeq
for $\alpha=t,\phi$, and 
\beq \label{Frnearcirc}
F^{r}_{\rm cons} \simeq F^{r}_{0}(r),\quad\quad
F^{r}_{\rm diss} \simeq \dot{r} F^{r}_{1}(r),
\eeq
where $F^{\alpha}_0$ and $F^{\alpha}_1$ are some functions of $r$ only.

Equations (\ref{Fcons})--(\ref{Frnearcirc}) are applicable, at leading order in $\eta$, even for an orbit that is slowly evolving under the GSF effect. In that case the GSF depends also on the instantaneous self-acceleration, but that dependence appears only at subleading order in $\eta$. At leading order, Eqs.\ (\ref{Fcons})--(\ref{Frnearcirc}) maintain their form at each point along the orbit.

The GSF integrals $\Delta E$ and $\Delta L$ can be related, in certain situations, to asymptotic fluxes of energy and angular momentum in gravitational waves. This was established rigorously in Ref.\ \cite{quinn} for a trajectory starting and ending at infinity.\footnote{The configuration considered in Ref.\ \cite{quinn} had no black hole in it, but the authors argue convincingly that a similar conclusion would hold also in the black hole case, if fluxes down the event horizon were accounted for in the balance equation.} A similar balance relation has been argued to hold also for adiabatic inspiral orbits around a black hole, subject to a suitable averaging over many orbital periods \cite{Detweiler:2008ft,Flanagan}. In both scenarios, the contribution from $F^{\alpha}_{\rm cons}$ to the integrals $\Delta E$ and $\Delta L$ (taken from $\tau=-\infty$ to $\tau=+\infty$) vanishes at leading order, by virtue of the orbital symmetry expressed in Eq.\ (\ref{Fcons}). This guarantees that the radiated fluxes balance the work done by the {\em dissipative} piece of the self-force alone, as expected.


 
\subsection{ADM energy and angular momentum}

Our analysis in the next section will require knowledge of the total, conserved ADM energy and angular momentum contents of the spacetime in the above setup. Specifically, we will need expressions for $E_{\rm ADM}$ and $L_{\rm ADM}$ in terms of $\Einf$ and $\Linf$ (and the two masses, $M$ and $\mu$), correct through $O(\mu^2)$. A subtlety is that ADM quantities are most conveniently evaluated in a ``center-of mass'' system (and, at the required order, would include a contribution from the black hole's ``recoil'' motion), whereas $\Einf$ and $\Linf$ are components of the particle's four-velocity, defined in a coordinate system centered around the black-hole.

In our setup, the ADM quantities are most easily evaluated on a hypersurface of constant $t\ll -M$, where the binary separation is $r\gg M$. In the limit $t\to -\infty$ ($r\to\infty$), the gravitational interaction energy vanishes and does not contribute to $E_{\rm ADM}$. Working at that limit, we assume that, for the purpose of calculating ADM quantities, the black hole--particle system may be replaced with that of two relativistic pointlike particles in flat spacetime. $E_{\rm ADM}$ is then simply the sum of the two relativistic energies in the center-of-mass frame, and $L_{\rm ADM}$ is similarly the sum of two angular momenta (with respect to the center of mass), plus the spin of the black hole. 

Appendix \ref{App:ADM} gives the details of this calculation, which is straightforward. The result is 
\beq\label{EADM}
E_{\rm ADM} = M\left[1+\eta \Einf -\frac{1}{2}\eta^2 (\Einf^2-1)\right] +o(\eta^2),
\eeq
\beq\label{LADM}
L_{\rm ADM} = M\left(a + \eta\Linf -\eta^2 \Linf\Einf\right) + o(\eta^2) .
\eeq

\subsection{Critical orbits}

In the geodesic case we have introduced the function $L_c(E)$, which we now interpret as the {\em critical value} of the angular momentum for a given energy: Geodesic orbits with $L>L_c(E)$ scatter back to infinity, while ones with $L<L_c(E)$ fall into the black hole. This type of critical behavior carries over to the GSF case, though radiation losses then introduce a subtlety, since orbits that are initially scattered may fall into the black hole at a subsequent approach. However, we may still speak of a critical threshold for an {\it immediate} capture, which separates (in the space of initial conditions) between orbits that scatter at first approach and orbits that do not.
A detailed analysis of this critical behavior was given in Ref.\ \cite{gund} for orbits in  Schwarzschild spacetime (working in the first-order GSF approximation, as here), and in the following discussion we assume the same qualitative behavior applies in the Kerr case too.  

In particular, we assume there exists a critical value $\Linf={\Linf}_{,c}(\Einf)$ that separates between the two possible outcomes. The initial conditions $\{\Einf,{\Linf}_{,c}(\Einf)\}$ thus define a one-parameter family of ``critical orbits''. Let us denote by $\hat E_c(\tau;\Einf)$ and $\hat L_c(\tau;\Einf)$ the functions $\hat E(\tau)$ and $\hat L(\tau)$ corresponding to a critical orbit with a given $\Einf$ [so that $\hat E_c(\tau\to-\infty;\Einf)=\Einf$ and $\hat L_c(\tau\to-\infty;\Einf)={\Linf}_{,c}(\Einf)$]. Unlike in the geodesic case where critical geodesics of different $E$ are disjoint, in the GSF case all critical orbits join a {\em global attractor}, which is the perfectly fine-tuned orbit that evolves radiatively along the sequence of unstable circular orbits starting at the light ring and ending at the ISCO, where it plunges into the black hole.  Figure 1 in Ref.\ \cite{gund} illustrates the evolution of the critical orbit along the attractor, and see also Fig.\ \ref{attractor} below. 

Let us define the ``GSF correction''
\begin{equation}\label{deltaL1}
\delta L_c(\tau;\Einf):=\hat L_c(\tau;\Einf)-L_c(\Einf),
\end{equation}
and then
\begin{equation}\label{deltaLdef}
\delta L_{\infty}(\Einf):= \delta L_c(\tau\to-\infty;\Einf).
\end{equation}
$\delta L_{\infty}$ is the GSF-induced shift in the critical value of $\Linf$ at a fixed $\Einf$. It may also be interpreted in terms of a GSF correction to the critical impact parameter. 
We assume that the difference $\delta L_c(\tau;\Einf)$ remains small [$O(\eta)$] during the approach, which should be the case in any reasonable gauge. However, clearly, that difference ceases to remain small as the critical orbit joins the global attractor and evolves along it; then the meaning of $\delta L_c(\tau;\Einf)$ as a small GSF correction is lost. 

For our analysis of overspinning orbits in the next section, we will require an explicit expression for $\delta L_{\infty}(\Einf)$ in terms of GSF quantities. It is instructive to derive this relation first with the dissipative piece of the GSF turned off, i.e.\ replacing the full GSF with its conservative piece (in which case the global attractor disappears, and critical orbits of different $\Einf$ remain disjoint). Let us call the resulting quantity $\delta L^{\rm cons}_{\infty}(\Einf)$. As a second step we will restore dissipation and consider its effect. 

\subsubsection{Conservative GSF effect}\label{subsubsec:cons}

With dissipation turned off, the critical orbit becomes exactly stationary at $\tau\to \infty$, where it joins an unstable (nongeodesic) circular orbit of radius $\hat R(\Einf)=R(\Einf)+\delta R$. Here $R(\Einf)$ is the geodesic relation given in Eq.\ (\ref{rhopar}), and $\delta R$ is a conservative GSF correction. To obtain $\delta L^{\rm cons}_{\infty}$ we first substitute $\hat E$ and $\hat L$ from Eq.\ (\ref{hatEL}) into the radial equation of motion (\ref{rdotGSF}), replacing $\Linf$ with $L_c(\Einf) + \delta\Linf^{\rm cons}(\Einf)$, where $L_c(\Einf)$ is the geodesic relation given in Eq.\ (\ref{LofE}). We then demand $dr/d\tau=0$ as well as $d^2r/d\tau^2=0$ at $r=\hat R$ through $O(\eta)$.  At leading order in $\eps$ this yields two algebraic equations for the two $O(\eta)$ unknowns $\delta\Linf^{\rm cons}$ and $\delta R$, given $\Einf$ and the GSF.  The solution is
\beq \label{deltaLcons1}
\delta {L}_{\infty}^{\rm cons}(\Einf)=2M\Delta E^{\rm cons}(\infty)-\Delta {L}^{\rm cons}(\infty) ,
\eeq
and $\delta R(\Einf)=O(\eps)O(\eta)$. Here $\Delta E^{\rm cons}$ and $\Delta L^{\rm cons}$ are the same as $\Delta E$ and $\Delta L$ of Eq.\ (\ref{DeltaEL}), but with $F_\alpha\to F_\alpha^{\rm cons}$, and with the GSF integrals evaluated along the critical orbit with energy-at-infinity $\Einf$. The precise dependence of $\delta R$ on the GSF will not be needed, but we note that the $O(\eps\eta)$ GSF correction to the radius of the critical circular orbit is reassuringly small compared to the $O(\eps)$ radial distance to the light ring.

To simplify the appearance of subsequent equations, let us from now on use units in which $M=1$. This, in particular, makes our ``tilde'' notation redundant (with $\tilde{L}=L$, etc.) and $\mu$ becomes interchangeable with $\eta$. 
Recalling our $W$ notation from Eq.\ (\ref{calE}), we rewrite Eq.\ (\ref{deltaLcons1}) as
\beq \label{deltaLcons2}
\delta{L}_{\infty}^{\rm cons}(\Einf)=\Delta W^{\rm cons}(\infty) ,
\eeq
where $\Delta W^{\rm cons}$ represents the conservative piece of
\begin{eqnarray}\label{DeltacalE}
\Delta W(\tau)&:=&2\Delta E(\tau) - \Delta L(\tau)
\nonumber\\ 
&=& -\eta^{-1} \int_{-\infty}^{\tau}\left(2F_t+ F_{\phi}\right)d\tau .
\end{eqnarray}
The quantity $\Delta W^{\rm cons}(\infty)$ is the limit $\tau\to\infty$ of $\Delta W^{\rm cons}(\tau)$. Does this limit actually exist? The answer is positive, since both $F_t^{\rm cons}$ and $F_\phi^{\rm cons}$ vanish exponentially fast in $\tau$ as the orbit approaches the limiting circular orbit at $\tau\to\infty$. 

To make this last statement more precise, let us split the $\tau$ integral into an ``approach'' piece, $\int_{-\infty}^{\tau_c}$, and a ``quasi-circular'' piece, $\int_{\tau_c}^{\infty}$, with $\tau_c$ chosen so that $\delta r(\tau_c)$, where $\delta r(\tau):=r(\tau)-\hat R$, is already very small. For a small $\delta r$ we have the form $F_t^{\rm cons}\simeq \dot{r} F_{1t}(r)$ [Eq.\ (\ref{Ftnearcirc})] and similarly for $F_\phi^{\rm cons}$. Thus $\int_{\tau_c}^{\infty}F_t^{\rm cons}d\tau\simeq -F_{1t}(\hat R)\delta r(\tau_c)$, and similarly for the $\phi$ component. 
A local analysis of Eq.\ (\ref{rdotGSF}) near the limiting circular orbit gives $\delta r\sim e^{-\lambda\tau}$, with a Lyapunov exponent $\lambda=M^{-1}(3\Einf^2-1)^{1/2}$ at leading order in $\eps$ (and ignoring the small effect of the GSF). 
The choice $\tau_c=-\lambda^{-1}\log\eta$, for example, gives $\delta r(\tau_c)\sim\eta$, and the quasi-circular piece of $\Delta W^{\rm cons}$ does not contribute to $\delta L_{\infty}^{\rm cons}$ at leading order in $\eta$. 


Our discussion assumes that $\Delta E(\tau_c)$ and $\Delta L(\tau_c)$ [hence also $\Delta W(\tau_c)$] are $O(\eta)$ quantities, i.e.\ that the accumulated GSF-induced positional shift in the orbit during the approach is a small, $O(\eta)$ quantity. This should be the case in any reasonable gauge. Under this assumption, we note that the value of the integral $\Delta W$ remains unchanged, at leading order in $\eta$, if in Eq.\ (\ref{DeltacalE}) we replace the integration along the actual, GSF-perturbed orbit, with an integration along the critical {\em geodesic} of energy $\Einf$. This can be exploited to simplify actual calculations: To compute $\delta L_{\infty}^{\rm cons}$ at leading order in $\eta$ requires only an evaluation of the GSF along a fixed geodesic, and there is no need to consider the back-reaction from the GSF on the orbital trajectory.

\subsubsection{Full GSF effect}

\begin{figure}
  \begin{center} 
    \includegraphics[scale=0.45]{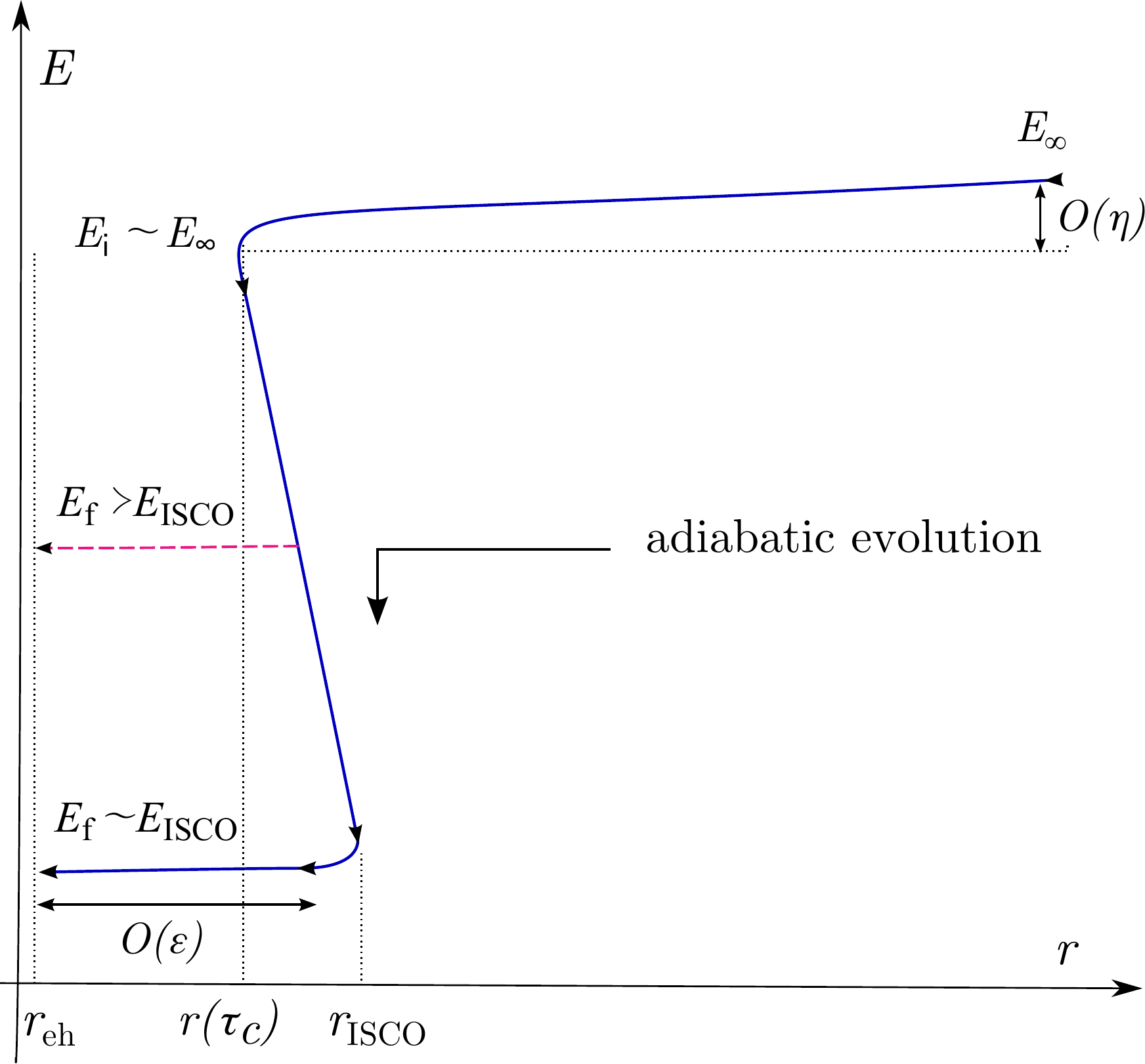}
    \caption{\label{attractor} 
Schematic illustration of the evolution of orbital energy along a perfectly fine-tuned critical orbit (solid blue curve). The orbit approaches from infinity, becomes trapped on an unstable circular orbit, and then evolves adiabatically in a quasi-circular fashion before transiting to plunge around the ISCO location. Radiative losses are small during the approach and plunge, but, through fine-tuning, the orbit can be made to lose ``all'' its energy during the quasi-circular stage. Intermediate values of the final energy $E_f$ may also be obtained by fine-tuning (dashed magenta line). Note the orbital radius {\it increases} through radiation losses during the quasi-circular stage. In the near-extremal case, $\eps\ll 1$, the quasi-circular evolution and final plunge occur within a small range of coordinate radii, $\Delta r= O(\eps)$.
 } 
  \end{center}
\end{figure}

Now restore dissipation. The fine-tuned critical orbit no longer settles into a strictly stationary motion, but rather it continues to evolve radiatively, in an adiabatic fashion, through a sequence of unstable circular orbits of decreasing energies (hence increasing radii). With a perfect fine-tuning, the orbit can reach the ISCO before plunging into the black hole---a scenario illustrated in Fig.\ \ref{attractor}. 
A relation between the degree of fine-tuning and the amount of energy loss was derived in Ref.\ \cite{gund} (for the Schwarzschild case): Rewriting their Eq.\ (124) in terms of angular momentum, we have the scaling $\delta L/L_{\infty,c}\sim \exp[(E_{\rm f}-E_{\rm i})/\eta]$, where $\delta L$ (not to be confused with the GSF shift $\delta\Linf$) is any small perturbation in the value of the initial angular momentum off the critical value $L_{\infty,c}$, and $E_{\rm f}-E_{\rm i}$ is the resulting change in specific energy as the orbit progresses along the global attractor. To achieve an $O(1)$ change in the specific energy requires an ``exponentially delicate'' fine-tuning, $\delta L/L_{\infty,c}\sim \exp(-1/\eta)$. 

For our analysis we do not require knowledge of the perfectly fine-tuned angular momentum at that level. We need $L_{\infty,c}$ through $O(\eta)$ only. Fine-tuning at $O(\eta)$ [corresponding to $\delta L=o(\eta)$] guarantees only $E_{\rm f}-E_{\rm i}=O(\eta\ln\eta)$. Therefore, for the purpose of determining $L_{\infty,c}$ through $O(\eta)$, it is sufficient to restrict attention to the early part of the critical orbit, where $\Delta E$ and $\Delta L$ (specific values) are still $O(\eta\ln\eta)$ at most, and have not yet accumulated $O(1)$ changes. This observation assists us in Appendix \ref{App:dL}, where we derive an expression for the leading-order full-GSF correction $\delta L_{\infty}(\Einf)$.



Our main result in Appendix \ref{App:dL} is that
\beq \label{deltaL2}
\delta {L}_{\infty}(\Einf) = \Delta W(\tau_c)+O(\eps)O(\eta\ln\eta),
\eeq
in analogy with the ``no dissipation'' case, Eq.\ (\ref{deltaLcons2}).
Here, $\Delta W(\tau_c)$ is the full-GSF integral shown in Eq.\ (\ref{DeltacalE}), evaluated along the orbit from infinity and up to the ``end of approach'' time $\tau_c$, when the orbit settles into a quasi-circular motion. Crucially, the contribution to $\Delta W$ from the quasi-circular part of the orbit is suppressed by a factor of $\eps$, so that the precise choice of $\tau_c$ does not affect the value of $\Delta W(\tau_c)$ at leading order. This assumes only that $\eps\ll |\ln\eta|^{-1}$, so that the error terms in Eq.\ (\ref{deltaL2}) are negligible compared to $\Delta W(\tau_c)=O(\eps^0)O(\eta)$.  
All we require of $\tau_c$ is to be sufficiently late that $|\dot{r}|$ is already very small [specifically, $\dot{r}(\tau_c)=O(\eta)$], but sufficiently early that $\Delta E$ is $O(\eta\ln\eta)$ at most. In practice, $\Delta W$ may again be evaluated along the critical {\it geodesic} of energy-at-infinity $\Einf$, with the integral in Eq.\ (\ref{deltaL2}) truncated after, say, 4--5 orbital revolutions. Truncating instead after (e.g.) 10 revolutions should change the result by a negligible amount of only $O(\eps)O(\eta)$.




The dissipative piece of $\Delta W(\tau_c)$ [call it $\Delta W^{\rm diss}$, defined by replacing $F_{\alpha}\to F_{\alpha}^{\rm diss}$ in Eq.\ (\ref{deltaL2})] may be expressed in terms of radiated quantities. Let $\calE_{\rm (apr)}$ and $\calL_{\rm (apr)}$ be the total energy and angular momentum in gravitational waves radiated out to infinity and down the black hole during the approach. We shall assume that the balance relation\footnote{The balance (\ref{balancecalE}) does not follow directly from the theorem of Ref.\ \cite{quinn}, because the approach part of the critical orbit does not end at infinity. It may be possible to construct a proof by considering a small outward deformation of the orbit (such that the new orbit starts at infinity and scatters back to infinity), then invoking the approximate symmetry about the periapsis, together with the $\eps$-suppression of the quasi-circular contribution to $\Delta W^{\rm diss}$.  We shall not endeavour to provide the details of such a proof here.}\textsuperscript{,}\footnote{One cannot expect to be able to similarly balance $\calE_{\rm (apr)}$ and $\calL_{\rm (apr)}$ individually, because the dissipative pieces of $\Delta E(\tau_c)$ and $\Delta L(\tau_c)$, unlike $\Delta W^{\rm diss}(\tau_c)$, {\em are} sensitive to the choice of $\tau_c$ already at leading order. However, such individual balance relations will not be needed in our analysis.}
\beq\label{balancecalE}
\calW_{\rm (apr)}:=2\calE_{\rm (apr)}-\calL_{\rm (apr)} =-\eta \Delta W^{\rm diss}(\tau_c)
\eeq
holds at leading order in $\eta$ and in $\eps$.
Equations (\ref{deltaLcons2}) and (\ref{deltaL2}) then lead to 
\beq \label{deltaLdiss2}
\delta{L}_{\infty}=\delta {L}_{\infty}^{\rm cons}
-\calW_{\rm (apr)}/\eta ,
\eeq
where subleading terms have been omitted.
This reexpresses $\delta{L}_{\infty}$ as a sum of conservative and radiative contributions, the motivation for which will become clear in the next section. 

Finally, let us further write $\calE_{\rm (apr)}=\calE_{\rm (apr)}^++\calE_{\rm (apr)}^-$ and similarly for $\calL_{\rm (apr)}$ and $\calW_{\rm (apr)}$, where hereafter superscripts `$+$' and `$-$' denote contributions from fluxes to infinity and down the black hole, respectively. The following argument, based on the first law of black hole mechanics, suggests that $\calW^-_{\rm (apr)}$ must vanish in the limit $\eps\to 0$. If we assume the black hole is not overspun during the approach, its horizon's area should increase by an amount $\delta A$ satisfying 
\beq\label{firstlaw}
\frac{\kappa}{8\pi}\delta A = \calE_{\rm (apr)}^- -\Omega_{H} \calL_{\rm (apr)}^-,
\eeq
where $\kappa=\eps/\sqrt{2}+O(\eps^2)$ is the horizon's surface gravity, and $\Omega_{\rm H}$ its angular velocity. Since $\Omega_{\rm H}=\frac{1}{2}+O(\eps)$, we identify the right-hand side of (\ref{firstlaw}) as $\calW^-_{\rm (apr)}$ at leading order in $\eps$. We thus have, in the extremal limit, $\calW^-_{\rm (apr)}\simeq \eps(c_1\delta A+c_2\calL^-_{\rm (apr)})$, with $c_1,c_2$ certain numerical coefficients. Since $\delta A$ and $\calL^-_{\rm (apr)}$ must remain bounded even in the extremal limit, we conclude that $\calW^-_{\rm (apr)}$ vanishes in that limit. 
Thus, at leading order in $\eps$, Eq.\ (\ref{deltaLdiss2}) becomes
\beq \label{deltaLdiss3}
\delta{L}_{\infty}=\delta {L}_{\infty}^{\rm cons}
-\calW^+_{\rm (apr)}/\eta ,
\eeq
which now features only outgoing fluxes. 

With this we have completed the necessary groundwork for our overspinning analysis, to be presented next.

\section{Overspinning with the full self-force}\label{Sec:OSwSF}

\subsection{General form of the censorship condition and reduction to near-critical orbits}

Starting with a near-extremal Kerr geometry with $a/M=1-\eps^2$, consider a particle sent in from infinity with specific energy $\Einf$ and specific angular-momentum $\Linf$ at $t\to -\infty$. The ADM mass and angular momentum of the spacetime are given in Eqs.\ (\ref{EADM}) and (\ref{LADM}) through $O(\eta^2)$. We assume the particle crosses the event horizon\footnote{More pedantically, we refer here to the crossing of a marginally trapped surface; spacetime need not contain a global horizon.} at some retarded time $u_h$, and we let $\calE^{+}$ and $\calL^{+}$ be the total energy and angular momentum radiated to null-infinity up until $u_h$ (with $u_h\to\infty$ if the post-capture geometry relaxes to a Kerr black hole). Then the Bondi mass and angular-momentum of spacetime at retarded time $u_h$ are $E_{\rm ADM}-\calE^{+}$ and $L_{\rm ADM}-\calL^{+}$, respectively. Overspinning is {\em avoided} if and only if\footnote{We do not know, and for our purpose do not need to know, the future evolution of spacetime beyond retarded time $u_h$ in the hypothetical case where (\ref{avoided}) is {\it not} satisfied. The likely scenario involves the formation of a naked singularity and a breakdown of predictability for $u>u_h$ \cite{wald}. If (\ref{avoided}) {\em is} satisfied, then, by ``no-hair'' theorems, geometry should relax to a Kerr black hole.}
\beq \label{avoided}
(E_{\rm ADM}-\calE^{+})^2 \geq  L_{\rm ADM}-\calL^{+} .
\eeq
To rule out the overspinning scenario, we need to show that this inequality holds for all $\Einf,\Linf$ and for all $\eta,\eps$. Since we work in first-order perturbation theory, we only demand that (\ref{avoided}) is satisfied at leading order in $\eta$. We also assume $\eps\ll 1$ and keep only leading terms in $\eps$, but we do not {\it a priori} restrict the magnitude of $\eps$ relative to that of $\eta$. 
We shall refer to the inequality (\ref{avoided}) as the {\it censorship condition}.\footnote{It may be argued that (\ref{avoided}) is guaranteed to hold (with a strong inequality) by virtue of the third law of black-hole mechanics \cite{Israel}, though it is clear that some of the third-law's assumptions are not satisfied within our model---see \cite{zimm} for a discussion. Even if that can be established, it is still of interest to explore the {\em physical mechanism} that enforces the third law in our setup, which is what our study aims to achieve.}

Substituting from Eqs.\ (\ref{eps}), (\ref{EADM}) and (\ref{LADM}), the censorship condition becomes 
\begin{multline}\label{OSGSF0}
\eps^2+\eta W_{\infty}+\eta^2(1+{L}_{\infty}\Einf-\Einf^2) 
\\
+(\eta\Einf-\calE^{+})^2
- {\calW}^{+} \geq 0 ,
\end{multline}
where $W_{\infty}:=2\Einf-{L}_{\infty}$, ${\calW}^{+}:=2\calE^{+}-\calL^{+}$, and we have omitted subleading terms of $o(\eta^2)$.
Note how the various terms here scale with $\eta$. The quantities $\Einf$ and $\Linf$ (hence also $W_{\infty}$) are specific values, thus {\it a priori} they are $O(\eta^0)$. The radiated energy $\calE^{+}$ is generically $O(\eta^2)$, but may accumulate at $O(\eta)$ for orbits that are fine-tuned to evolve along the global attractor; it is to allow for such orbits that we have kept the terms $2\eta\Einf\calE^{+}$ and $(\calE^{+})^2$ in Eq.\ (\ref{OSGSF0}). The quantity $\calW^{+}$ is likewise $O(\eta^2)$ generically and up to $O(\eta)$ with fine-tuning, but, as will be shown below, in the latter case the $O(\eta)$ term is also proportional to $\eps$.

Inspecting Eq.\ (\ref{OSGSF0}), we observe that, for all captured orbits that are not sufficiently close to criticality, the term $\eta W_\infty$ is $O(\eta)$ and positive, so the censorship condition (\ref{OSGSF0}) is trivially  satisfied at leading order in $\eta$ and $\eps$. Violation of  (\ref{OSGSF0}) (hence overspinning) may only be achieved, potentially, if $\Linf$ is tuned so that $\Linf=2\Einf+O(\eps,\eta)$, giving $W_\infty=O(\eps,\eta)$.
It is therefore sufficient to restrict attention to this class of orbits, to be referred to in what follows as ``near-critical''. Formulated on near-critical orbits, the censorship condition takes the sufficient form 
\begin{equation}\label{OSGSF01}
\eps^2+\eta W_{\infty}+\eta^2(1+\Einf^2) 
+(\eta\Einf-\calE^{+})^2
- {\calW}^{+} \geq 0 ,
\end{equation}
where we have dropped $O(\eps\eta^2,\eta^3)$ terms. This is required to hold for each member of the reduced two-parameter family $\{\Einf,\Linf\}$ with $\Linf-2\Einf=O(\eps,\eta)$. 

To proceed, we need to make more precise the distinction between ``fine-tuned'' orbits and ``generic'' near-critical orbits that are not fine-tuned. Referring to Fig.\ \ref{attractor}, let $E_f$ be the final value of the specific energy with which the particle plunges into the hole; and let $L_{\infty,c}(\Einf)$ be the perfectly fine-tuned value of $\Linf$, for which the orbit joins the global attractor and evolves along it to the ISCO. Assuming the universal scaling $\Linf-L_{\infty,c}(\Einf)\sim \exp[(E_f-\Einf)/\eta]$ \cite{gund}, near-critical orbits as defined above generically have $E_f-\Einf=O(\eta\ln\eta\eps)$ [here we neglect the $O(\eta)$ difference between $\Einf$ and $E_i$]. Calibrating $\Linf$ at higher order in $\eta,\eps$ [so that $\Linf-L_{\infty,c}=O(\eta^n,\eps^k)$ with some $n,k>1$] does not qualitatively change this generic scaling of $E_f-\Einf$. To achieve $E_f-\Einf=O(1)$ requires an exponentially accurate tuning, i.e.\ $\Linf-L_{\infty,c}\sim \exp(-1/\eta)$.
In what follows we use the $\eta$ scaling of $E_f-\Einf$ to distinguish between {\it generic} and {\it fine-tuned} members of the near-critical family: The former admit $E_f-\Einf=O(\eta\ln\eta\eps)$, and the latter $E_f-\Einf=O(1)$. This distinction can also be formulated in terms of the radiated quantities $\calE^+$ or $\calL^+$:
\begin{eqnarray}
\calE^+,\calL^+ &=& O(\eta^2\ln\eta\eps)\quad \text{(``generic'')}, \label{scaling_generic}
\\
\calE^+,\calL^+ &=& O(\eta)\quad\quad\ \text{(``fine-tuned'')}.  \label{scaling_FT}
\end{eqnarray}

\subsection{Further reduction to critical orbits}

The inequality (\ref{OSGSF01}) is still a condition on a two-parameter family of orbits. Ignoring fine-tuned orbits for now, it is possible---and beneficial---to reduce it further to a sufficient condition formulated on a one-parameter family. We achieve this by minimizing the left-hand side of Eq.\ (\ref{OSGSF01}) over all near-critical orbits for a given $\Einf$. We argue that the minimizing orbit is one with $\Linf$ tuned to its critical value at least through $O(\eta,\eps)$, namely 
\beq\label{Lcrit}
\Linf
=2\Einf+\eps\sqrt{6\Einf^2-2}+\delta{L}_{\infty}(\Einf)+o(\eta,\eps),
\eeq
where we have recalled Eq.\ (\ref{LofE}), and $\delta{L}_{\infty}(\Einf)$ is the $O(\eta)$ GSF term derived in the previous section.
To see this, note first that $W_{\infty}=2\Einf-\Linf$ is trivially minimized by $\Linf=L_{\infty,c}(\Einf)$, since $L_{\infty,c}(\Einf)$ maximizes $\Linf$ (over all captured orbits of a fixed $\Einf$) by definition of a critical orbit. This means that, to minimize $W_{\infty}$ through $O(\eta,\eps)$ (higher orders are irrelevant in our approximation) it suffices to demand $\Linf-L_{\infty,c}(\Einf)=o(\eta,\eps)$. Then also note that the two radiative terms on the left-hand side of (\ref{OSGSF01}) are insensitive, at relevant order, to variations of $\Linf$ within the family of nearly-critical orbits for a fixed $\Einf$. For generic orbits, the term $(\eta\Einf-\calE^{+})^2$ is simply $\eta^2\Einf^2$ at leading order, recalling Eq.\ (\ref{scaling_generic}). As for the term $-{\calW}^{+}$, we note that the contribution to that term from the approach part of the orbit, which is already $O(\eta^2)$, is not sensitive, at that order, to $O(\eta,\eps)$ variations in $\Linf$. Meanwhile, the contribution to ${\calW}^{+}$ from the quasi-circular part of the orbit is of $O(\eps)O(\eta^2\ln\eta\eps)$ at most (the occurrence of the factor $\eps$ will be explained below) and hence negligible in Eq.\ (\ref{OSGSF01}), assuming only $\eps\ll |\ln\eta|^{-1}$. 


Thus, discounting fine-tuned orbits, we find that the entire left-hand side of Eq.\ (\ref{OSGSF01}) is minimized by $\Linf$ as given in Eq.\ (\ref{Lcrit}). A new sufficient version of the censorship condition may therefore be written as
\begin{multline}\label{OSGSF1}
\eps^2-\eta\eps\sqrt{6\Einf^2-2}-\eta\, \delta{L}_{\infty}+\eta^2(1+\Einf^2) 
\\
+(\eta\Einf-\calE^{+})^2
- \calW^{+} \geq 0 ,
\end{multline}
which, at the relevant, leading order, is a condition on the {\it one}-parameter family of (generic) critical orbits parametrized by $\Einf$ alone. 

It should now be noted that the condition (\ref{OSGSF1}) also applies to fine-tuned orbits [whether or not they minimize the left-hand size of (\ref{OSGSF01})], simply because such orbits always satisfy Eq.\ (\ref{Lcrit}). However, for fine-tuned orbits the condition still involves {\em two} parameters, conveniently chosen as $\Einf$ and $E_f$. Different values of $E_f$ correspond to a fine-tuning of $\Linf$ at an exponential level. In principle, any value of $E_f$ in the range $E_{\rm isco}\lesssim E_f\lesssim \Einf$ may be obtained this way. To rule out overspinning by fine-tuned orbits, the censorship condition (\ref{OSGSF1}) must hold for all $\{\Einf,E_f\}$ with $E_f$ in the above range.

Observe that in Eq.\ (\ref{OSGSF1}) we have
\begin{eqnarray}
\eta\delta{L}_{\infty}+\calW^{+}
&=&  \eta\,\delta{L}_{\infty}^{\rm cons} +\calW^{+}-\calW^{+}_{\rm (apr)}
\nonumber\\
&=&\eta{L}^{\rm cons}_{\infty}+\calW_{\rm (qc)}^{+}+\calW_{\rm (end)}^{+},
\end{eqnarray}
where in the first line we have recalled Eq.\ (\ref{deltaLdiss3}), 
$\calW_{\rm (qc)}^{+}$ is the piece of $\calW^{+}$ coming from the evolution along the quasi-circular part of the orbit, and $\calW_{\rm (end)}^{+}$ is the piece coming from the transition to a final plunge into the black hole and from the plunge itself. It follows that only the {\it conservative} piece of the shift $\delta{L}_{\infty}$ actually enters our condition:
\begin{multline}\label{OSGSF2}
\eps^2-\eta\eps\sqrt{6\Einf^2-2}-\eta\,\delta{L}^{\rm cons}_{\infty}+\eta^2(1+\Einf^2) 
\\
+(\eta\Einf-\calE^{+})^2
- \calW_{\rm (qc)}^{+} -\calW_{\rm (end)}^{+} \geq 0 .
\end{multline}

In this last form, conservative and dissipative terms of the GSF feature separately. The former are associated with the approach leg of the orbit, and the latter accumulate during the adiabatic evolution along the attractor. In Appendix \ref{App:plunge} we combine results by Ori and Thorne \cite{orithor}, Kesden \cite{Kesden} and Mino and Brink \cite{Mino}, to argue that the term $\calW_{\rm (end)}^{+}$ is always subdominant and negligible in Eq.\ (\ref{OSGSF2}). We shall therefore omit that term in the rest of our discussion.

In subsection \ref{subsec:fine-tuned} below we will show that the radiative term $\calW_{\rm (qc)}^{+}$ scales as $O(\eps)O[\eta(\Einf-E_f)]$. This term can thus feature at leading order in Eq.\ (\ref{OSGSF2}) only for fine-tuned orbits, for which $\Einf-E_f=O(1)$. Likewise, terms involving $\calE^{+}$ feature only for fine-tuned orbits and are negligible otherwise. On the other hand, the conservative term $\eta{L}^{\rm cons}_{\infty}$ is always $O(\eta^2)$, featuring in the censorship condition regardless of fine-tuning. An important consequence is that {\em dissipative effects of the GSF enter the censorship condition only for fine-tuned orbits}.  This seems consistent with suggestions made in earlier analyses \cite{bck2,Kesden,Harada} (in which fine-tuning has not been considered).


Below we further simplify the condition (\ref{OSGSF2}), and reformulate it explicitly in terms of $\Einf$ alone (for generic orbits) or $\Einf$ and $E_f$ alone (for fine-tuned ones), without reference to $\eta$ and $\eps$. We shall consider separately the cases of generic and fine-tuned orbits, starting with the former, simpler case.

\subsection{Censorship condition for generic orbits}

As mentioned above (and shown in the next subsection), without fine-tuning the radiative terms $\calE^{+}$ and $\calW^{+}_{\rm (qc)}$ become subdominant in Eq.\ (\ref{OSGSF2}) and drop out of it. The censorship condition then reduces to
\beq\label{OSGSF3}
\eps^2+\eta\eps F +\eta^2 H \geq 0 ,
\eeq
with
\begin{eqnarray}
F&:=&-\sqrt{6\Einf^2-2} \, ,  \label{F}
\\   
\label{H}
H&:=& 1+2\Einf^2 - \delta\breve{L}_{\infty}^{\rm cons} .
\end{eqnarray}
Here we have made the $\eta$-scaling of $\delta{L}_\infty^{\rm cons}$ explicit by introducing the shift-per-eta
\beq\label{brevedL}
\delta\breve{L}_{\infty}^{\rm cons}:=\eta^{-1}\,\delta{L}_\infty^{\rm cons} ,
\eeq
which should have a finite (nonzero) limit $\eta\to 0$.

For the overspinning scenario to be ruled out, the inequality (\ref{OSGSF3}) must be satisfied for all $\eta,\eps>0$ and all $\Einf\geq 1$. The condition can be written in the equivalent form $\Phi:=\alpha^2+\alpha F+H\geq 0$, with $\alpha:=\eps/\eta>0$. At fixed $\Einf$, $\Phi$ is quadratic in $\alpha$, with a minimum value $\Phi(\alpha=-F/2)=H-F^2/4$. To guarantee $\Phi\geq 0$ for all $\Einf$ and all $\alpha>0$ (hence all $\eta,\eps>0$) we must demand $H\geq F^2/4$; if $H< F^2/4$ for some $\Einf$, then for that $\Einf$ there exist $\eta,\eps$ values for which overspinning is achieved. In that way, $H\geq F^2/4$ is both sufficient and necessary for overspinning to be avoided. Inserting the values of $F$ and $H$, the censorship condition takes the simple form 
\beq \label{OSfinal}
 \delta\breve{L}_{\infty}^{\rm cons} \leq \frac{1}{2}(\Einf^2+3).
\eeq
Overspinning is averted (for orbits that are not fine-tuned) if and only if (\ref{OSfinal}) is satisfied for each member of the one-parameter family of critical orbits with $\Einf\geq 1$, in the limit $\eta,\eps\to 0$.  

Equation (\ref{OSfinal}) states our final result for generic orbits. As already mentioned, it involves only {\it conservative} GSF effects, specifically the shift in the critical value of the angular-momentum-at-infinity (at fixed $\Einf$) due to the conservative piece of the GSF.
For easy reference, we give here the explicit formula for $\delta\breve{L}_{\infty}^{\rm cons}$ in terms of GSF components:
\beq \label{dLfinal}
\delta\breve{L}_{\infty}^{\rm cons}(\Einf) = 
\lim_{\eps\to 0}\frac{1}{\mu^2} \int^{\infty}_{R_\eps}\left(2MF_t^{\rm cons}+F_{\phi}^{\rm cons}\right)dr/\dot{r},
\eeq
where we have recalled Eqs.\ (\ref{deltaLcons2}) and (\ref{DeltacalE}). The integration is carried out along the critical geodesic of specific energy $\Einf$ on a background with spin $a/M=1-\eps^2$, starting at the unstable circular orbit of radius $R_\eps=R_\eps(\eps,\Einf)$ and ending at infinity.

Inspecting Eq.\ (\ref{OSfinal}), it may seem peculiar that overspinning may be averted even for some {\em positive} values of $\delta{L}_\infty^{\rm cons}$: A positive $\delta{L}_\infty^{\rm cons}$ would seem to mean that the GSF {\em increases} the critical impact parameter, allowing in particles that would otherwise be scattered away. However, we must recall that the shift $\delta{L}_\infty^{\rm cons}$ is defined not with respect to the physical, ADM angular momentum, but with respect to the quantity $\hat L=\hat u_{\phi}$, which (while convenient to work with in practice) does not have a clear invariant meaning beyond the geodesic approximation. To rewrite (\ref{OSfinal}) in a more physically insightful way, let us, then, recast it in terms of ADM quantities, as follows. 

First, let us introduce the specific quantities $E^{\rm p}_{\rm ADM}$ and $L^{\rm p}_{\rm ADM}$ defined through
\begin{eqnarray}\label{ADMspec}
\mu E^{\rm p}_{\rm ADM}&:=&E_{\rm ADM}-M,
\nonumber\\
\mu L^{\rm p}_{\rm ADM}&:=&L_{\rm ADM}-Ma,
\end{eqnarray}
which may be thought of as the particle's contributions to the total ADM energy and angular momentum of the system. Then, denote by $\delta L^{\rm p}_{\rm ADM}(E^{\rm p}_{\rm ADM})$ the shift, due to the conservative GSF, in the critical value of $L^{\rm p}_{\rm ADM}$ for a fixed $E^{\rm p}_{\rm ADM}$. A short calculation, based on Eqs.\ (\ref{EADM}) and (\ref{LADM}), gives
\beq\label{deltaLADM}
\delta L^{\rm p}_{\rm ADM}(E^{\rm p}_{\rm ADM}) = \delta\Linf^{\rm cons}(\Einf)-\eta(\Einf^2+1)+O(\eta^2).
\eeq
Thus, in terms of $\delta\breve{L}^{\rm p}_{\rm ADM}:=\eta^{-1}\,\delta L^{\rm p}_{\rm ADM}$, the censorship condition (\ref{OSfinal}) becomes
\beq \label{OSfinalADM}
 \delta\breve{L}^{\rm p}_{\rm ADM}\leq \frac{1}{2}(1-\Einf^2),
\eeq
where on the right-hand side $\Einf$ may be replaced with $E^{\rm p}_{\rm ADM}$ at relevant order.

The alternative form (\ref{OSfinalADM}) is now more intuitive: For unbound orbits ($\Einf\geq 1$), the GSF averts overspinning if it shifts the critical value of the (ADM-related) angular momentum by a sufficiently negative amount, which depends only on $\Einf$. In the marginal case of $\Einf=1$ (where overspinning is marginally prevented already in the geodesic case), the shift $\delta\breve{L}_{\rm ADM}^{\rm p}$ need only be nonpositive.  
We are not aware of any {\it a priori} argument to suggest that $\delta\breve{L}_{\rm ADM}^{\rm p}$ must necessarily be nonpositive for all $\Einf\geq 1$. Verifying this would need to await a numerical calculation. Any counterexample would imply a direct violation of cosmic censorship.   

Let us make a few more points about the condition (\ref{OSfinal}). First, in its form (\ref{OSfinalADM}) it is manifestly gauge invariant (within a class of suitable asymptotically flat gauges) despite the gauge dependence of the local GSF featuring in $\delta\breve{L}_{\infty}^{\rm cons}$ [Eq.\ (\ref{dLfinal})]. The condition involves only quantities that are defined and evaluated at infinite separation, namely the specific energy $\Einf$ (or $E^{\rm p}_{\rm ADM}$) and angular-momentum shift $\delta\breve{L}_{\rm ADM}^{\rm p}$, each having a clear gauge-invariant physical meaning. The evident invariance of our final condition is reassuring.

Second, as already mentioned, the condition that (\ref{OSfinal}) is satisfied for all $\Einf\geq 1$ is both sufficient and necessary for overspinning to be avoided within the scenario considered here. It is a {\it sufficient} condition only in the sense that it guarantees no overspinning occurs {\em for sufficiently small mass-ratio $\eta$}; since we work in the first-order self-force approximation, we cannot make the statement any stronger. Equation (\ref{OSfinal}) describes a {\it necessary} condition in the sense that its violation for any $\Einf$ would mean there exist (small) $\eta$ values for which overspinning occurs.  

Finally, the condition (\ref{OSfinal}) involves the single parameter $\Einf$, and the task of testing whether it is satisfied amounts to evaluating a single function of $\Einf$, namely $\delta\breve{L}_{\infty}^{\rm cons}(\Einf)$. The perturbative parameters themselves, $\eta$ and $\eps$, do not feature in the final condition. This is expected, given our first-order perturbative treatment and the fact that GSF effects (including ADM terms) appear in the overspinning condition already at leading order. It is precisely because of this ``order mixing'' that one cannot neglect the GSF in considering the overspinning problem, and why there is no sense in which the geodesic limit may be said to provide a useful approximation here.

\subsection{Censorship condition for fine-tuned orbits}\label{subsec:fine-tuned}

It is not {\it a priori} clear whether fine-tuning favours the overspinning scenario  or disfavours it: The answer depends on the details of the radiative evolution along the attractor. Indeed, for fine-tuned orbits the radiative terms $\calE^{+}$ and $\calW_{\rm (qc)}^{+}$ feature already at leading order in Eq.\ (\ref{OSGSF2}), and cannot be neglected.
We may again write Eq.\ (\ref{OSGSF2}) in the form (\ref{OSGSF3}), with $F$ and $H$ replaced with,  respectively, 
\begin{eqnarray}
\bar F&=&-\sqrt{6\Einf^2-2}-\breve{\calW}^+_{\rm (qc)} \, ,  \label{FH_ft}
\nonumber \\    
\bar H&=& 1+\Einf^2 +(\Einf-\breve{\calE}^+)^2 - \delta\breve{L}_{\infty}^{\rm cons} .
\end{eqnarray}
Here we have introduced the rescaled quantities 
\beq\label{breve}
\breve{\calE}^{\pm}:=\eta^{-1}\calE^{\pm},
\quad\quad
\breve{\calW}_{\rm (qc)}^{\pm}:= (\eps\eta)^{-1}{\calW}_{\rm (qc)}^{\pm},
\eeq
which should have finite (nonzero) limits $\eps,\eta\to 0$ for fine-tuned orbits [that ${\calW}_{\rm (qc)}^{\pm}=O(\eps)O(\eta)$ will be discussed in the next two paragraphs]. 

It will prove beneficial to reexpress $\bar F$ and $\bar H$ in terms of the absorption-related quantities $\breve{\calE}^-$ and $\breve{\calW}^-_{\rm (qc)}$, in place of $\breve{\calE}^+$ and $\breve{\calW}^+_{\rm (qc)}$. This is easily done for $\bar H$, noting $\Einf-\breve{\calE}^+ = E_f + \breve{\calE}^-$ at the relevant, leading order. As for $\bar F$, we start by writing ${\calW}^+_{\rm (qc)}={\calW}_{\rm (qc)}-{\calW}^-_{\rm (qc)}$, where, under the assumption of adiabaticity, the total ${\calW}_{\rm (qc)}$ may be expressed as an integral over the local GSF:
\beq\label{calEradint}
\calW_{\rm (qc)}=\int_{\Einf}^{E_f}\left(2F_t^{\rm diss}+F_{\phi}^{\rm diss}\right)dE/\dot{E}.
\eeq
Here we have used Eq.\ (\ref{DeltacalE}), changing the integration variable from $\tau$ to specific energy $E$, and assumed a balance relation as in Eq.\ (\ref{balancecalE}).  We have also neglected the subdominant [$O(\eta^2)$] amount of radiated energy during the approach, replacing the initial energy of the quasi-circular motion with $\Einf$. Then, following the method of Appendix \ref{App:dL} [cf.\  Eq.\ (\ref{ort2})], we use $u^{\alpha}F_{\alpha}^{\rm diss}=0$ to obtain 
\beq\label{ort}
F_t^{\rm diss}+F_{\phi}^{\rm diss}/2=-(u^r/u^t)F^{\rm diss}_r -\frac{3E}{\sqrt{6E^2-2}}\,\eps F_{t}^{\rm diss} ,
\eeq 
where subdominant terms in $\eps$ have been omitted. The contribution from the term $\propto F^{\rm diss}_r$ to the integral in (\ref{calEradint}) can be evaluated following the same steps as in Appendix \ref{App:dL} [see the paragraph containing Eq.\ (\ref{DeltacalEr})], and shown to be of only $O(\eps)O(\eta^2)$ (or smaller) ---hence negligible. The contribution from the term $\propto F^{\rm diss}_t$ can be evaluated explicitly upon replacing $F_t=\mu\dot{E}$, giving
\beq\label{calErad}
\calW_{\rm (qc)}=
-\eta\eps\left(\sqrt{6\Einf^2-2}-\sqrt{6E_f^2-2}\right).
\eeq
Thus, Eqs.\ (\ref{FH_ft}) are obtained in their alternative form 
\begin{eqnarray}
\bar F&=&-\sqrt{6E_f^2-2}+\breve{\calW}^-_{\rm (qc)} \, ,  \label{FH_ft2}
\nonumber \\   
\bar H&=& 1+\Einf^2 +(E_f+\breve{\calE}^-)^2 - \delta\breve{L}_{\infty}^{\rm cons} .
\end{eqnarray}

We note that Eq.\ (\ref{calErad}) establishes the scaling $\calW_{\rm (qc)}=O(\eps)O(\eta)$ for fine-tuned orbits. The first-law argument used in the previous section [refer to the discussion around Eq.\ (\ref{firstlaw})] can also be used to show $\calW^-_{\rm (qc)}=O(\eps)O(\eta)$. This then establishes the scaling $\calW_{\rm (qc)}^+=O(\eps)O(\eta)$ assumed above.

In both Eqs.\ (\ref{FH_ft}) and (\ref{FH_ft2}), the radiative quantities $\breve{\calE}^\pm$ and $\breve{\calW}^\pm_{\rm (qc)}$ should be thought of as functions of $\Einf$ and $E_f$ only. While $\breve{\calE}^+$ is necessarily positive, the absorbed energy $\breve{\calE}^-$ may be either positive, or---due to superradiance---negative, depending on $\Einf$ and $E_f$. Circular equatorial geodesics are superradiant for $\Omega<\Omega_H$, which, in the extremal limit, translates to $E<\frac{2}{\sqrt{3}}$. Thus, $\breve{\calE}^-$ is necessarily negative for any $E_f<\Einf\leq\frac{2}{\sqrt{3}}$. The sign (and magnitude) of $\breve{\calE}^-$ for other values of $\Einf$ and $E_f$, as well as the sign (and magnitude) of $\breve{\calW}^-_{\rm (qc)}$, remain to be determined numerically. The quantity $\breve{\calW}^+_{\rm (qc)}$, on the other hand, is easily shown to be negative definite. In fact,  Eqs.\ (\ref{calEplus}) and (\ref{calWplus}), given below, imply 
\beq
-\breve{\calW}^+_{\rm (qc)}>\breve{\calE}^+>0.
\eeq
Note this means that $\bar F$ in Eq.\ (\ref{FH_ft}) may change sign, depending on $\Einf,E_f$.



To proceed, we once again write the condition (\ref{OSGSF3}) (for the barred quantities) in the form $\bar\Phi:=\alpha^2+\alpha \bar F+\bar H\geq 0$, with $\alpha=\eps/\eta$. Here, however, the sign of $\bar F$ is not known {\it a priori}, which somewhat complicates matters. For $\bar F<0$, $\bar\Phi$ has its minimum at $\bar\Phi(\alpha=-\bar F/2)=\bar H-\bar F^2/4$, so the condition becomes $\bar H\geq \bar F^2/4$ as before. However, for $\bar F\geq 0$ the condition $\bar\Phi\geq 0$ is satisfied trivially for all $\bar H\geq 0$, and violated trivially for all $\bar H<0$ (by choosing a sufficiently small $\alpha$). In that case, therefore, a necessary and sufficient condition for $\bar\Phi\geq 0$ to hold for any $\eta,\eps$ is $\bar H\geq 0$. In summary, we obtain 
\beq \label{OSfinalFT}
 \bar H\geq (\min\{\bar F/2,0\})^2 
\eeq
as a necessary and sufficient condition for overspinning to be averted for all $\eta,\eps$. In this condition, $\bar F$ and $\bar H$ are both functions of the two independent parameters $\Einf$ and $E_f$.
To rule out overspinning we must require that (\ref{OSfinalFT}) is satisfied for all $\Einf>E_f(>E_{\rm isco})$. 

Evaluation of the condition (\ref{OSfinalFT}) requires knowledge of the radiative quantities $\breve{\calE}^\pm$ and $\breve{\calW}^\pm_{\rm (qc)}$ (in addition to $\delta\breve{L}_{\infty}^{\rm cons}$). To conclude our discussion, we now give convenient expressions for these two quantities in terms of 
a single function of one variable, namely the ratio
\beq
\calR(E):=\frac{\dot{\calE}^-(E)}{\dot{\calE}^+(E)},
\eeq
where $\dot{\calE}^{+/-}(E)$ are the outgoing/incoming fluxes of energy in gravitational waves sourced by a particle on a circular geodesic, evaluated in the extremal limit at a fixed specific energy $E$. [In deviation from our notation elsewhere, here and in the next two paragraphs an overdot denotes differentiation with respect to (any suitable) coordinate time.] We note $\calR<0$ for $E<\frac{2}{\sqrt{3}}$, the superradiance regime in the extremal limit.

First, we use the specific energy $E$ as a parameter along the global attractor, to write
\beq
\breve\calE^+ = \int_{\Einf}^{E_f}\frac{\dot{\calE}^+}{\eta\dot{E}}\,dE
=
-\int_{\Einf}^{E_f}\frac{\dot{\calE}^+}{\dot{\calE}^+ +\dot{\calE}^-}\, dE,
\eeq
where we assumed the balance relation $\eta\dot{E}=-(\dot{\calE}^+ +\dot{\calE}^-)$ applies during the adiabatic evolution along the attractor. Thus,
\beq\label{calEplus}
\breve{\calE}^+ (\Einf,E_f)=
-\int_{\Einf}^{E_f}\frac{dE}{1+\calR(E)},
\eeq
and, similarly,
\beq\label{calEminus}
\breve\calE^- (\Einf,E_f) = -\int_{\Einf}^{E_f}\frac{\calR(E)}{1+\calR(E)}\, dE ,
\eeq
which should be evaluated in the extremal limit, $\eps\to 0$. Note $\dot{\calE}^\pm\to 0$ in the extremal limit \cite{bck2}, while the ratio $\calR$ admits a finite, nonzero limit \cite{Colleoni_etal, priv_comm}. Thus, by writing $\breve{\calE}^\pm$ as in Eqs.\ (\ref{calEplus}) and (\ref{calEminus}) we have made it possible for the limit $\eps\to 0$ to be taken before the integration, which is advantageous in practice. 

As for $\calW^\pm_{\rm (qc)}$, we start by writing 
\beq\label{Wdot}
\dot{\calW}^\pm_{\rm (qc)} :=2\dot{\calE}^\pm -\dot{\calL}^\pm =-\eps b(E)\dot{\calE}^\pm ,
\eeq
where $\dot{\calL}^\pm$ are the angular-momentum fluxes corresponding to $\dot{\calE}^\pm$, and 
\beq\label{bofE}
b(E):=\frac{6E}{\sqrt{6E^2-2}}.
\eeq
To derive the second equality in (\ref{Wdot}), which is valid to leading order in $\eps$, we have used the small-$\eps$ expansion of the orbital angular velocity at fixed $E$, 
\beq\label{OmegaExpansion}
\Omega = \frac{1}{2} - \frac{1}{4}b(E)\eps +O(\eps^2),
\eeq
together with the general relation $\dot{\calE}^\pm =\Omega\dot{\calL}^\pm$ applicable to the radiation from any circular orbit \cite{eomegal}. 
Thus, proceeding as with $\breve{\calE}^+$, we obtain 
\beq\label{calWplus}
\breve\calW^+_{\rm (qc)} = \lim_{\eps\to 0} 
\int_{\Einf}^{E_f} \frac{\dot{\calW}^+}{\eps\eta\dot{E}}\,dE
=\int_{\Einf}^{E_f}\frac{b(E)}{1+\calR(E)}\,dE,
\eeq
and, similarly,
\beq\label{calWminus}
\breve\calW^-_{\rm (qc)}  
=\int_{\Einf}^{E_f}\frac{b(E)\calR(E)}{1+\calR(E)}\,dE.
\eeq
Equations (\ref{calEplus}), (\ref{calEminus}), (\ref{calWplus}) and (\ref{calWminus}) express $\breve\calE^\pm$ and $\breve\calW^\pm_{\rm (qc)}$ in terms of the single function $\calR(E)$, left to be determined numerically.

In Sec.\ \ref{Sec:num} we will assess the numerical task of evaluating our censorship conditions, review the status of relevant existing GSF codes, and comment on how they would need to be modified in order to provide the necessary data. But first, in the next section, we take a short detour to explore an alternative approach to the determination of $\delta\breve{L}_{\infty}^{\rm cons}$, which offers a practical advantage. 

\section{Reformulation in terms of redshift variable}  \label{Sec:redshift}

Our final overspinning conditions (\ref{OSfinal}) and (\ref{OSfinalFT}) feature the critical angular-momentum shift $\delta{L}_{\infty}^{\rm cons}$, whose evaluation, through equation (\ref{dLfinal}), requires an integration of the GSF from infinity along critical geodesics. As we discuss in the next section, this step is the main stumbling block when it comes to evaluating the conditions using currently available GSF codes. The integration from infinity comes about, essentially, because of the need to relate the local properties $\hat E$ and $\hat L$ of the particle just before it falls into the  black hole, to ADM properties of spacetime defined at infinity. This would have been unnecessary if we had available explicit formulas for $E_{\rm ADM}$ and $L_{\rm ADM}$ (or for the corresponding Bondi quantities $E_{\rm ADM}-\calE^+$ and $L_{\rm ADM}-\calL^+$), correct through $O(\eta^2)$, for the configuration of a particle in an unstable circular orbit around a Kerr black hole. Furthermore, given such formulas we would have been able to relax the requirement that the particle is sent in from infinity, and explore the possibility of overspinning with ``bound'' orbits. (We recall our result from Sec.\ \ref{Sec:geodesics} that bound {\em geodesics} cannot overspin; however, in principle, there remains the possibility that GSF effects change this situation.)

By good fortune, suitable formulas have been proposed very recently, in Ref.\ \cite{isoyama14}. The expressions, to be presented below, were obtained using (and in agreement between) two independent frameworks. One is the Hamiltonian approach of Isoyama and collaborators \cite{hami}, in which the conservative portion of the orbital dynamics is described (through first order in $\eta$ beyond the geodesic approximation) in terms of geodesic motion in a certain effective smooth spacetime. The other is based on the recently proposed ``first law of binary black-hole mechanics'' \cite{alt1,alt3} (itself a limiting case of the generalized law established in \cite{Friedman:2001pf}), which relates ADM properties of a helically-symmetric binary system of post-Newtonian particles to the so-called ``redshift'' of the particles (see below). Neither frameworks is {\em a priori} guaranteed to correctly describe the strong-field dynamics in the black-hole--particle system relevant to us, but some evidence suggests that they might (we return to discuss this point at the end of the section). 

The said results, as they are stated in \cite{isoyama14}, apply to a particle in a circular equatorial orbit (stable or unstable) around a Kerr black hole, ignoring the dissipative piece of the gravitational interaction (or, more precisely, time-symmetrizing the gravitational perturbation, so that spacetime admits a global helical symmetry). They express the Bondi\footnote{First-law literature \cite{alt1,alt2,alt3} usually alludes to ADM properties, which are defined even in helical symmetry within the PN context in which these works operate. In the context of black hole perturbation theory, the first-law results should be interpreted as referring to {\it Bondi} properties. See also \cite{Gralla:2012dm}, where first-law results are formulated directly in terms of Bondi quantities for a black-hole--particle system.}
energy and angular momentum of that configuration, through $O(\eta^2)$, in terms of Detweiler's redshift variable \cite{Detweiler:2008ft}
\beq
\hat z:= (\hat{u}^t)^{-1} ,
\eeq
where ${u}^t$ is the $t$ component of the four-velocity on the circular orbit, and overhats, recall, denote properties of the GSF-corrected orbit. The usefulness of such relations is in the fact that a computation of $\hat z$ requires only GSF information for circular orbits, and there is no need to integrate from infinity. Such information is essentially accessible to existing GSF codes.

Following \cite{isoyama14}, let us formally expand  the redshift $\hat z$ in powers of $\eta$, in the form 
\beq
\hat z = z_0(\Omega) + \eta z_1(\Omega) +O(\eta^2) ,
\eeq
where $\Omega(=d\phi/dt)$ is the circular orbit's angular velocity,
\beq
z_0=(1-a\Omega)^{1/2}\left[1+a\Omega-3(M\Omega)^{2/3}(1-a\Omega)^{1/3}\right]^{1/2}
\eeq
is the geodesic limit of $\hat z$, and $\eta z_1(\Omega)$ is the $O(\eta)$ GSF correction, defined for a fixed value of $\Omega$. According to Ref.\ \cite{isoyama14}, the Bondi energy and angular momentum of the circular-orbit binary are given, through $O(\eta^2)$, by 
\beq \label{ELADMtotal}
E_{\rm B}^{\rm sym}=M+\mu E^{\rm p}_{\rm B},\quad\quad L_{\rm B}^{\rm sym}=Ma+\mu L^{\rm p}_{\rm B},
\eeq
where 
\beq \label{ELBondi}
E^{\rm p}_{\rm B}=\tilde z-\Omega \frac{d\tilde z}{d\Omega},
\quad\quad
L^{\rm p}_{\rm B}=-\frac{d\tilde z}{d\Omega} ,
\eeq
with
\beq\label{z}
\tilde z(\Omega)=z_0(\Omega)+\frac{1}{2}\eta z_1(\Omega)+ O(\eta^2).
\eeq
The label `sym' is to remind us that these Bondi properties are defined in a time-symmetrized (``half-retarded-plus-half-advanced'') spacetime. 
The function $z_1(\Omega)$ explicitly determines $E_{\rm B}^{\rm sym}$ and $L_{\rm B}^{\rm sym}$ through $O(\eta^2)$.

We are now reaching the crux of our discussion.
Consider a critical orbit, subject to the conservative GSF alone (dissipation ignored), which asymptotes to a certain unstable circular orbit at $\tau\to\infty$. Let $E^{\rm sym}_{\rm B}(u)$ and $L^{\rm sym}_{\rm B}(u)$ be the Bondi energy and angular momentum of the corresponding time-symmetrized spacetime, with $u$ a suitable retarded-time coordinate. At $u\to\infty$, these quantities must approach the corresponding Bondi quantities of the asymptotic circular-orbit configuration, as given in Eq.\ (\ref{ELADMtotal}). Furthermore,
\beq
E^{\rm sym}_{\rm B}(u\to\infty)=E_{\rm ADM}, \quad\quad L^{\rm sym}_{\rm B}(u\to\infty)=L_{\rm ADM},
\eeq
where on the right-hand side we have the ADM properties of the physical (``retarded'') critical-orbit spacetime. [That this must be the case follows from $E^{\rm sym}_{\rm B}(u\to\infty)=E^{\rm sym}_{\rm ADM}-{\cal F}^+=E^{\rm sym}_{\rm ADM}-{\cal F}^-=E_{\rm ADM}$, where ${\cal F}^+$ and ${\cal F}^-$ are the total energies flowing, respectively, outward at future null-infinity and inward at past null-infinity, in the time-symmetrized setup where ${\cal F}^+={\cal F}^-$. A similar argument applies to the angular momentum.] As a result, we can write $E_{\rm ADM}=M+\mu E^{\rm p}_{\rm ADM}$ and $L_{\rm ADM}=M+\mu L^{\rm p}_{\rm ADM}$ [as in Eqs.\ (\ref{ADMspec})], with 
\beq \label{ELADM}
E^{\rm p}_{\rm ADM}=\tilde z-\Omega \frac{d\tilde z}{d\Omega},
\quad\quad
L^{\rm p}_{\rm ADM}=-\frac{d\tilde z}{d\Omega} .
\eeq
These expressions relate the ADM properties of the physical critical-orbit configuration to the redshift of the asymptotic circular orbit when dissipation is ignored.


The conservative GSF shift $\delta L^{\rm p}_{\rm ADM}(E^{\rm p}_{\rm ADM})$ [recall Eq.\ (\ref{deltaLADM})] may now be obtained simply by considering the $O(\eta)$ piece of $L^{\rm p}_{\rm ADM}$ in Eq.\ (\ref{ELADM}), for a fixed $E^{\rm p}_{\rm ADM}$. Equations (\ref{ELADM}) with (\ref{z}) immediately give us the $O(\eta)$ piece of $L^{\rm p}_{\rm ADM}$ for a fixed {\it angular velocity}: $\delta^{(\Omega)}L^{\rm p}_{\rm ADM}=-(\eta/2)dz_1/d\Omega$, where we introduced the operator $\delta^{(X)}$ to denote a linear variation with respect to $\eta$ at fixed $X$. To obtain the shift at fixed {\it energy},  $\delta L^{\rm p}_{\rm ADM}\equiv \delta^{(E)} L^{\rm p}_{\rm ADM}$, we write
\begin{eqnarray}
\delta L^{\rm p}_{\rm ADM}
&=& \delta^{(\Omega)} L^{\rm p}_{\rm ADM} +\frac{d L^{\rm p}_{\rm ADM}}{d\Omega}\, \delta^{(E)} \Omega
\nonumber\\
&=& \delta^{(\Omega)} L^{\rm p}_{\rm ADM} -\frac{d L^{\rm p}_{\rm ADM}}{dE^{\rm p}_{\rm ADM}}\, \delta^{(\Omega)}E^{\rm p}_{\rm ADM}
\nonumber\\
&= & 
\delta^{(\Omega)} L^{\rm p}_{\rm ADM} -\Omega^{-1}\delta^{(\Omega)}E^{\rm p}_{\rm ADM} ,
\end{eqnarray}
where in the second line we used $\delta^{(E)} \Omega=- (d\Omega/dE^{\rm p}_{\rm ADM})\delta^{(\Omega)}E^{\rm p}_{\rm ADM}$, and in the third line we applied $dL^{\rm p}_{\rm ADM}/dE^{\rm p}_{\rm ADM}=\Omega^{-1}$, which is valid for any circular geodesic (omitting subdominant terms in $\eta$).  From Eqs.\ (\ref{ELADM}) and (\ref{z}) we find $\delta^{(\Omega)}E^{\rm p}_{\rm ADM}=(\eta/2)\left[z_1-\Omega (dz_1/d\Omega)\right]$, and substituting this with the above result for $\delta^{(\Omega)} L^{\rm p}_{\rm ADM}$, we arrive at the simple expression
\beq\label{deltaLzgeneral}
\delta L^{\rm p}_{\rm ADM}=-\frac{\eta}{2\Omega}\, z_1.
\eeq

Note that in the analysis leading to Eq.\ (\ref{deltaLzgeneral}) we have not assumed anything about the spin $a$ of the central black hole, so the result should apply in general (and suggests an interesting new interpretation of $z_1$ in terms of a shift in the critical value of the angular momentum). In the extremal case, $\Omega=1/2 +O(\eps)$, so at leading order in $\eps$ we obtain 
\beq\label{deltaLADMz}
\delta L^{\rm p}_{\rm ADM}(E)=-\eta Z_1(E),
\eeq
where 
\beq \label{Z1}
Z_1(E):=\lim_{\eps\to 0} z_1(\Omega(E;\eps),\eps),
\eeq
with the limit taken {\em at fixed energy} $E$. 
Here, for clarity, we have made explicit the functional dependence of $z_1$ and $\Omega$ on $\eps$, and have parametrized the circular orbits by their geodesic energy $E$, noting that the difference between $E$ and $E^{\rm p}_{\rm ADM}$ is subdominant in Eq.\ (\ref{deltaLADMz}). Indeed, in practice, $Z_1(E)$ may be evaluated by considering a sequence of circular {\em geodesics} of diminishing $\eps$ (and a fixed $E$).

Equation (\ref{deltaLADMz}) may now be used with Eq.\ (\ref{deltaLADM}) to obtain the sought-for relation
\beq\label{deltaLz}
\delta\breve{L}_{\infty}^{\rm cons}(E)= E^2 + 1 - Z_1(E),
\eeq
which may then be used in place of (\ref{dLfinal}) in both conditions (\ref{OSfinal}) and (\ref{OSfinalFT}). The relation (\ref{deltaLz}) relieves us from the need to restrict attention to particles coming from infinity, which is why we have used in it the argument $E$ in place of $\Einf$. The energy may now take any value $E>E_{\rm isco}(=1/\sqrt{3})$, and the conditions (\ref{OSfinal}) and (\ref{OSfinalFT}) may be evaluated for all corresponding orbits. 

We may also use Eq.\ (\ref{deltaLADMz}) directly in conjunction with Eq.\ (\ref{OSfinalADM}), to write the censorship condition (in the generic case) in the remarkably simple form 
\beq\label{OSfinalZ}
Z_1(E)\geq \frac{1}{2}(E^2-1).
\eeq
Overspinning is averted if and only if this inequality holds for all $E>E_{\rm isco}$. The evaluation of the condition (\ref{OSfinalZ}) requires only redshift information on unstable circular orbits, evaluated at the extremal limit $\eps\to 0$ with $E$ held fixed.

It should be emphasized that the applicability of the theoretical framework underpinning Eq.\ (\ref{ELBondi}) is yet to be rigorously established within our black-hole-perturbative context. There is, however, accumulating evidence to suggest it. The first-law framework has been tested exhaustively against results from post-Newtonian theory in both the Schwarzschild \cite{alt2} and Kerr \cite{isoyama14} cases, and, more importantly for us, it has been shown to precisely reproduce certain rigorous GSF results in the strong-field regime, at least in the Schwarzschild case \cite{alt2}. Further reassurance is provided, in the Kerr case, by the agreement between the first-law framework and the perturbative Hamiltonian one \cite{isoyama14}. Ideally, $\delta\breve{L}_{\infty}^{\rm cons}$ should be evaluated via both Eq.\ (\ref{deltaLz}) and the more rigorous GSF formula (\ref{dLfinal}). Indeed, if nothing else, a demonstrated agreement between these two expressions would lend a strong support to the validity of both first-law and Hamiltonian frameworks in the strong-field regime.

\section{Assessment of numerical task}  \label{Sec:num}



A direct evaluation of the censorship condition (\ref{OSfinal}) requires the function $\delta\Linf^{\rm cons}(\Einf)$, which involves the conservative piece of the GSF along critical equatorial geodesics on a Kerr background, in the extremal limit. To explore the case of fine-tuned orbits, via the condition (\ref{OSfinalFT}), would require, in addition, a calculation of the radiative fluxes (either at infinity or down the event horizon) for unstable circular equatorial geodesics, again in the extremal limit. Finally, to evaluate the overspinning condition in its alternative form (\ref{OSfinalZ}) demands redshift data for these same unstable circular geodesics. To the best of our knowledge, these numerical tasks are beyond the capability of existing GSF codes---though, perhaps, not by much. We think that custom-built codes to produce the necessary data could be developed through relatively mild adaptations of existing codes. In this section we review relevant numerical methods that are currently available, and discuss how they should be customized. 



There now exist two computational frameworks for strong-field GSF calculations in Kerr geometry. One, by Dolan and collaborators \cite{Dolan}, tackles the metric perturbation equations in the time-domain (TD), via a direct numerical evolution of the hyperbolic set of linearized Einstein's equations in the Lorenz gauge. A judiciously designed ``puncture'' scheme is applied to extract the correct regular piece of the metric perturbation, from which the local GSF is calculated along the orbit. The second framework, due to Shah and collaborators \cite{GSFradgauge,Shah:2012gu}, is based on Teukolsky's perturbation formalism: The relevant perturbation equations are decomposed into Fourier-harmonic modes and tackled numerically mode-by-mode, in the frequency domain (FD). The GSF is then reconstructed using a recently formulated mode-sum regularization procedure \cite{PMB}. Both methods are underpinned by (the same) rigorous theory, and have been tested against each other and against results from post-Newtonian theory. All applications so far considered the GSF on fixed geodesic orbits, neglecting back-reaction on the orbit.  

The above two methods are somewhat complementary with respect to the range of problems they can tackle. The FD method does best with bound-orbit configurations, where the perturbation field admits a discrete spectrum. It is not immediately clear how to apply the method to orbits that come from infinity. 
The TD method, on the other hand, can handle all types of orbits equally well. The special case of {\it circular} orbits can be tackled by both methods, but much more efficiently in the FD, thanks to the simple spectrum of radiation from such orbits.


Let us consider first the more straightforward of the aforementioned numerical tasks: calculation of energy and angular-momentum fluxes from circular orbits (for the fine-tuning case). This is a standard calculation, most efficiently performed by solving Teukolsky's equation in the FD, either numerically \cite{hughes} or via the semi-analytical method of Mano, Suzuki and Takasugi \cite{MST}. Usually calculations focus on {\it stable} circular orbits as the source of radiation, but exactly the same techniques should be applicable without change to unstable orbits below the ISCO; the stability properties of the orbit are immaterial here.  
The only potential complication comes from the need to evaluate the fluxes in the extremal limit.
Care would need to be taken in correctly identifying an inner ``wave zone'' in which to impose the boundary conditions, for each finite value $\eps\ll 1$. It may prove convenient to work with a rescaled radial coordinate [such as $\bar r:=(r-R_{\rm eh})/\eps$] in order to better resolve the near-horizon wave dynamics. However, we do not see any issues of principle to hinder such a calculation, and it could be based on any of the existing platforms, such as the one by Shah {\it et al.} \cite{Shah:2012gu}.


We next turn to the calculation of $\delta\Linf^{\rm cons}(E)$ using the form (\ref{deltaLz}). This requires an evaluation of the redshift function $z_1(\Omega;\eps)$ [recall Eq.\ (\ref{Z1})] on unstable circular orbits, at the limit $\eps\to 0$. Calculations of $z_1$ for {\it stable} (or otherwise near-ISCO) geodesic orbits have been performed using both FD \cite{Shah:2012gu,isoyama14} and TD \cite{isoyama14} methods, in a non-extremal Kerr geometry. In the Schwarzschild case, such a calculation was performed (in the FD) even for unstable orbits, reaching very near the light ring \cite{Akcay}. The challenge is to extend these calculations to the near-extremal Kerr case. The issues here are similar to the ones affecting fluxes. In the FD, one would need to carefully set inner boundary conditions, and also carefully monitor the convergence of the multipole mode-sum, particularly at large energy (lessons can be learned from the Schwarzschild analysis of \cite{Akcay}). Some development, tests, and a careful error analysis would be required, but the problem seems to us perfectly tractable. 

If one is satisfied with the level of rigour provided by the first-law and Hamiltonian formulations, then no further calculations would be needed: The question of overspinning, for both generic and fine-tuned orbits, can be answered based on numerical data pertaining to circular orbits only. However, to establish a full confidence in the results, a direct evaluation of $\delta\Linf^{\rm cons}(E)$ via Eq.\ (\ref{dLfinal}) would be required. Since this involves GSF data on unbound orbits, a TD treatment is preferable. So far, TD calculations of the GSF in Kerr have been limited to circular (and equatorial) orbits in a non-extremal geometry \cite{dandb}. However, it should not be hard to implement orbits that arrive from infinity, including critical orbits in the extremal case, at least in principle. To achieve this, certain technical details would need to be addressed. In the rest of this section we give an assessment of the challenges involved. 

First, to probe the extremal limit, we would need GSF data for a sequence of spacetimes approaching extremality. In the current code \cite{dandb}, accuracy has been observed to degrade with increasing $a$. In the circular-orbit case the code can handle spins up to $a=0.7$--$0.8M$, beyond which the loss of accuracy is rapid. The cause of the accuracy degradation is not quite clear yet, but work towards a full diagnosis of the problem, and towards its resolution, is now in progress. It appears likely that the code could be pushed to very high spins with only moderate effort. 

Second, one would need to implement plunging (geodesic, critical) orbits. In a TD framework this would entail only a minor coding effort; the basic code architecture should remain intact. Two of the technical details that would need addressing are (1) the handling of the auxiliary worldtube (see \cite{Dolan} for details), which would now be moving to track the radial motion of the particle; and (2) the treatment of ``junk radiation'' (again, see \cite{Dolan} for details), whose problematic effect is expected to be more pronounced than for circular orbits. A method for tackling the latter problem has been developed and successfully implemented in the electromagnetic case, by Zimmerman {\it et al.}\ \cite{zimm}. It remains to test its performance in the gravitational case.  

Perhaps the most significant remaining problem is that of the $m=1$ mode instability---an issue identified and thoroughly analyzed in \cite{Dolan}. In numerical experiments, this particular azimuthal mode of the Lorenz-gauge metric perturbation appeared to develop a linear instability at late time, which so far could not be cured. The seed of the instability appears to be a certain non-radiative, Lorenz-gauge mode, which is perfectly regular on each time slice. Various methods have been tried in attempt to filter that mode out of the numerical solutions, so far without much success. A simple filter can be applied in the circular-orbit case \cite{Dolan}, giving a satisfactory ad-hoc solution, but the method would not work for non-circular orbits.   We are aware of at least two parallel efforts, by two groups, to obtain a more fundamental solution to the problem, and we remain optimistic that the issue will be resolved soon.

The problem of $m=1$ mode instability is entirely avoided within a third computational framework, now being developed by a collaboration involving one of the authors (LB). The new method combines the simplicity of the Teukolsky formulation with the utility and flexibility of the TD approach. It essentially involves a TD implementation of the Teukolsky equation, together with a scheme for constructing the GSF in a certain (non-Lorenz) gauge. When completed, the code will offer a most natural tool for performing the calculation required here. 

\section{Summary and concluding remarks} \label{Sec:conclusions}

We studied the scenario in which a massive (spinless) particle is dropped into a nearly-extremal Kerr black hole on an equatorial-plane trajectory. For this scenario, we presented a systematic analysis of the censorship condition at first order in the self-force approximation. One of our main goals was to determine what GSF information, precisely, would be needed in order to provide a definitive answer to the question of whether an over-extremal black hole was a possible outcome (within the classical theory). We achieved this by formulating concrete, necessary and sufficient conditions for overspinning to be averted; these are given explicitly in terms of GSF quantities, ready for numerical evaluation.

Along the way, we have established several interesting results:
\begin{itemize}
\item When the GSF is ignored, overspinning can be achieved in a certain open domain of the parameter space, mapped here precisely for the first time. 
\item Overspinning is possible (when the GSF is ignored) only with particles thrown in from infinity. For any value of the initial energy-at-infinity, overspinning can be achieved by choosing the particle's rest mass and angular momentum from within certain open intervals, as prescribed in the last sentence of Sec.\ \ref{Sec:geodesics}, below Eq.\ (\ref{calEmp}).
\item In the full-GSF case, a sufficient and necessary censorship condition for ``generic'' orbits may be formulated on the one-parameter family of critical geodesics. That condition is sensitive only to the {\em conservative} piece of the GSF. 
\item A more general condition may be formulated to encompass also fine-tuned orbits (ones whose parameters are exponentially fine-tuned to produce an adiabatic evolution along the global attractor). That condition involves also the radiative fluxes from unstable circular geodesics.    
\item The conservative GSF effect may be reformulated in terms of the ``redshift'' variable. This results in an alternative form of the censorship conditions, which involves only perturbative quantities (redshift and radiative fluxes) calculated on unstable circular geodesics.  
\end{itemize}

Our main results are expressed in Eqs.\ (\ref{OSfinal}), (\ref{OSfinalFT}) and (\ref{OSfinalZ}). Equation (\ref{OSfinal}) [with (\ref{dLfinal})] is the censorship condition for generic orbits. It required as input $\delta\breve{L}_{\infty}^{\rm cons}(\Einf)$, the conservative GSF correction to the critical value of the angular momentum at a fixed $\Einf$. The condition is a 
{\it sufficient} one in the sense that its validity for all $\Einf\geq 1$ would imply that censorship is protected in our scenario, at least for sufficiently small values of the particle's rest mass (this last caveat is to remind us that our analysis and conditions are formulated within the first-order self-force approximation). Equation (\ref{OSfinal}) is also a {\it necessary} condition, in the sense that its violation for any value $\Einf\geq 1$ would mean a direct infringement of cosmic censorship.

Equation (\ref{OSfinalFT}) [with (\ref{FH_ft}) or (\ref{FH_ft2})] is the more general censorship condition that covers also fine-tuned orbits. Its evaluation requires, in addition to $\delta\breve{L}_{\infty}^{\rm cons}$, also the radiative quantities $\breve\calE^+$ and $\breve\calW^+_{\rm (qc)}$ (or $\breve\calE^-$ and $\breve\calW^-_{\rm (qc)}$) associated with the adiabatic evolution along the global attractor. Without fine-tuning, $\breve\calE^\pm$ and $\breve\calW^\pm_{\rm (qc)}$ vanish at relevant order, and (\ref{OSfinalFT}) reduces to the generic condition (\ref{OSfinal}). With fine-tuning, the condition (\ref{OSfinalFT}) should be evaluated on the two-parameter space of initial and final energies, $\{\Einf,E_f\}$. It is both  sufficient and necessary, in the same sense as (\ref{OSfinal}).

Finally, Eq.\ (\ref{OSfinalZ}) is a reformulation of (\ref{OSfinal}) in terms of the redshift variable $Z_1(E)$, calculated on unstable circular geodesics in the extremal limit (taken with fixed $E$). The more general condition (\ref{OSfinalFT}) may also be formulated in terms of the redshift, by substituting for $\delta\breve{L}_{\infty}^{\rm cons}$ using Eq.\ (\ref{deltaLz}). This alternative form is more readily amenable to numerical evaluation, because it requires only circular-orbit information. An additional advantage is that (\ref{OSfinalZ}) is applicable to any value $\Einf>E_{\rm isco}$ of the initial energy, without the restriction that particles have to be sent from infinity. 
However, the redshift formulation relies on some layers of non-rigorous theory, so a direct evaluation of the conservative GSF effect, via Eq.\ (\ref{dLfinal}), would be desirable as a check. 


We are unable to predict, in advance of a numerical calculation, whether our censorship conditions (\ref{OSfinal}) or (\ref{OSfinalZ}) hold. We are not familiar with an argument to suggest even the {\em signs} of the terms $\delta\breve{L}_{\infty}^{\rm cons}$ and $Z_1$ appearing in these conditions.
However, if we {\em assume} that the generic-orbit conditions (\ref{OSfinal}) [or (\ref{OSfinalZ})] are satisfied, then it is possible to conclude that censorship is protected also for fine-tuned orbits [i.e., the inequality (\ref{OSfinalFT}) holds], assuming only a certain plausible lower bound on the flux ratio $\calR(E)$. Specifically, one can show, with the help of Eqs.\ (\ref{calEplus}) and (\ref{calWplus}), that (\ref{OSfinalFT}) is satisfied for all $\Einf>E_f\geq E_{\rm isco}$ if $\calR(E)\geq-\frac{1}{3}$. This lower bound lies comfortably below the ISCO value $\calR(E_{\rm isco})\sim -0.1$ estimated in Ref.\ \cite{Kapadia}, and we expect unstable circular orbits to be {\em less} superradiant than the ISCO [i.e, have $\calR(E)>\calR(E_{\rm isco})$] on account of their frequency being larger than the ISCO frequency (recall also $\calR>0$ for $E>\frac{2}{\sqrt{3}}$). If a numerical calculation of $\calR(E)$ confirms this expectation, it would mean that {\em fine-tuning disfavours overspinning}. In that case, establishing the simple inequality (\ref{OSfinal}) [or (\ref{OSfinalZ})] would suffice for ruling out the overspinning scenario.

\section*{Acknowledgements}

We have benefited from discussions with many colleagues, including Enrico Barausse, Vitor Cardoso, Carsten Gundlach, Soichiro Isoyama, Alexandre Le Tiec, Maarten van de Meent, Amos Ori, Eric Poisson, Adam Pound, Norichika Sago and Takahiro Tanaka.
We are particularly grateful to Maarten van de Meent, Abhay Shah and Niels Warburton for providing unpublished numerical data to test certain aspects of our analysis; and to Maarten van de Meent and Adam Pound for their valuable comments on a draft of this manuscript. We gratefully acknowledge support from the European Research Council under the European Union's Seventh Framework Programme FP7/2007-2013/ERC, Grant No.\ 304978.  LB acknowledges additional support from STFC through grant number PP/E001025/1.    
         					
\appendix

\section{ADM energy and angular momentum}\label{App:ADM}

This appendix gives a detailed derivation of Eqs.\ (\ref{EADM}) and (\ref{LADM}). We consider two relativistic point particles in flat space, representing the black hole--particle system at an infinite separation. We are given the two rest masses, $M$ and $\mu$, and the particle's energy $\mu\Einf(>\mu)$ and angular momentum $\mu\Linf$ in a reference frame attached to the mass $M$. The mass $M$ has an intrinsic spin $M a$ perpendicular to the orbital plane, and the mass $\mu$ is spinless.  Our goal is to obtain the system's total energy and angular momentum in the center-of-mass (CoM) frame.   

First, we note that, in the limit of infinite separation, both black-hole frame and CoM frame are inertial, and they are related via a simple Lorentz boost. Let $x^{\alpha}=(t,x,y,z)$ be a Cartesian frame centered at $M$, so that the spin of $M$ is aligned with the $z$ direction, and the particle's orbit lies in the $x$--$y$ plane. Denote the four-momenta of $\mu$ and $M$ in the black-hole frame by
\begin{eqnarray}
p^{\alpha}_{(\mu)}&=&(\mu\Einf,p_{(\mu)}^x,p_{(\mu)}^y,0), \nonumber\\
p^{\alpha}_{(M)}&=&(M,0,0,0).
\end{eqnarray}
The magnitude of particle's 3-momentum satisfies
\beq\label{norm}
|\mathbf{p}_{(\mu)}|^2=(p_{(\mu)}^x)^2+(p_{(\mu)}^y)^2 = \mu^2(\Einf^2-1).
\eeq
The CoM system $\tilde x^{\alpha}$ is related to $x^{\alpha}$ via a Lorentz boost $\Lambda^{\alpha}_{\ \beta}$ in the $x$--$y$ plane. Let $\tilde p^{\alpha}_{(\mu)}=\Lambda^{\alpha}_{\ \beta}p^{\beta}_{(\mu)}$ and $\tilde p^{\alpha}_{(M)}=\Lambda^{\alpha}_{\ \beta}p^{\beta}_{(M)}$ denote the 4-momenta of $\mu$ and $M$ in $\tilde x^{\alpha}$. The CoM condition,
\beq
\mathbf{\tilde p}_{(M)}+\mathbf{\tilde p}_{(\mu)} = 0 ,
\eeq
yields two nontrivial equations for the two boost parameters $\beta^x=v^x/c$ and $\beta^y=v^y/c$ (where $v^x$ and $v^y$ are components of the boost velocity). One finds $\beta^i=p^i_{(\mu)}(M+\mu\Einf)^{-1}$ for $i=x,y$. Thus, using (\ref{norm}),
\beq\label{beta}
\beta=[(\beta^x)^2+(\beta^y)^2]^{1/2}=\frac{\eta(\Einf^2-1)^{1/2}}{1+\eta\Einf},
\eeq
where $\eta=\mu/M$.

Now that we have at hand the boost $\Lambda^{\alpha}_{\ \beta}(\beta)$, the relativistic energies of $\mu$ and $M$ in the CoM system are obtained as $\tilde p^{0}_{(\mu)}=\Lambda^{0}_{\ \beta}p^{\beta}_{(\mu)}$ and $\tilde p^{0}_{(M)}=\Lambda^{0}_{\ \beta}p^{\beta}_{(M)}$. The sum of these two energies is the total, ADM energy of the system. A short calculation gives
\beq
E_{\rm ADM}=\tilde p^{0}_{(\mu)}+\tilde p^{0}_{(M)}=
M(1+2\eta \Einf+\eta^2)^{1/2}.
\eeq
This result is valid for any mass ratio $\eta$. For $\eta\ll 1$, an expansion in $\eta$ through $O(\eta)$ gives Eq.\ (\ref{EADM}).  

To obtain the total angular momentum in the CoM system, we need first to relate the particle's CoM position $\mathbf {\tilde x}_{(\mu)}$ and black hole's CoM position $\mathbf {\tilde x}_{(M)}$ to their separation ${\mathbf x}$ in the black-hole frame. This is achieved by solving the CoM condition 
\beq
\mathbf {\tilde x}_{(\mu)} \tilde p^{0}_{(\mu)}
+ \mathbf{\tilde x}_{(M)} \tilde p^{0}_{(M)} = 0,
\eeq 
simultaneously with 
$|\mathbf{\tilde x}_{(\mu)}|+|\mathbf{\tilde x}_{(M)}| = |\mathbf{\tilde x}|$,
where $\tilde x^i = \Lambda^{i}_{\ \beta}x^{\beta}$ ($i=x,y$) is the separation in the CoM frame. The particle's CoM orbital angular momentum is then given by $\tilde L_{(\mu)}=\tilde x_{(\mu)}\tilde p^{y}_{(\mu)}-\tilde y_{(\mu)}\tilde p^{x}_{(\mu)}$, and similarly for the mass $M$. The total, ADM angular momentum of the system (with respect to the CoM) is the sum of $\tilde L_{(\mu)}$, $\tilde L_{(M)}$ and the spin angular momentum. A short calculation, using Eqs.\ (\ref{norm}) and (\ref{beta}) and the relation $\mu\Linf=x p_{(\mu)}^y-y p_{(\mu)}^x$, gives
\beq 
L_{\rm ADM}=M\left(a + \mu\Linf/E_{\rm ADM}\right) .
\eeq
This is valid for any $\eta$. Substituting for $E_{\rm ADM}$ and expanding through $O(\eta^2)$ produces Eq.\ (\ref{LADM}).

\section{Shift in the critical value of the angular momentum due to the full self-force}\label{App:dL}

In this appendix we derive Eq.\ (\ref{deltaL2}), which describes the GSF-induced shift $\delta\Linf$ in the critical value of the angular-momentum-at-infinity for a given energy-at-infinity, at leading order in $\eta$ and in $\eps$. In Section \ref{subsubsec:cons} we obtained an expression for $\delta\Linf$ [Eq.\ (\ref{deltaLcons2})] under the (non-physical) simplifying assumption that the GSF has no dissipative piece. Here we restore dissipation and calculate the shift $\delta\Linf$ caused by the full GSF.

In our analysis we only require the leading order [$O(\eta)$] of $\delta\Linf$. As explained in the text, for that purpose it is sufficient to ignore fine-tuning and assume the changes in (specific) energy and angular momentum along the critical orbit are $O(\eta\ln\eta)$ at most.

Consider, then, a critical orbit parametrized by $\Einf(\geq 1)$, and an arbitrary moment $\tau=\tau_0$ after the orbit had settled into quasi-circular motion, but before $\Delta E(\tau_0)$ and $\Delta L(\tau_0)$ have accumulated $O(1)$ changes [specifically, assume $\Delta E(\tau_0),\Delta L(\tau_0)=O(\eta\ln\eta)$ at most]. 
Evaluate the full-GSF radial equation of motion (\ref{rdotGSF}) at time $\tau_0$, subject to the near-circularity condition $dr/d\tau=O(\eta)$, substituting for $\hat E(\tau_0)$ and $\hat L(\tau_0)$ from Eq.\ (\ref{hatEL}), and replacing $\Linf$ with $L_c(\Einf) + \delta\Linf(\Einf)$.
At leading order in $\eta$ and in $\eps$ one obtains the solution
\beq \label{deltaL4}
\delta \Linf(\Einf)=\Delta W(\tau_0) ,
\eeq
where $\Delta W$ is the GSF integral defined in Eq.\ (\ref{DeltacalE}). 
This result can only make sense if (i) the expression on the right-hand side is in fact independent of $\tau_0$ at leading order; and (ii) the quantity $\Delta W(\tau_0)=2\Delta E(\tau_0)-\Delta L(\tau_0)$ remains of $O(\eta)$ even for $\tau_0$ large enough that the individual terms $\Delta E(\tau_0)$ and $\Delta L(\tau_0)$ are already of $O(\eta\ln\eta)$. We now argue that both conditions are satisfied. 

To make the argument, let us split $\Delta W(\tau_0)$ into an ``approach'' piece $\Delta W(\tau_c)$, and a ``quasi-circular'' piece
\begin{eqnarray}\label{DeltacalEwhirl}
\Delta W(\tau_c,\tau_0)&:=&\Delta W(\tau_0)-\Delta W(\tau_c)
\nonumber\\
&=&
-\eta^{-1} \int_{\tau_c}^{\tau_0}\left(2F_t+ F_{\phi}\right)d\tau .
\end{eqnarray}
Here the end-of-approach time $\tau_c$ can be taken to be any moment $\tau_c<\tau_0$ after the orbit had settled into quasi-circular motion, in the sense that $\dot{r}(\tau_c)=O(\eta)$ at most. 
We will now show that  
\beq\label{scaling}
\Delta W(\tau_c,\tau_0)=O(\eps)O(\eta\ln\eta)
\eeq
at most, for any choice of $\tau_0$ and of $\tau_c$. Assuming $\eps|\ln\eta|\ll 1$, this would mean that $\Delta W(\tau_0)$ is dominated by its approach piece $\Delta W(\tau_c)$, so that (i) $\Delta W(\tau_0)$ does not depend on $\tau_0$ at leading order, and (ii) $\Delta W(\tau_0)\simeq \Delta W(\tau_c)=O(\eta)$ as argued above.


To establish the scaling in (\ref{scaling}), use the orthogonality relation $\hat u^{\alpha}F_{\alpha}=0$ to write $F_t+\Omega F_{\phi}=-(u^r/u^t)F_r$, where $\Omega=u^{\phi}/u^t$, and the replacement $\hat u^{\alpha}\to u^{\alpha}$ does not affect the expression at leading order in $\eta$. We wish to evaluate this relation for $\tau_c\leq \tau\leq \tau_0$, when the orbit is quasi-circular. At leading order in $\eta$, $\Omega=\Omega(E;\eps)$ is then equal to the angular velocity of an unstable circular geodesic of specific energy $E=\hat E(\tau)$. At fixed energy, the angular velocity admits the small-$\eps$ expansion $\Omega=\frac{1}{2}-\frac{1}{4}b(E)\eps +O(\eps^2)$ with $b=6E(6E^2-2)^{-1/2}$ [see Eqs.\ (\ref{bofE}) and (\ref{OmegaExpansion})].
Thus, omitting terms that are subdominant in $\eps$, the integrand in Eq.\ (\ref{DeltacalEwhirl}) is
$2F_t+ F_{\phi}=-2(u^r/u^t)F_r +\frac{1}{2}b\eps F_{\phi}$,
or, equivalently,
\beq\label{ort2}
2F_t+ F_{\phi}=-2(u^r/u^t)F_r -b\eps F_{t}.
\eeq
Let us denote the contributions to $\Delta W(\tau_0,\tau_c)$ from the first and second terms on the right-hand side of (\ref{ort2}) by $\Delta W_{(r)}$ and $\Delta W_{(t)}$, respectively.
In what follows we consider each of the two contributions in turn.

Start with  $\Delta W_{(r)}$, given by 
\beq \label{DeltacalEr}
\Delta W_{(r)}=
(2/\eta) \int_{r(\tau_c)}^{r(\tau_0)}\left(F_r/u^t\right)dr .
\eeq
Note $r(\tau_0)-r(\tau_c)=O(\eps)$, since both radii belong to unstable circular geodesics. From Eq.\ (\ref{Frnearcirc}) we recall that for $|\dot{r}|\ll 1$ the radial component $F_r$ is dominated by its conservative piece, $F_{r}^{\rm cons} \simeq F_{0r}(r)$, which is approximately constant within the integration domain and may therefore (as we only keep track of the leading term in $\eps$) be pulled out of the integral. A simple calculation gives $1/u^t\propto \eps$ (at fixed $E$), and this factor can likewise be taken out of the integral. We thus obtain the scaling $\Delta W_{(r)}\sim \eps^2 F_{0r}/\eta$,
where $F_{0r}$ is evaluated, e.g., at $r=r(\tau_c)$, the end of approach.
It remains to determine the $\eps$-scaling of $F_{0r}$.
Numerical evidence, to be presented elsewhere \cite{Colleoni_etal}, suggests the scaling $F_{0r}\sim \eps^{-1}$. This is consistent with what one would obtain by assuming that the GSF components in a normalized coordinate basis are finite: $F_{0r}=F_{0\hat{r}}(g^{rr})^{-1/2}\sim \eps^{-1}$ (or smaller), assuming the normalized component $F_{0\hat{r}}$ is finite and noting $g^{rr}\sim\eps^2$. 
Assuming, therefore, that  $F_{0r}$ does not diverge faster than $\sim \eps^{-1}$, and recalling $F_{0r}\propto\eta^2$ as usual, we arrive at
\beq\label{calEr_scaling}
\Delta W_{(r)}=O(\eps)O(\eta)
\eeq
(or smaller).
 

Next, consider the contribution $\Delta W_{(t)}$:
\begin{eqnarray}
\Delta W_{(t)}&=&
(\eps/\eta)\int_{\tau_0}^{\tau_c}b(E) F_{t}d\tau
=-\eps \int_{E(\tau_c)}^{E(\tau_0)}b(E)dE
\nonumber\\
&=&
-\left.\eps\sqrt{6E^2-2}\right|_{E(\tau_c)}^{E(\tau_0)}.
\end{eqnarray}
In the second equality we have used $F_{t}=-\mu\, dE/d\tau$, and in the third we have substituted for $b(E)$ and integrated explicitly. Since the energy difference $E(\tau_0)-E(\tau_c)$ is at most of $O(\eta\ln\eta)$, we conclude that 
\beq\label{calEt_scaling}
 \Delta W_{(t)} =  O(\eps)O(\eta\ln\eta)
\eeq
(or smaller).

The combination of Eqs.\ (\ref{calEr_scaling}) and (\ref{calEt_scaling}) leads to the scaling stated in Eq.\ (\ref{scaling}). The upshot is that the contribution to $\Delta W(\tau_0)$ from the quasi-circular part, $\Delta W(\tau_c,\tau_0)$, is negligible compared to the contribution from the approach part, $\Delta W(\tau_c)=O(\eps^0)O(\eta)$ (assuming $\eps|\ln\eta|\ll 1$). In other words, the GSF integral   $\Delta W(\tau_0)$ in Eq.\ (\ref{deltaL4}) may be truncated at the end-of-approach time $\tau_c$, with the latter taken to be any instance after the orbit had settled into quasi-circular motion [but before the specific energy has accumulated $O(1)$ changes]. This establishes Eq.\ (\ref{deltaL2}) in the main text.

\section{Radiation from transition to plunge and final plunge}\label{App:plunge}

In this appendix we argue that the term $\calW^+_{\rm (end)}$ in Eq.\ (\ref{OSGSF2}) is subdominant for $\eta,\eps\ll 1$ and may therefore be dropped within our leading-order analysis. Recall $\calW^+_{\rm (end)}=2\calE^+_{\rm (end)}-\calL^+_{\rm (end)}$, where $\calE^+_{\rm (end)}$ and $\calL^+_{\rm (end)}$ are the energy and angular momentum radiated out to infinity during the transition from adiabatic inspiral along the attractor to a final plunge into the black hole, and during the plunge itself. 

Critical orbits may transit into plunge in one of two ways: If the orbit is perfectly fine-tuned, the transition will occur around the location of the ISCO, and will then be similar to the transition at the end of a physical adiabatic inspiral (on a {\it stable} orbit) that has already been studied in detail \cite{orithor,Buonanno:2000ef,Kesden,Harada}. With any less than perfect fine-tuning, the particle will slide off the peak of the effective potential and into the black hole before the ISCO is reached (cf.\ Fig.\ \ref{attractor}). It is reasonable to expect the former scenario (transition through the ISCO) to yield the maximal radiation output, because (i) orbits linger much longer around the ISCO location, where the potential is very flat, than they do around the peak of the potential; and (ii) the remaining distance to the horizon is maximal when the transition is at the ISCO.  Below we start by looking at the ``worse case'' scenario of transition through the ISCO, and argue that, even in that case, $\calW^+_{\rm (end)}$ is negligible in Eq.\ (\ref{OSGSF2}). We then also examine the more generic scenario of a transition from an unstable orbit, and, as expected, arrive at a similar conclusion.

\subsection{Plunge from the ISCO}

For a transition through the ISCO, we use results by Ori and Thorne \cite{orithor}, who studied radiation from the transition regime at the end of a quasi-circular inspiral. An inspection of their analysis reveals that the main results are insensitive to whether the particle arrives the transition regime along stable or unstable orbits. In particular, their equation (3.26) for the deficits in (specific) energy and angular moment over the entire transition should hold in either case. We write it here in the form 
\begin{eqnarray}\label{trans1}
(\Delta L)_{\rm trans} &=& - {\cal A}(a)\eta^{4/5} , \nonumber\\
(\Delta E)_{\rm trans} &=& - \Omega_{\rm isco}{\cal A}(a)\eta^{4/5},
\end{eqnarray}
where ${\cal A}(a)$ is a certain (positive) function of the spin $a$ only, given explicitly in \cite{orithor}. These expressions hold for any $a$, at leading order in $\eta\ll 1$. The factor $\eta^{4/5}$ arises, essentially, from the fact that the transiting orbit spends an amount of proper time $\propto \eta^{-1/5}$ whirling around the ISCO location on a nearly circular orbit [cf.\ Eq.\ (3.20) of \cite{orithor}].

The function ${\cal A}(a)$ involves a certain $a$-dependent dimensionless factor, denoted $\dot\calE$ in \cite{orithor} (not to be confused with our $\calE^{\pm}$), which describes corrections to the leading-order quadrupole emission formula for circular orbits, and is to be determined numerically. Kesden \cite{Kesden} used numerical data by Hughes (from a code presented in \cite{Hughes:1999bq}) to estimate, in the near-extremal case,  $\dot\calE\propto \eps^{2/3}$, a scaling previously suggested by Chrzanowski \cite{Chrzanowski}. We have confirmed this scaling using much more accurate numerical results by M.\ van de Meent \cite{priv_comm}, to be presented elsewhere \cite{Colleoni_etal}. We also note that the scaling $\dot\calE\propto \eps^{2/3}$ follows simply from a regularity assumption, namely that $dE/d\tau$ must remain bounded (and nonzero) even in the limit $\eps\to 0$: Noting that the Ori-Thorne function $\dot\calE$ is defined with respect to coordinate time $t$ (not proper time $\tau$), and that $(d\tau/dt)_{\rm isco}\propto \eps^{2/3}$, we obtain $\dot\calE\propto (dE/d\tau)(d\tau/dt)_{\rm isco}\propto \eps^{2/3}$. Assuming this scaling, and expanding the remaining $a$-dependence of ${\cal A}(a)$ in $\eps$, we obtain, at leading order in $\eps$,
\begin{eqnarray}\label{trans2}
(\Delta L)_{\rm trans} &=&- a_0 \eps^{-2/15}\eta^{4/5} ,\nonumber\\
(\Delta E)_{\rm trans} &=& - a_0 \Omega_{\rm isco}\eps^{-2/15}\eta^{4/5},
\end{eqnarray}
with some (positive) numerical coefficient $a_0$ whose explicit value will not be needed here.

Refs.\ \cite{Kesden,Harada} discuss the reason for the non-physical divergence of $(\Delta E)_{\rm trans}$ and $(\Delta L)_{\rm trans}$ when $\eps\to 0$ is taken with a fixed $\eta$, but this will not concern us here. We are instead interested in the combination $(\Delta W)_{\rm trans}:=2(\Delta E)_{\rm trans}-(\Delta L)_{\rm trans}$, which, noting $\Omega_{\rm isco}=\frac{1}{2}+O(\eps^{2/3})$ and keeping only the leading term, reads
\begin{eqnarray}\label{trans3}
(\Delta W)_{\rm trans} = b_0 \eps^{8/15}\eta^{4/5},
\end{eqnarray}
with some (positive) numerical coefficient $b_0$.
Assuming the usual balance between the local dissipative GSF and the flux of energy and angular momentum in gravitational waves emitted during the transition, we have $\calW^+_{\rm trans}=-\eta(\Delta W)_{\rm trans}$ and thus
\begin{eqnarray}\label{trans3}
\calW^+_{\rm trans} =-b_0 \eps^{8/15}\eta^{9/5}.
\end{eqnarray}

Now examine the magnitude of $\calW^+_{\rm trans}$ compared to that of other terms in the censorship condition (\ref{OSGSF2}). If $\eps>\eta$, we have $\eps^{8/15}\eta^{9/5}<\eps^{7/3}$, which, for $\eps\ll 1$, is much smaller than the $\eps^2$ term in Eq.\ (\ref{OSGSF2}). If instead we have $\eps\leq \eta$, then $\eps^{8/15}\eta^{9/5}\leq \eta^{7/3}$, which, for $\eta\ll 1$, is much smaller than the $O(\eta^2)$ terms in that equation. The conclusion is that the contribution to $\calW^+_{\rm (end)}$ from the transition regime, $\calW^+_{\rm trans}$,  is always subdominant in Eq.\ (\ref{OSGSF2}) for $\eta,\eps\ll 1$.

It remains to assess the contribution to $\calW^+_{\rm plng}$ from the final plunge into the hole. In general (when the black hole is not near-extremal) one identifies a final stage, extending smoothly from the transition regime, where radiation reaction is negligible and the orbit plunges into the black hole on a nearly geodesic trajectory \cite{orithor,Mino}. The picture may change a little in the near-extremal case, because the radial velocity remains small, $|\dot{r}|\propto\eps\ll 1$, all the way to the horizon. This can mean that the conditions that define the transition regime never quite break down before the the horizon is reached. In other words, there is a possibility that the particle crosses the horizon while still in the transition regime.\footnote{This possibility was studied in some detail in Ref.\ \cite{Kesden}; see, in particular, Fig.\ 5 therein, in which `$\delta$' is equivalent to our $\eps^2$.}

That possibility can be assessed using Eq.\ (3.20) of \cite{orithor}, according to which the radial extent $(\Delta r)_{\rm trans}$ of the transition regime, in the near-extremal case, is $\propto \eps^{4/15}\eta^{2/5}$, with a coefficient of order unity. This should be compared with the radial distance from the ISCO to the horizon, $\Delta r\simeq (2\eps)^{2/3}$ [recall Eqs.\ (\ref{reh}) and (\ref{risco})]. It follows that for $\eps/\eta$ smaller than a number of order unity, the transition regime extends to the horizon. In such cases, the amplitude of $\calW^+_{\rm trans}$ in Eq.\ (\ref{trans3}) serves as an upper bound for the amplitude of $\calW^+_{\rm (end)}$, and it follows immediately that the entire term $\calW^+_{\rm (end)}$ is negligible in Eq.\ (\ref{OSGSF2}). 

Let us then consider the case where the transition ends before the horizon is reached, so that a plunge regime is identifiable. In the plunge regime, by definition, the motion is very nearly geodesic, and the near-horizon analysis of Mino and Brink \cite{Mino} should be applicable. Ref.\ \cite{Mino} obtained (among other things) an analytic expression for the energy output from the final plunge, by analyzing solutions to Teukolsky's equation in the near-horizon, low-frequency, quadrupole approximations. They find (in our notation)\footnote{See Eq.\ (4.3) of \cite{Mino}, noting their $\kappa$ is $\sqrt{2}\, \eps$ in our notation. The discussion following that equation seems to ignore the $\eps$ dependence implicit in the factors $\exp[-\kappa(T-t_0)/r_+]\sim (r_{\rm init}-r_+)/r_+$ and $(E-\Omega_H L)$.}\textsuperscript{,}\footnote{It is not clear to us whether the low-frequency and quadrupole approximations, introduced in \cite{Mino} to enable analytic calculation, are justifiable for near-horizon orbits in the near-extremal case (where the dimensionless angular velocity is $\sim 1/2$). We prefer to regard the form of Eq.\ (\ref{Eplunge}) as indicative only, but this should suffice for our purpose.}
\beq\label{Eplunge}
\calE^+_{\rm plung} \propto \eta^2 \eps^5 (r_{\rm init}-r_{\rm eh})(E-\Omega_H L)^{-2}
\eeq 
at leading order in $\eps$, where $r_{\rm init}$ is the radius at the start of the plunge, and $r_{\rm eh}$ is the horizon's radius (denoted $R_{\rm eh}$ in the main text). For a plunge following a transition at the ISCO, $r_{\rm init}-r_{\rm eh}=O(\eps^{2/3})$ and $E-\Omega_H L=O(\eps)$, giving $\calE^+_{\rm plung} =O( \eta^2 \eps^{11/3})$. Thus $\calE^+_{\rm plung}$ is strongly suppressed at small $\eps$. 
Since the motion is nearly circular even during the plunge, we have $\calL^+_{\rm plung}\simeq \Omega^{-1}\calE^+_{\rm plung}\simeq 2\calE^+_{\rm plung}$ and the radiated angular momentum is similarly suppressed. The combination $\calW^+_{\rm plung}:=2\calE^+_{\rm plung}-\calL^+_{\rm plung}$ is even more strongly suppressed at small $\eps$, and clearly contributes negligibly to $\calW^+_{\rm (end)}$. 

In conclusion, we have found that, for a transition through the ISCO, $\calW^+_{\rm (end)}=\calW^+_{\rm trans}+\calW^+_{\rm plunge}\simeq \calW^+_{\rm trans}$ is always negligible in Eq.\ (\ref{OSGSF2}) for $\eta,\eps\ll 1$. That radiation from the transition to plunge should have a negligible effect on the conditions for overspinning was previously suggested by Kesden \cite{Kesden} and Harada and Kimura \cite{Harada}.

\subsection{Plunge from an unstable circular orbit}

This scenario is rather different from---and much simpler than---a transition through the ISCO. As the orbit is perturbed away from unstable equilibrium, its subsequent evolution is almost immediately controlled by the ``geosedic'' radial force (proportional to the derivative of the effective potential), and back-reaction corrections become negligible. Let us state this point more precisely. Suppose $r=r_{\rm end}$ is the radius at which the particle leaves the attractor (for concreteness, this may be chosen as the radius of the last turning point along the attractor). Note that the radial acceleration due to the geodesic potential is $\propto (r-r_{\rm end})$, with a coefficient of order unity [since the second derivative of the effective potential at $r_{\rm end}$ is of $O(\eps^0)$]. Since $r_{\rm end}-r_{\rm eh}=O(\eps)$, the magnitude of the geodesic radial acceleration is of $O(\eps)$ throughout much of the plunge. This should be compared with the magnitude of the radial self-acceleration, which is of $O(\eta\eps)$.\footnote{This can be seen in one of two ways. First, by noting that the shift in the radial location of the unstable equilibrium due to the GSF is of $O(\eps\eta)$ [see the discussion below Eq.\ (\ref{deltaLcons1})]; and, second, by using the regularity argument presented below Eq.\ (\ref{DeltacalEr}) to show that $F^r$ must be $\propto\eps$ (in addition to being $\propto\eta^2$).} 
We may {\it define} the onset of plunge as the point where the geodesic acceleration takes over from the GSF in controlling the motion; this happens near a radius $r=r_{\rm plng}$ satisfying $r_{\rm plng}-r_{\rm end}=O(\eta\eps)$. For $r\lesssim r_{\rm plng}$, the motion is governed by the geodesic equation of motion (\ref{rdot}) to a good approximation.

We wish to bound the magnitude of $\calW^+_{\rm (end)}$ sufficiently well to show that it contributes negligibly in Eq.\ (\ref{OSGSF2}). Let us first consider the contribution to $\calW^+_{\rm (end)}$ from the pre-plunge orbital segment $r_{\rm plng}\leq r\leq r_{\rm end}$. The proper-time interval along this segment is $\Delta\tau\sim (r_{\rm end}-r_{\rm plng})/\dot{r}=O(\eta^0\eps^0)$ (at most), since $\dot{r}=O(\eta\eps)$ along the attractor, where the evolution is driven by radiation reaction. Hence, the change experienced by the specific energy and angular momentum along this segment is $\Delta E\simeq\Omega\Delta L=O(\eta)$. Recalling $\Omega=1/2+O(\eps)$, this gives $\Delta W=O(\eps\eta)$ and thus a contribution of $O(\eps\eta^2)$ to $\calW^+_{\rm (end)}$, negligible compared to the $O(\eta^2)$ terms in Eq.\ (\ref{OSGSF2}).

Next consider the contribution to $\calW^+_{\rm (end)}$ from the plunge segment $r_{\rm eh}\leq r< r_{\rm plng}$. Using the geodesic equation of motion (\ref{rdot}), one finds that the proper-time interval along that segment is $\propto \ln[(r_{\rm end}-r_{\rm eh})/(r_{\rm end}-r_{\rm plng})]\sim\ln\eta$. The corresponding change in specific energy is $\Delta E=O(\eta\ln\eta)$. Since the radial velocity remains small, $\dot{r}=O(\eps)$, throughout the entire plunge, we have $\Delta L\simeq\Omega^{-1}\Delta E$, giving $\Delta W=O(\eps\eta\log\eta)$ and a contribution of $O(\eps\eta^2\log\eta)$ to $\calW^+_{\rm (end)}$. Once again, this is negligible compared to the $O(\eta^2)$ terms in Eq.\ (\ref{OSGSF2}), assuming only $\eps\ll 1/|\ln\eta|$.

We conclude that, whether the plunge from the attractor occurs at the ISCO or earlier, the term $\calW^+_{\rm (end)}$ in Eq.\ (\ref{OSGSF2}) is always sub-dominant and negligible for $\eps,\eta\ll 1$. Within our approximation, the energy the particle carries with it as it crosses the horizon is the energy with which it has left the attractor.

\bibliography{biblio}

\begin{thebibliography}{56}%
\makeatletter
\providecommand \@ifxundefined [1]{%
 \@ifx{#1\undefined}
}%
\providecommand \@ifnum [1]{%
 \ifnum #1\expandafter \@firstoftwo
 \else \expandafter \@secondoftwo
 \fi
}%
\providecommand \@ifx [1]{%
 \ifx #1\expandafter \@firstoftwo
 \else \expandafter \@secondoftwo
 \fi
}%
\providecommand \natexlab [1]{#1}%
\providecommand \enquote  [1]{``#1''}%
\providecommand \bibnamefont  [1]{#1}%
\providecommand \bibfnamefont [1]{#1}%
\providecommand \citenamefont [1]{#1}%
\providecommand \href@noop [0]{\@secondoftwo}%
\providecommand \href [0]{\begingroup \@sanitize@url \@href}%
\providecommand \@href[1]{\@@startlink{#1}\@@href}%
\providecommand \@@href[1]{\endgroup#1\@@endlink}%
\providecommand \@sanitize@url [0]{\catcode `\\12\catcode `\$12\catcode
  `\&12\catcode `\#12\catcode `\^12\catcode `\_12\catcode `\%12\relax}%
\providecommand \@@startlink[1]{}%
\providecommand \@@endlink[0]{}%
\providecommand \url  [0]{\begingroup\@sanitize@url \@url }%
\providecommand \@url [1]{\endgroup\@href {#1}{\urlprefix }}%
\providecommand \urlprefix  [0]{URL }%
\providecommand \Eprint [0]{\href }%
\providecommand \doibase [0]{http://dx.doi.org/}%
\providecommand \selectlanguage [0]{\@gobble}%
\providecommand \bibinfo  [0]{\@secondoftwo}%
\providecommand \bibfield  [0]{\@secondoftwo}%
\providecommand \translation [1]{[#1]}%
\providecommand \BibitemOpen [0]{}%
\providecommand \bibitemStop [0]{}%
\providecommand \bibitemNoStop [0]{.\EOS\space}%
\providecommand \EOS [0]{\spacefactor3000\relax}%
\providecommand \BibitemShut  [1]{\csname bibitem#1\endcsname}%
\let\auto@bib@innerbib\@empty
\bibitem [{\citenamefont {Penrose}(1969)}]{wcc}%
  \BibitemOpen
  \bibfield  {author} {\bibinfo {author} {\bibfnamefont {R.}~\bibnamefont
  {Penrose}},\ }\href@noop {} {\bibfield  {journal} {\bibinfo  {journal}
  {Rivista del Nuovo Cimento}\ }\textbf {\bibinfo {volume} {1}},\ \bibinfo
  {pages} {242} (\bibinfo {year} {1969})}\BibitemShut {NoStop}%
\bibitem [{\citenamefont {Choptuik}(1993)}]{chop}%
  \BibitemOpen
  \bibfield  {author} {\bibinfo {author} {\bibfnamefont {M.~W.}\ \bibnamefont
  {Choptuik}},\ }\href@noop {} {\bibfield  {journal} {\bibinfo  {journal}
  {Phys. Rev. Lett}\ }\textbf {\bibinfo {volume} {70}},\ \bibinfo {pages} {9}
  (\bibinfo {year} {1993})}\BibitemShut {NoStop}%
\bibitem [{\citenamefont {Wald}()}]{waldrev}%
  \BibitemOpen
  \bibfield  {author} {\bibinfo {author} {\bibfnamefont {R.}~\bibnamefont
  {Wald}},\ }\href@noop {} {}\bibinfo {howpublished}
  {arXiv:gr-qc/9710068}\BibitemShut {NoStop}%
\bibitem [{\citenamefont {Wald}(1974)}]{wald}%
  \BibitemOpen
  \bibfield  {author} {\bibinfo {author} {\bibfnamefont {R.}~\bibnamefont
  {Wald}},\ }\href@noop {} {\bibfield  {journal} {\bibinfo  {journal} {Ann.
  Phys.}\ }\textbf {\bibinfo {volume} {82}},\ \bibinfo {pages} {548} (\bibinfo
  {year} {1974})}\BibitemShut {NoStop}%
\bibitem [{\citenamefont {Hubeny}(1999)}]{hub}%
  \BibitemOpen
  \bibfield  {author} {\bibinfo {author} {\bibfnamefont {V.~E.}\ \bibnamefont
  {Hubeny}},\ }\href@noop {} {\bibfield  {journal} {\bibinfo  {journal} {Phys.
  Rev.~D}\ }\textbf {\bibinfo {volume} {59}},\ \bibinfo {pages} {064013}
  (\bibinfo {year} {1999})}\BibitemShut {NoStop}%
\bibitem [{\citenamefont {Jacobson}\ and\ \citenamefont {Sotiriou}(2009)}]{js}%
  \BibitemOpen
  \bibfield  {author} {\bibinfo {author} {\bibfnamefont {T.}~\bibnamefont
  {Jacobson}}\ and\ \bibinfo {author} {\bibfnamefont {T.~P.}\ \bibnamefont
  {Sotiriou}},\ }\href@noop {} {\bibfield  {journal} {\bibinfo  {journal}
  {Phys. Rev.~Lett.}\ }\textbf {\bibinfo {volume} {103}},\ \bibinfo {pages}
  {141101} (\bibinfo {year} {2009})}\BibitemShut {NoStop}%
\bibitem [{\citenamefont {Saa}\ and\ \citenamefont {Santarelli}(2011)}]{Saa}%
  \BibitemOpen
  \bibfield  {author} {\bibinfo {author} {\bibfnamefont {A.}~\bibnamefont
  {Saa}}\ and\ \bibinfo {author} {\bibfnamefont {R.}~\bibnamefont
  {Santarelli}},\ }\href@noop {} {\bibfield  {journal} {\bibinfo  {journal}
  {Phys. Rev. D}\ }\textbf {\bibinfo {volume} {84}},\ \bibinfo {pages} {027501}
  (\bibinfo {year} {2011})}\BibitemShut {NoStop}%
\bibitem [{\citenamefont {Isoyama}\ \emph {et~al.}(2011)\citenamefont
  {Isoyama}, \citenamefont {Sago},\ and\ \citenamefont {Tanaka}}]{soich}%
  \BibitemOpen
  \bibfield  {author} {\bibinfo {author} {\bibfnamefont {S.}~\bibnamefont
  {Isoyama}}, \bibinfo {author} {\bibfnamefont {N.}~\bibnamefont {Sago}}, \
  and\ \bibinfo {author} {\bibfnamefont {T.}~\bibnamefont {Tanaka}},\
  }\href@noop {} {\bibfield  {journal} {\bibinfo  {journal} {Phys. Rev.~D}\
  }\textbf {\bibinfo {volume} {84}},\ \bibinfo {pages} {124024} (\bibinfo
  {year} {2011})}\BibitemShut {NoStop}%
\bibitem [{\citenamefont {Zimmerman}\ \emph {et~al.}(2013)\citenamefont
  {Zimmerman}, \citenamefont {Vega},\ and\ \citenamefont {Poisson}}]{zimm}%
  \BibitemOpen
  \bibfield  {author} {\bibinfo {author} {\bibfnamefont {P.}~\bibnamefont
  {Zimmerman}}, \bibinfo {author} {\bibfnamefont {I.}~\bibnamefont {Vega}}, \
  and\ \bibinfo {author} {\bibfnamefont {E.}~\bibnamefont {Poisson}},\
  }\href@noop {} {\bibfield  {journal} {\bibinfo  {journal} {Phys. Rev.~D}\
  }\textbf {\bibinfo {volume} {87}},\ \bibinfo {pages} {041501} (\bibinfo
  {year} {2013})}\BibitemShut {NoStop}%
\bibitem [{\citenamefont {Zimmerman}\ and\ \citenamefont
  {Poisson}(2014)}]{zimmerman}%
  \BibitemOpen
  \bibfield  {author} {\bibinfo {author} {\bibfnamefont {P.}~\bibnamefont
  {Zimmerman}}\ and\ \bibinfo {author} {\bibfnamefont {E.}~\bibnamefont
  {Poisson}},\ }\href@noop {} {\bibfield  {journal} {\bibinfo  {journal} {Phys.
  Rev. D}\ }\textbf {\bibinfo {volume} {90}},\ \bibinfo {pages} {084030}
  (\bibinfo {year} {2014})}\BibitemShut {NoStop}%
\bibitem [{\citenamefont {Linz}\ \emph {et~al.}(2014)\citenamefont {Linz},
  \citenamefont {Friedman},\ and\ \citenamefont {Wiseman}}]{tlinz}%
  \BibitemOpen
  \bibfield  {author} {\bibinfo {author} {\bibfnamefont {T.~M.}\ \bibnamefont
  {Linz}}, \bibinfo {author} {\bibfnamefont {J.~L.}\ \bibnamefont {Friedman}},
  \ and\ \bibinfo {author} {\bibfnamefont {A.~G.}\ \bibnamefont {Wiseman}},\
  }\href@noop {} {\bibfield  {journal} {\bibinfo  {journal} {Phys. Rev. D}\
  }\textbf {\bibinfo {volume} {90}},\ \bibinfo {pages} {084031} (\bibinfo
  {year} {2014})}\BibitemShut {NoStop}%
\bibitem [{\citenamefont {Barausse}\ \emph {et~al.}(2010)\citenamefont
  {Barausse}, \citenamefont {Cardoso},\ and\ \citenamefont {Khanna}}]{bck1}%
  \BibitemOpen
  \bibfield  {author} {\bibinfo {author} {\bibfnamefont {E.}~\bibnamefont
  {Barausse}}, \bibinfo {author} {\bibfnamefont {V.}~\bibnamefont {Cardoso}}, \
  and\ \bibinfo {author} {\bibfnamefont {G.}~\bibnamefont {Khanna}},\
  }\href@noop {} {\bibfield  {journal} {\bibinfo  {journal} {Phys. Rev.~Lett.}\
  }\textbf {\bibinfo {volume} {105}},\ \bibinfo {pages} {261102} (\bibinfo
  {year} {2010})}\BibitemShut {NoStop}%
\bibitem [{\citenamefont {Barausse}\ \emph {et~al.}(2011)\citenamefont
  {Barausse}, \citenamefont {Cardoso},\ and\ \citenamefont {Khanna}}]{bck2}%
  \BibitemOpen
  \bibfield  {author} {\bibinfo {author} {\bibfnamefont {E.}~\bibnamefont
  {Barausse}}, \bibinfo {author} {\bibfnamefont {V.}~\bibnamefont {Cardoso}}, \
  and\ \bibinfo {author} {\bibfnamefont {G.}~\bibnamefont {Khanna}},\
  }\href@noop {} {\bibfield  {journal} {\bibinfo  {journal} {Phys. Rev.~D}\
  }\textbf {\bibinfo {volume} {84}},\ \bibinfo {pages} {104006} (\bibinfo
  {year} {2011})}\BibitemShut {NoStop}%
\bibitem [{\citenamefont {Jacobson}()}]{Jacobson}%
  \BibitemOpen
  \bibfield  {author} {\bibinfo {author} {\bibfnamefont {T.}~\bibnamefont
  {Jacobson}},\ }\href@noop {} {}\bibinfo {howpublished}
  {arXiv:1107.5081v2}\BibitemShut {NoStop}%
\bibitem [{\citenamefont {Hod}(2002)}]{hod}%
  \BibitemOpen
  \bibfield  {author} {\bibinfo {author} {\bibfnamefont {S.}~\bibnamefont
  {Hod}},\ }\href@noop {} {\bibfield  {journal} {\bibinfo  {journal} {Phys.
  Rev.~D}\ }\textbf {\bibinfo {volume} {66}},\ \bibinfo {pages} {024016}
  (\bibinfo {year} {2002})}\BibitemShut {NoStop}%
\bibitem [{\citenamefont {Isoyama}\ \emph {et~al.}(2014)\citenamefont
  {Isoyama}, \citenamefont {Barack}, \citenamefont {Dolan}, \citenamefont
  {Le~Tiec}, \citenamefont {Nakano}, \citenamefont {Shah}, \citenamefont
  {Tanaka},\ and\ \citenamefont {Warburton}}]{isoyama14}%
  \BibitemOpen
  \bibfield  {author} {\bibinfo {author} {\bibfnamefont {S.}~\bibnamefont
  {Isoyama}}, \bibinfo {author} {\bibfnamefont {L.}~\bibnamefont {Barack}},
  \bibinfo {author} {\bibfnamefont {S.~R.}\ \bibnamefont {Dolan}}, \bibinfo
  {author} {\bibfnamefont {A.}~\bibnamefont {Le~Tiec}}, \bibinfo {author}
  {\bibfnamefont {H.}~\bibnamefont {Nakano}}, \bibinfo {author} {\bibfnamefont
  {A.~G.}\ \bibnamefont {Shah}}, \bibinfo {author} {\bibfnamefont
  {T.}~\bibnamefont {Tanaka}}, \ and\ \bibinfo {author} {\bibfnamefont
  {N.}~\bibnamefont {Warburton}},\ }\href@noop {} {\bibfield  {journal}
  {\bibinfo  {journal} {Phys. Rev. Lett.}\ }\textbf {\bibinfo {volume} {113}},\
  \bibinfo {pages} {161101} (\bibinfo {year} {2014})}\BibitemShut {NoStop}%
\bibitem [{\citenamefont {Levin}\ and\ \citenamefont
  {Perez-Giz}(2009)}]{Levin}%
  \BibitemOpen
  \bibfield  {author} {\bibinfo {author} {\bibfnamefont {J.}~\bibnamefont
  {Levin}}\ and\ \bibinfo {author} {\bibfnamefont {G.}~\bibnamefont
  {Perez-Giz}},\ }\href@noop {} {\bibfield  {journal} {\bibinfo  {journal}
  {Phys. Rev. D}\ }\textbf {\bibinfo {volume} {79}},\ \bibinfo {pages} {124013}
  (\bibinfo {year} {2009})}\BibitemShut {NoStop}%
\bibitem [{\citenamefont {Glampedakis}\ and\ \citenamefont
  {Kennefick}(2002)}]{glamp}%
  \BibitemOpen
  \bibfield  {author} {\bibinfo {author} {\bibfnamefont {K.}~\bibnamefont
  {Glampedakis}}\ and\ \bibinfo {author} {\bibfnamefont {D.}~\bibnamefont
  {Kennefick}},\ }\href@noop {} {\bibfield  {journal} {\bibinfo  {journal}
  {Phys. Rev.~D}\ }\textbf {\bibinfo {volume} {66}},\ \bibinfo {pages} {044002}
  (\bibinfo {year} {2002})}\BibitemShut {NoStop}%
\bibitem [{\citenamefont {Bardeen}\ \emph {et~al.}(1972)\citenamefont
  {Bardeen}, \citenamefont {Press},\ and\ \citenamefont {Teukolsky}}]{bard}%
  \BibitemOpen
  \bibfield  {author} {\bibinfo {author} {\bibfnamefont {J.~M.}\ \bibnamefont
  {Bardeen}}, \bibinfo {author} {\bibfnamefont {W.~H.}\ \bibnamefont {Press}},
  \ and\ \bibinfo {author} {\bibfnamefont {S.~A.}\ \bibnamefont {Teukolsky}},\
  }\href@noop {} {\bibfield  {journal} {\bibinfo  {journal} {Astrophys. J.}\
  }\textbf {\bibinfo {volume} {178}},\ \bibinfo {pages} {347} (\bibinfo {year}
  {1972})}\BibitemShut {NoStop}%
\bibitem [{\citenamefont {Gralla}\ and\ \citenamefont {Wald}(2008)}]{Gralla}%
  \BibitemOpen
  \bibfield  {author} {\bibinfo {author} {\bibfnamefont {S.~E.}\ \bibnamefont
  {Gralla}}\ and\ \bibinfo {author} {\bibfnamefont {R.~M.}\ \bibnamefont
  {Wald}},\ }\href@noop {} {\bibfield  {journal} {\bibinfo  {journal} {Class.
  Quant. Grav.}\ }\textbf {\bibinfo {volume} {25}},\ \bibinfo {pages} {205009}
  (\bibinfo {year} {2008})}\BibitemShut {NoStop}%
\bibitem [{\citenamefont {Pound}(2010)}]{Pound:2009sm}%
  \BibitemOpen
  \bibfield  {author} {\bibinfo {author} {\bibfnamefont {A.}~\bibnamefont
  {Pound}},\ }\href@noop {} {\bibfield  {journal} {\bibinfo  {journal} {Phys.
  Rev. D}\ }\textbf {\bibinfo {volume} {81}},\ \bibinfo {pages} {024023}
  (\bibinfo {year} {2010})}\BibitemShut {NoStop}%
\bibitem [{\citenamefont {Harte}(2012)}]{Harte}%
  \BibitemOpen
  \bibfield  {author} {\bibinfo {author} {\bibfnamefont {A.~I.}\ \bibnamefont
  {Harte}},\ }\href@noop {} {\bibfield  {journal} {\bibinfo  {journal} {Class.
  Quant. Grav.}\ }\textbf {\bibinfo {volume} {29}},\ \bibinfo {pages} {055012}
  (\bibinfo {year} {2012})}\BibitemShut {NoStop}%
\bibitem [{\citenamefont {Poisson}\ \emph {et~al.}(2011)\citenamefont
  {Poisson}, \citenamefont {Pound},\ and\ \citenamefont {Vega}}]{Poisson}%
  \BibitemOpen
  \bibfield  {author} {\bibinfo {author} {\bibfnamefont {E.}~\bibnamefont
  {Poisson}}, \bibinfo {author} {\bibfnamefont {A.}~\bibnamefont {Pound}}, \
  and\ \bibinfo {author} {\bibfnamefont {I.}~\bibnamefont {Vega}},\ }\href@noop
  {} {\bibfield  {journal} {\bibinfo  {journal} {Living Rev.\ Rel.}\ }\textbf
  {\bibinfo {volume} {14}},\ \bibinfo {pages} {1} (\bibinfo {year}
  {2011})}\BibitemShut {NoStop}%
\bibitem [{\citenamefont {Harte}()}]{Harte:2014wya}%
  \BibitemOpen
  \bibfield  {author} {\bibinfo {author} {\bibfnamefont {A.~I.}\ \bibnamefont
  {Harte}},\ }\href@noop {} {\enquote {\bibinfo {title} {Motion in classical
  field theories and the foundations of the self-force problem},}\ }\bibinfo
  {howpublished} {arXiv:1405.5077 [gr-qc]}\BibitemShut {NoStop}%
\bibitem [{\citenamefont {Barack}(2009)}]{barack}%
  \BibitemOpen
  \bibfield  {author} {\bibinfo {author} {\bibfnamefont {L.}~\bibnamefont
  {Barack}},\ }\href@noop {} {\bibfield  {journal} {\bibinfo  {journal} {Class.
  Quant. Grav.}\ }\textbf {\bibinfo {volume} {26}},\ \bibinfo {pages} {213001}
  (\bibinfo {year} {2009})}\BibitemShut {NoStop}%
\bibitem [{\citenamefont {Barack}\ and\ \citenamefont {Ori}(2001)}]{gsfandgt}%
  \BibitemOpen
  \bibfield  {author} {\bibinfo {author} {\bibfnamefont {L.}~\bibnamefont
  {Barack}}\ and\ \bibinfo {author} {\bibfnamefont {A.}~\bibnamefont {Ori}},\
  }\href@noop {} {\bibfield  {journal} {\bibinfo  {journal} {Phys. Rev. D}\
  }\textbf {\bibinfo {volume} {64}},\ \bibinfo {pages} {124003} (\bibinfo
  {year} {2001})}\BibitemShut {NoStop}%
\bibitem [{\citenamefont {Quinn}\ and\ \citenamefont {Wald}(1999)}]{quinn}%
  \BibitemOpen
  \bibfield  {author} {\bibinfo {author} {\bibfnamefont {T.~C.}\ \bibnamefont
  {Quinn}}\ and\ \bibinfo {author} {\bibfnamefont {R.~M.}\ \bibnamefont
  {Wald}},\ }\href@noop {} {\bibfield  {journal} {\bibinfo  {journal} {Phys.
  Rev.~D}\ }\textbf {\bibinfo {volume} {60}},\ \bibinfo {pages} {064009}
  (\bibinfo {year} {1999})}\BibitemShut {NoStop}%
\bibitem [{\citenamefont {Detweiler}(2008)}]{Detweiler:2008ft}%
  \BibitemOpen
  \bibfield  {author} {\bibinfo {author} {\bibfnamefont {S.}~\bibnamefont
  {Detweiler}},\ }\href@noop {} {\bibfield  {journal} {\bibinfo  {journal}
  {Phys. Rev. D}\ }\textbf {\bibinfo {volume} {77}},\ \bibinfo {pages} {124026}
  (\bibinfo {year} {2008})}\BibitemShut {NoStop}%
\bibitem [{\citenamefont {Flanagan}\ \emph {et~al.}(2014)\citenamefont
  {Flanagan}, \citenamefont {Hughes},\ and\ \citenamefont
  {Ruangsri}}]{Flanagan}%
  \BibitemOpen
  \bibfield  {author} {\bibinfo {author} {\bibfnamefont {E.~E.}\ \bibnamefont
  {Flanagan}}, \bibinfo {author} {\bibfnamefont {S.~A.}\ \bibnamefont
  {Hughes}}, \ and\ \bibinfo {author} {\bibfnamefont {U.}~\bibnamefont
  {Ruangsri}},\ }\href@noop {} {\bibfield  {journal} {\bibinfo  {journal}
  {Phys. Rev. D}\ }\textbf {\bibinfo {volume} {89}},\ \bibinfo {pages} {084028}
  (\bibinfo {year} {2014})}\BibitemShut {NoStop}%
\bibitem [{\citenamefont {Gundlach}\ \emph {et~al.}(2012)\citenamefont
  {Gundlach}, \citenamefont {Akcay}, \citenamefont {Barack},\ and\
  \citenamefont {Nagar}}]{gund}%
  \BibitemOpen
  \bibfield  {author} {\bibinfo {author} {\bibfnamefont {C.}~\bibnamefont
  {Gundlach}}, \bibinfo {author} {\bibfnamefont {S.}~\bibnamefont {Akcay}},
  \bibinfo {author} {\bibfnamefont {L.}~\bibnamefont {Barack}}, \ and\ \bibinfo
  {author} {\bibfnamefont {A.}~\bibnamefont {Nagar}},\ }\href@noop {}
  {\bibfield  {journal} {\bibinfo  {journal} {Phys. Rev.~D}\ }\textbf {\bibinfo
  {volume} {86}},\ \bibinfo {pages} {084022} (\bibinfo {year}
  {2012})}\BibitemShut {NoStop}%
\bibitem [{\citenamefont {Israel}(1986)}]{Israel}%
  \BibitemOpen
  \bibfield  {author} {\bibinfo {author} {\bibfnamefont {W.}~\bibnamefont
  {Israel}},\ }\href@noop {} {\bibfield  {journal} {\bibinfo  {journal} {Phys.
  Rev. Lett.}\ }\textbf {\bibinfo {volume} {57}},\ \bibinfo {pages} {397}
  (\bibinfo {year} {1986})}\BibitemShut {NoStop}%
\bibitem [{\citenamefont {Ori}\ and\ \citenamefont {Thorne}(2000)}]{orithor}%
  \BibitemOpen
  \bibfield  {author} {\bibinfo {author} {\bibfnamefont {A.}~\bibnamefont
  {Ori}}\ and\ \bibinfo {author} {\bibfnamefont {K.~S.}\ \bibnamefont
  {Thorne}},\ }\href@noop {} {\bibfield  {journal} {\bibinfo  {journal} {Phys.
  Rev. D}\ }\textbf {\bibinfo {volume} {62}},\ \bibinfo {pages} {124022}
  (\bibinfo {year} {2000})}\BibitemShut {NoStop}%
\bibitem [{\citenamefont {Kesden}(2011)}]{Kesden}%
  \BibitemOpen
  \bibfield  {author} {\bibinfo {author} {\bibfnamefont {M.}~\bibnamefont
  {Kesden}},\ }\href@noop {} {\bibfield  {journal} {\bibinfo  {journal} {Phys.
  Rev. D}\ }\textbf {\bibinfo {volume} {83}},\ \bibinfo {pages} {104011}
  (\bibinfo {year} {2011})}\BibitemShut {NoStop}%
\bibitem [{\citenamefont {Mino}\ and\ \citenamefont {Brink}(2008)}]{Mino}%
  \BibitemOpen
  \bibfield  {author} {\bibinfo {author} {\bibfnamefont {Y.}~\bibnamefont
  {Mino}}\ and\ \bibinfo {author} {\bibfnamefont {J.}~\bibnamefont {Brink}},\
  }\href@noop {} {\bibfield  {journal} {\bibinfo  {journal} {Phys. Rev. D}\
  }\textbf {\bibinfo {volume} {78}},\ \bibinfo {pages} {124015} (\bibinfo
  {year} {2008})}\BibitemShut {NoStop}%
\bibitem [{\citenamefont {Harada}\ and\ \citenamefont {Kimura}(2011)}]{Harada}%
  \BibitemOpen
  \bibfield  {author} {\bibinfo {author} {\bibfnamefont {T.}~\bibnamefont
  {Harada}}\ and\ \bibinfo {author} {\bibfnamefont {M.}~\bibnamefont
  {Kimura}},\ }\href@noop {} {\bibfield  {journal} {\bibinfo  {journal} {Phys.
  Rev. D}\ }\textbf {\bibinfo {volume} {84}},\ \bibinfo {pages} {124032}
  (\bibinfo {year} {2011})}\BibitemShut {NoStop}%
\bibitem [{\citenamefont {Colleoni}\ \emph {et~al.}()\citenamefont {Colleoni},
  \citenamefont {Shah}, \citenamefont {van~de Meent},\ and\ \citenamefont
  {Barack}}]{Colleoni_etal}%
  \BibitemOpen
  \bibfield  {author} {\bibinfo {author} {\bibfnamefont {M.}~\bibnamefont
  {Colleoni}}, \bibinfo {author} {\bibfnamefont {A.}~\bibnamefont {Shah}},
  \bibinfo {author} {\bibfnamefont {M.}~\bibnamefont {van~de Meent}}, \ and\
  \bibinfo {author} {\bibfnamefont {L.}~\bibnamefont {Barack}},\ }\href@noop {}
  {}\bibinfo {howpublished} {in preparation}\BibitemShut {NoStop}%
\bibitem [{\citenamefont {van~de Meent}()}]{priv_comm}%
  \BibitemOpen
  \bibfield  {author} {\bibinfo {author} {\bibfnamefont {M.}~\bibnamefont
  {van~de Meent}},\ }\href@noop {} {}\bibinfo {howpublished} {(private
  communication)}\BibitemShut {NoStop}%
\bibitem [{\citenamefont {Kennefick}(1998)}]{eomegal}%
  \BibitemOpen
  \bibfield  {author} {\bibinfo {author} {\bibfnamefont {D.}~\bibnamefont
  {Kennefick}},\ }\href@noop {} {\bibfield  {journal} {\bibinfo  {journal}
  {Phys. Rev. D}\ }\textbf {\bibinfo {volume} {58}},\ \bibinfo {pages} {064012}
  (\bibinfo {year} {1998})}\BibitemShut {NoStop}%
\bibitem [{\citenamefont {Isoyama}\ \emph {et~al.}()\citenamefont {Isoyama},
  \citenamefont {Fujita}, \citenamefont {Le~Tiec}, \citenamefont {Nakano},
  \citenamefont {Sago},\ and\ \citenamefont {Tanaka}}]{hami}%
  \BibitemOpen
  \bibfield  {author} {\bibinfo {author} {\bibfnamefont {S.}~\bibnamefont
  {Isoyama}}, \bibinfo {author} {\bibfnamefont {R.}~\bibnamefont {Fujita}},
  \bibinfo {author} {\bibfnamefont {A.}~\bibnamefont {Le~Tiec}}, \bibinfo
  {author} {\bibfnamefont {H.}~\bibnamefont {Nakano}}, \bibinfo {author}
  {\bibfnamefont {N.}~\bibnamefont {Sago}}, \ and\ \bibinfo {author}
  {\bibfnamefont {T.}~\bibnamefont {Tanaka}},\ }\href@noop {} {}\bibinfo
  {howpublished} {in preparation}\BibitemShut {NoStop}%
\bibitem [{\citenamefont {Le~Tiec}\ \emph
  {et~al.}(2012{\natexlab{a}})\citenamefont {Le~Tiec}, \citenamefont
  {Blanchet},\ and\ \citenamefont {Whiting}}]{alt1}%
  \BibitemOpen
  \bibfield  {author} {\bibinfo {author} {\bibfnamefont {A.}~\bibnamefont
  {Le~Tiec}}, \bibinfo {author} {\bibfnamefont {L.}~\bibnamefont {Blanchet}}, \
  and\ \bibinfo {author} {\bibfnamefont {B.~F.}\ \bibnamefont {Whiting}},\
  }\href@noop {} {\bibfield  {journal} {\bibinfo  {journal} {Phys. Rev.~D}\
  }\textbf {\bibinfo {volume} {85}},\ \bibinfo {pages} {064039} (\bibinfo
  {year} {2012}{\natexlab{a}})}\BibitemShut {NoStop}%
\bibitem [{\citenamefont {Blanchet}\ \emph {et~al.}(2013)\citenamefont
  {Blanchet}, \citenamefont {Buonanno},\ and\ \citenamefont {Le~Tiec}}]{alt3}%
  \BibitemOpen
  \bibfield  {author} {\bibinfo {author} {\bibfnamefont {L.}~\bibnamefont
  {Blanchet}}, \bibinfo {author} {\bibfnamefont {A.}~\bibnamefont {Buonanno}},
  \ and\ \bibinfo {author} {\bibfnamefont {A.}~\bibnamefont {Le~Tiec}},\
  }\href@noop {} {\bibfield  {journal} {\bibinfo  {journal} {Phys. Rev.~D}\
  }\textbf {\bibinfo {volume} {87}},\ \bibinfo {pages} {024030} (\bibinfo
  {year} {2013})}\BibitemShut {NoStop}%
\bibitem [{\citenamefont {Friedman}\ \emph {et~al.}(2002)\citenamefont
  {Friedman}, \citenamefont {Ury\ifmmode~\bar{u}\else \={u}\fi{}},\ and\
  \citenamefont {Shibata}}]{Friedman:2001pf}%
  \BibitemOpen
  \bibfield  {author} {\bibinfo {author} {\bibfnamefont {J.~L.}\ \bibnamefont
  {Friedman}}, \bibinfo {author} {\bibfnamefont {K.}~\bibnamefont
  {Ury\ifmmode~\bar{u}\else \={u}\fi{}}}, \ and\ \bibinfo {author}
  {\bibfnamefont {M.}~\bibnamefont {Shibata}},\ }\href@noop {} {\bibfield
  {journal} {\bibinfo  {journal} {Phys. Rev. D}\ }\textbf {\bibinfo {volume}
  {65}},\ \bibinfo {pages} {064035} (\bibinfo {year} {2002})}\BibitemShut
  {NoStop}%
\bibitem [{\citenamefont {Le~Tiec}\ \emph
  {et~al.}(2012{\natexlab{b}})\citenamefont {Le~Tiec}, \citenamefont
  {Barausse},\ and\ \citenamefont {Buonanno}}]{alt2}%
  \BibitemOpen
  \bibfield  {author} {\bibinfo {author} {\bibfnamefont {A.}~\bibnamefont
  {Le~Tiec}}, \bibinfo {author} {\bibfnamefont {E.}~\bibnamefont {Barausse}}, \
  and\ \bibinfo {author} {\bibfnamefont {A.}~\bibnamefont {Buonanno}},\
  }\href@noop {} {\bibfield  {journal} {\bibinfo  {journal} {Phys. Rev.~Lett.}\
  }\textbf {\bibinfo {volume} {108}},\ \bibinfo {pages} {131103} (\bibinfo
  {year} {2012}{\natexlab{b}})}\BibitemShut {NoStop}%
\bibitem [{\citenamefont {Gralla}\ and\ \citenamefont
  {Le~Tiec}(2013)}]{Gralla:2012dm}%
  \BibitemOpen
  \bibfield  {author} {\bibinfo {author} {\bibfnamefont {S.~E.}\ \bibnamefont
  {Gralla}}\ and\ \bibinfo {author} {\bibfnamefont {A.}~\bibnamefont
  {Le~Tiec}},\ }\href@noop {} {\bibfield  {journal} {\bibinfo  {journal} {Phys.
  Rev. D}\ }\textbf {\bibinfo {volume} {88}},\ \bibinfo {pages} {044021}
  (\bibinfo {year} {2013})}\BibitemShut {NoStop}%
\bibitem [{\citenamefont {Dolan}\ and\ \citenamefont {Barack}(2013)}]{Dolan}%
  \BibitemOpen
  \bibfield  {author} {\bibinfo {author} {\bibfnamefont {S.~R.}\ \bibnamefont
  {Dolan}}\ and\ \bibinfo {author} {\bibfnamefont {L.}~\bibnamefont {Barack}},\
  }\href@noop {} {\bibfield  {journal} {\bibinfo  {journal} {Phys. Rev. D}\
  }\textbf {\bibinfo {volume} {87}},\ \bibinfo {pages} {084066} (\bibinfo
  {year} {2013})}\BibitemShut {NoStop}%
\bibitem [{\citenamefont {Keidl}\ \emph {et~al.}(2010)\citenamefont {Keidl},
  \citenamefont {Shah}, \citenamefont {Friedman}, \citenamefont {Kim},\ and\
  \citenamefont {Price}}]{GSFradgauge}%
  \BibitemOpen
  \bibfield  {author} {\bibinfo {author} {\bibfnamefont {T.~S.}\ \bibnamefont
  {Keidl}}, \bibinfo {author} {\bibfnamefont {A.~G.}\ \bibnamefont {Shah}},
  \bibinfo {author} {\bibfnamefont {J.~L.}\ \bibnamefont {Friedman}}, \bibinfo
  {author} {\bibfnamefont {D.-H.}\ \bibnamefont {Kim}}, \ and\ \bibinfo
  {author} {\bibfnamefont {L.~R.}\ \bibnamefont {Price}},\ }\href@noop {}
  {\bibfield  {journal} {\bibinfo  {journal} {Phys. Rev. D}\ }\textbf {\bibinfo
  {volume} {82}},\ \bibinfo {pages} {124012} (\bibinfo {year}
  {2010})}\BibitemShut {NoStop}%
\bibitem [{\citenamefont {Shah}\ \emph {et~al.}(2012)\citenamefont {Shah},
  \citenamefont {Friedman},\ and\ \citenamefont {Keidl}}]{Shah:2012gu}%
  \BibitemOpen
  \bibfield  {author} {\bibinfo {author} {\bibfnamefont {A.~G.}\ \bibnamefont
  {Shah}}, \bibinfo {author} {\bibfnamefont {J.~L.}\ \bibnamefont {Friedman}},
  \ and\ \bibinfo {author} {\bibfnamefont {T.~S.}\ \bibnamefont {Keidl}},\
  }\href@noop {} {\bibfield  {journal} {\bibinfo  {journal} {Phys. Rev. D}\
  }\textbf {\bibinfo {volume} {86}},\ \bibinfo {pages} {084059} (\bibinfo
  {year} {2012})}\BibitemShut {NoStop}%
\bibitem [{\citenamefont {Pound}\ \emph {et~al.}(2014)\citenamefont {Pound},
  \citenamefont {Merlin},\ and\ \citenamefont {Barack}}]{PMB}%
  \BibitemOpen
  \bibfield  {author} {\bibinfo {author} {\bibfnamefont {A.}~\bibnamefont
  {Pound}}, \bibinfo {author} {\bibfnamefont {C.}~\bibnamefont {Merlin}}, \
  and\ \bibinfo {author} {\bibfnamefont {L.}~\bibnamefont {Barack}},\
  }\href@noop {} {\bibfield  {journal} {\bibinfo  {journal} {Phys. Rev. D}\
  }\textbf {\bibinfo {volume} {89}},\ \bibinfo {pages} {024009} (\bibinfo
  {year} {2014})}\BibitemShut {NoStop}%
\bibitem [{\citenamefont {Hughes}(2000{\natexlab{a}})}]{hughes}%
  \BibitemOpen
  \bibfield  {author} {\bibinfo {author} {\bibfnamefont {S.~A.}\ \bibnamefont
  {Hughes}},\ }\href@noop {} {\bibfield  {journal} {\bibinfo  {journal} {Phys.
  Rev. D}\ }\textbf {\bibinfo {volume} {61}},\ \bibinfo {pages} {084004}
  (\bibinfo {year} {2000}{\natexlab{a}})}\BibitemShut {NoStop}%
\bibitem [{\citenamefont {Mano}\ \emph {et~al.}(1996)\citenamefont {Mano},
  \citenamefont {Suzuki},\ and\ \citenamefont {Takasugi}}]{MST}%
  \BibitemOpen
  \bibfield  {author} {\bibinfo {author} {\bibfnamefont {S.}~\bibnamefont
  {Mano}}, \bibinfo {author} {\bibfnamefont {H.}~\bibnamefont {Suzuki}}, \ and\
  \bibinfo {author} {\bibfnamefont {E.}~\bibnamefont {Takasugi}},\ }\href@noop
  {} {\ \textbf {\bibinfo {volume} {95}},\ \bibinfo {pages} {1079} (\bibinfo
  {year} {1996})}\BibitemShut {NoStop}%
\bibitem [{\citenamefont {Akcay}\ \emph {et~al.}(2012)\citenamefont {Akcay},
  \citenamefont {Barack}, \citenamefont {Damour},\ and\ \citenamefont
  {Sago}}]{Akcay}%
  \BibitemOpen
  \bibfield  {author} {\bibinfo {author} {\bibfnamefont {S.}~\bibnamefont
  {Akcay}}, \bibinfo {author} {\bibfnamefont {L.}~\bibnamefont {Barack}},
  \bibinfo {author} {\bibfnamefont {T.}~\bibnamefont {Damour}}, \ and\ \bibinfo
  {author} {\bibfnamefont {N.}~\bibnamefont {Sago}},\ }\href@noop {} {\bibfield
   {journal} {\bibinfo  {journal} {Phys. Rev. D}\ }\textbf {\bibinfo {volume}
  {86}},\ \bibinfo {pages} {104041} (\bibinfo {year} {2012})}\BibitemShut
  {NoStop}%
\bibitem [{\citenamefont {Dolan}\ and\ \citenamefont {Barack}()}]{dandb}%
  \BibitemOpen
  \bibfield  {author} {\bibinfo {author} {\bibfnamefont {S.}~\bibnamefont
  {Dolan}}\ and\ \bibinfo {author} {\bibfnamefont {L.}~\bibnamefont {Barack}},\
  }\href@noop {} {}\bibinfo {howpublished} {in preparation}\BibitemShut
  {NoStop}%
\bibitem [{\citenamefont {Kapadia}\ \emph {et~al.}(2013)\citenamefont
  {Kapadia}, \citenamefont {Kennefick},\ and\ \citenamefont
  {Glampedakis}}]{Kapadia}%
  \BibitemOpen
  \bibfield  {author} {\bibinfo {author} {\bibfnamefont {S.~J.}\ \bibnamefont
  {Kapadia}}, \bibinfo {author} {\bibfnamefont {D.}~\bibnamefont {Kennefick}},
  \ and\ \bibinfo {author} {\bibfnamefont {K.}~\bibnamefont {Glampedakis}},\
  }\href@noop {} {\bibfield  {journal} {\bibinfo  {journal} {Phys. Rev. D}\
  }\textbf {\bibinfo {volume} {87}},\ \bibinfo {pages} {044050} (\bibinfo
  {year} {2013})}\BibitemShut {NoStop}%
\bibitem [{\citenamefont {Buonanno}\ and\ \citenamefont
  {Damour}(2000)}]{Buonanno:2000ef}%
  \BibitemOpen
  \bibfield  {author} {\bibinfo {author} {\bibfnamefont {A.}~\bibnamefont
  {Buonanno}}\ and\ \bibinfo {author} {\bibfnamefont {T.}~\bibnamefont
  {Damour}},\ }\href@noop {} {\bibfield  {journal} {\bibinfo  {journal} {Phys.
  Rev. D}\ }\textbf {\bibinfo {volume} {62}},\ \bibinfo {pages} {064015}
  (\bibinfo {year} {2000})}\BibitemShut {NoStop}%
\bibitem [{\citenamefont {Hughes}(2000{\natexlab{b}})}]{Hughes:1999bq}%
  \BibitemOpen
  \bibfield  {author} {\bibinfo {author} {\bibfnamefont {S.~A.}\ \bibnamefont
  {Hughes}},\ }\href@noop {} {\bibfield  {journal} {\bibinfo  {journal} {Phys.
  Rev. D}\ }\textbf {\bibinfo {volume} {61}},\ \bibinfo {pages} {084004}
  (\bibinfo {year} {2000}{\natexlab{b}})}\BibitemShut {NoStop}%
\bibitem [{\citenamefont {Chrzanowski}(1976)}]{Chrzanowski}%
  \BibitemOpen
  \bibfield  {author} {\bibinfo {author} {\bibfnamefont {P.~L.}\ \bibnamefont
  {Chrzanowski}},\ }\href@noop {} {\bibfield  {journal} {\bibinfo  {journal}
  {Phys. Rev. D}\ }\textbf {\bibinfo {volume} {13}},\ \bibinfo {pages} {806}
  (\bibinfo {year} {1976})}\BibitemShut {NoStop}%
\end{thebibliography}%

\end{document}